\documentstyle[aps,multicol,epsf]{revtex}
\epsfverbosetrue

\makeatletter
\renewcommand{\theequation}{\arabic{section}.\arabic{equation}}
\@addtoreset{equation}{section}
\makeatother

\newcounter{figurenumber}
\renewenvironment{figure}%
  {\begin{enumerate}\stepcounter{figurenumber}}%
  {\end{enumerate}}
\newcommand{\figcaption}[1]%
{\setcounter{enumi}{\value{figurenumber}}\item FIG.~\arabic{enumi}. #1
\vskip 10pt}

\newfont{\bg}{cmr10 scaled\magstep 4}
\newcommand{\bigzerol}{\raisebox{-0.7ex}{\bg 0}}
\newcommand{\bigzerou}{\raisebox{0.7ex}{\bg 0}}
\font\mibsmall=cmmib10 scaled 900
\font\mibscriptsize=cmmib10 scaled 700
\font\bfscriptsize=cmbx7

\begin{document}

\draft

\title{Vortex Pinning and Non-Hermitian Quantum Mechanics}
\author{Naomichi Hatano\thanks{Present Address:
Theoretical Divisions, Los Alamos National Laboratory,
Los Alamos, New Mexico 87545}}
\address{Lyman Laboratory of Physics, Harvard University,
Cambridge, Massachusetts 02138\\
and
Department of Physics, University of Tokyo,
Hongo, Bunkyo, Tokyo 113, Japan}
\author{David R. Nelson}
\address{Lyman Laboratory of Physics, Harvard University,
Cambridge, Massachusetts 02138}
\date{Submitted May 15, 1997}
\maketitle
\begin{abstract}
A delocalization phenomenon is studied in a class of non-Hermitian random 
quantum-mechanical problems.
Delocalization arises in response to a sufficiently large constant 
imaginary vector potential.
The transition is related to depinning of flux lines from 
extended defects in type-II superconductors
subject to a tilted external magnetic field.
The physical meaning of the complex eigenvalues and currents of the 
non-Hermitian system is elucidated in terms of properties of tilted 
vortex lines.
The singular behavior of the penetration length describing stretched 
exponential screening of a perpendicular magnetic field  
(transverse Meissner effect), the surface transverse magnetization, 
and the trapping length are determined near the flux-line depinning point.
\end{abstract}

\begin{multicols}{2}

\section{Introduction}
\label{sec-intro}

Although Hamiltonians must be Hermitian in conventional
quantum mechanics, non-Hermitian operators do appear in other physical 
contexts:
the time evolution of non-Hermitian Liouville operators can describe 
various nonequilibrium processes;\cite{Kadanoff68,Fogedby95,Kim95}
the transfer matrix of two-dimensional asymmetric vertex models 
leads to non-Hermitian Hamiltonians for quantum spin 
chains.\cite{McCoy68}

In the present paper, we investigate localization phenomena in 
an especially simple class of {\em random} non-Hermitian Hamiltonians.
Although non-Hermitian, our problem is sufficiently close to 
conventional quantum mechanics that it will be convenient to use a 
quantum language to describe the results.
Specifically, we show that a delocalization transition occurs 
(even in one and two dimensions)
in the following one-body Hamiltonians in $d$ dimensions:
first, the Hamiltonian in continuum space,
\begin{equation}\label{12020}
{\cal H}
\equiv\frac{(\mbox{\boldmath $p$}+i\mbox{\boldmath $g$})^2}{2m}
+V(\mbox{\boldmath $x$}),
\end{equation}
where $\mbox{\boldmath $p$}=(\hbar/i)\partial/\partial\mbox{\boldmath $x$}$
is the momentum operator and $V(\mbox{\boldmath $x$})$ is a random 
potential; 
second, the second-quantized lattice Hamiltonian, namely the 
non-Hermitian Anderson model on a hypercubic lattice,
\begin{eqnarray}\label{41010}
{\cal H}&\equiv&-\frac{t}{2}
\sum_{\mbox{\mibscriptsize x}}
\sum_{\nu=1}^d \left(
e^{\mbox{\mibscriptsize g}\cdot\mbox{\mibscriptsize e}_\nu/\hbar}
b^\dagger_{\mbox{\mibscriptsize x}+\mbox{\mibscriptsize e}_\nu}
b_{\mbox{\mibscriptsize x}}
+e^{-\mbox{\mibscriptsize g}\cdot\mbox{\mibscriptsize e}_\nu/\hbar}
b^\dagger_{\mbox{\mibscriptsize x}}
b_{\mbox{\mibscriptsize x}+\mbox{\mibscriptsize e}_\nu}
\right)
\nonumber\\
&&+\sum_{\mbox{\mibscriptsize x}}
V_{\mbox{\mibscriptsize x}} b^\dagger_{\mbox{\mibscriptsize x}}
b_{\mbox{\mibscriptsize x}},
\end{eqnarray}
where the vectors $\{\mbox{\boldmath $e$}_\nu\}$
are the unit lattice vectors, 
the $\{b^\dagger_{\mbox{\mibscriptsize x}},b_{\mbox{\mibscriptsize x}}\}$
are (boson) creation and annihilation operators,
$V_{\mbox{\mibscriptsize x}}$ is a random potential.
In both of the Hamiltonians, $\mbox{\boldmath $g$}$ is a
non-Hermitian external field.
Although the bulk of our discussion will concentrate on the properties 
of the single-particle Hamiltonian~(\ref{12020}) and~(\ref{41010}), 
many of our results will be relevant for {\em interacting} 
many-body boson problems, provided that we forbid double 
occupancy of eigenstates in the localized regime.\cite{Hatano96}
Interaction effects in both the localized and delocalized phases will 
be discussed in Sec.~\ref{sec-interaction}.

We can regard the non-Hermitian field as an imaginary vector potential.
Models with a real gauge field $\mbox{\boldmath $A$}$ would be 
written in the Hermitian forms
\begin{equation}\label{41032}
{\cal H}'=\frac{(\mbox{\boldmath $p$}-e\mbox{\boldmath $A$})^2}{2m}
+V(\mbox{\boldmath $x$})
\end{equation}
and 
\begin{eqnarray}\label{41034}
{\cal H}'&=&-\frac{t}{2}
\sum_{\mbox{\mibscriptsize x}}
\sum_{\nu=1}^d \left(
e^{ie\mbox{\mibscriptsize A}\cdot\mbox{\mibscriptsize e}_\nu/\hbar}
b^\dagger_{\mbox{\mibscriptsize x}+\mbox{\mibscriptsize e}_\nu}
b_{\mbox{\mibscriptsize x}}
+e^{-ie\mbox{\mibscriptsize A}\cdot\mbox{\mibscriptsize e}_\nu/\hbar}
b^\dagger_{\mbox{\mibscriptsize x}}
b_{\mbox{\mibscriptsize x}+\mbox{\mibscriptsize e}_\nu}
\right)
\nonumber\\
&&+\sum_{\mbox{\mibscriptsize x}}
V_{\mbox{\mibscriptsize x}} b^\dagger_{\mbox{\mibscriptsize x}}
b_{\mbox{\mibscriptsize x}}.
\end{eqnarray}
In two dimensions with spatially varying 
$\mbox{\boldmath $A$}=\mbox{\boldmath $A$}(\mbox{\boldmath $x$})$, 
these Hamiltonians describe the quantum Hall system, 
where some of the localized states of the 
$\mbox{\boldmath $A$}=\mbox{\bf 0}$ case 
are delocalized in the presence of the gauge field.\cite{Prange90}
We obtain the non-Hermitian Hamiltonians~(\ref{12020}) 
and~(\ref{41010}) by replacing 
$-e\mbox{\boldmath $A$}(\mbox{\boldmath $x$})$ with a constant,
$i\mbox{\boldmath $g$}$.
In this non-Hermitian case, we show that {\em all} eigenstates can be
delocalized (even in one dimension) for large $\mbox{\boldmath $g$}$.

\epsfxsize=3.375in
\epsfbox{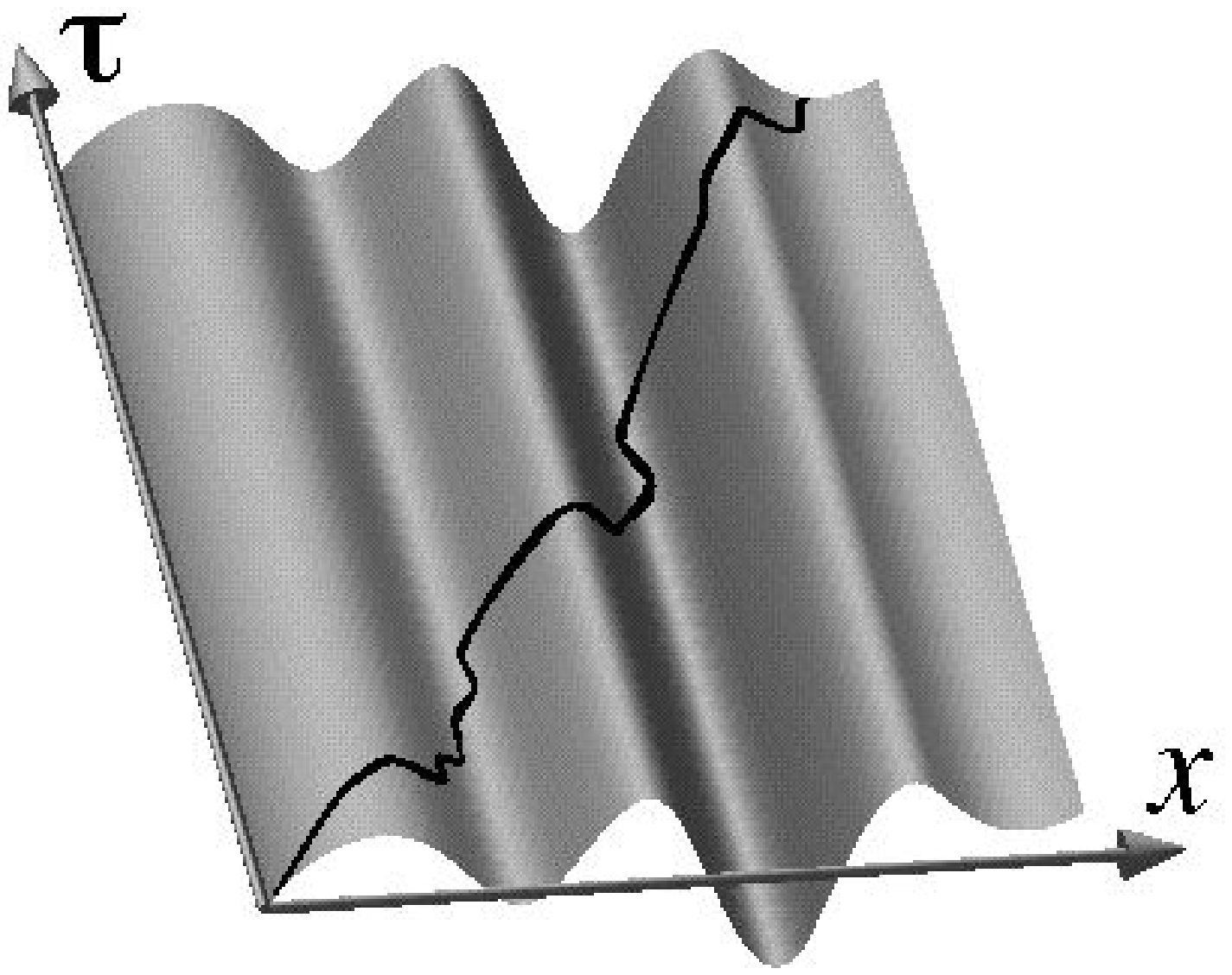}
\begin{figure}
\figcaption{
A situation described by the classical 
Hamiltonian~(\protect\ref{00000}) in $(1+1)$ dimensions.
The wavy line indicates a tilted string with single-valued trajectory
$x(\tau)$ subject to thermal fluctuations.
}
\label{fig-classical}
\end{figure}

This problem has direct physical relevance when we map the non-Hermitian 
quantum system in $d$ dimensions to a problem of classical 
equilibrium statistical mechanics in $(d+1)$ dimensions.
Consider a matrix element of the imaginary-time-evolution operator,
\begin{equation}\label{11030}
{\cal Z}=\left\langle\psi^f\left|
e^{-L_{\tau}{\cal H}/\hbar}
\right|\psi^i\right\rangle,
\end{equation}
where the Hamiltonian ${\cal H}$ is given by~(\ref{12020}).
The bra and ket vectors are the initial and final states, respectively. 
Using the standard path-integral scheme, we can express 
the above matrix element as the partition function 
\begin{equation}\label{11020-1}
{\cal Z}=\int{\cal D}\mbox{\boldmath $x$}
e^{-{E_{\rm cl}[\mbox{\mibscriptsize x}(\tau)]}/{\hbar}},
\end{equation}
where $\int{\cal D}\mbox{\boldmath $x$}$ denotes summation over all possible
world-line configurations $\mbox{\boldmath $x$}(\tau)$, and 
$E_{\rm cl}$ is the energy of a classical 
elastic string in $(d+1)$ dimensions, given by
\begin{equation}\label{00000}
E_{\rm cl}[\mbox{\boldmath $x$}(\tau)]
\equiv\int_0^{L_{\tau}}d\tau
\left[
\frac{m}{2}
\left(\frac{d\mbox{\boldmath $x$}}{d\tau}\right)^2
-\mbox{\boldmath $g$}\cdot\frac{d\mbox{\boldmath $x$}}{d\tau}
+V(\mbox{\boldmath $x$})
\right].
\end{equation}
The energy $E_{\rm cl}$ is of course the imaginary-time action for 
the equivalent quantum problem.
The Planck parameter $\hbar$ is interpreted as the
temperature of the classical system.

The above classical system expresses the following physics; see 
Fig.~\ref{fig-classical}.
Consider a thermally fluctuating string put on a {\em random} 
washboard potential.
The first term of the Eq.~(\ref{00000}) is optimized 
when the string is tilted from the $\tau$ direction by the 
angle $\theta\equiv\tan^{-1}(g/m)$.\linebreak

\epsfxsize=3.375in
\epsfbox{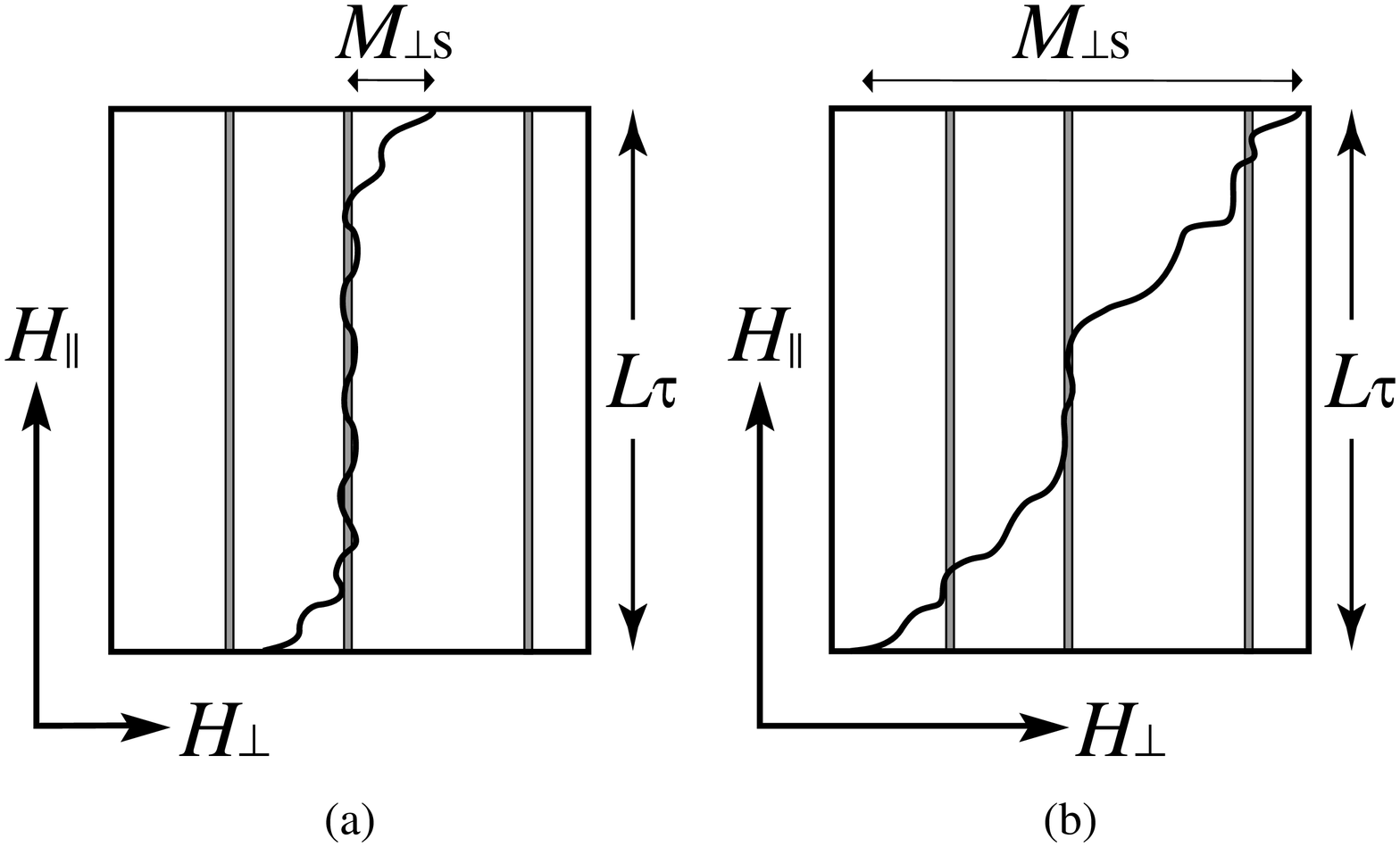}
\begin{figure}
\figcaption{
Flux-line depinning from a columnar defect in a 
superconductor: (a) the bulk part of the flux line is pinned by the 
defect, and the transverse magnetization has contribution only from
the {\em surface} transverse magnetization $M_{\perp s}$; 
(b) the flux line is depinned, and the \lq\lq surface'' transverse
magnetization now has a macroscopic value $M_{\perp s}\propto L_{\tau}$,
where $L_{\tau}$ is the thickness of the sample.
}
\label{fig-supercon}
\end{figure}
On the other hand, the potential tries to make the string
parallel to the $\tau$ direction, to take advantage of
particularly deep valleys in the random potential.
We might thus expect a depinning transition of the string from the 
random washboard potential with increasing $g$.

We suspect that there are many physical realizations of the above 
statistical-mechanical problem.
For an application to population biology, see Ref.~\cite{Shnerb97}.
The present study is motivated particularly by a depinning 
transition of magnetic flux lines in type-II superconductors
with quenched disorder.
It has been recognized that impurities and defects 
play an essential role in the mixed phase of type-II superconductors.
Flux lines, subject to electromagnetic forces due to applied currents,
can move and dissipate energy, thus
destroying the superconductivity, 
unless they are pinned by impurities or defects.
In particular, extended defects such as columnar defects and twin 
boundaries pin flux lines parallel to them efficiently, 
and thereby expand the region of zero resistance dramatically; 
see Refs.~\cite{Civale91}-\cite{Nelson93a}.

The situation described by the classical Hamiltonian~(\ref{00000}) emerges
when the external magnetic field generating the flux lines is not 
parallel to the extended defects; see Fig.~\ref{fig-supercon}.
We may expect that the bulk part of the flux line remains pinned by 
the extended defects when the transverse component of the magnetic 
field, $\mbox{\boldmath $H$}_{\perp}$, is weak.
This phenomenon is called {\it transverse Meissner effect};\cite{Nelson93a}
the system exhibits perfect bulk diamagnetism against 
$\mbox{\boldmath $H$}_{\perp}$ for a slightly tilted magnetic field.
When we increase $\mbox{\boldmath $H$}_{\perp}$, 
{\it i.e.}\ tilt the magnetic field further 
relative to the extended defects, a depinning transition takes place 
at a certain strength of the transverse field, namely $H_{\perp c}$.
We note that randomness in the potential is essential for the 
transverse Meissner effect to work at finite\linebreak

\epsfxsize=3.375in
\epsfbox{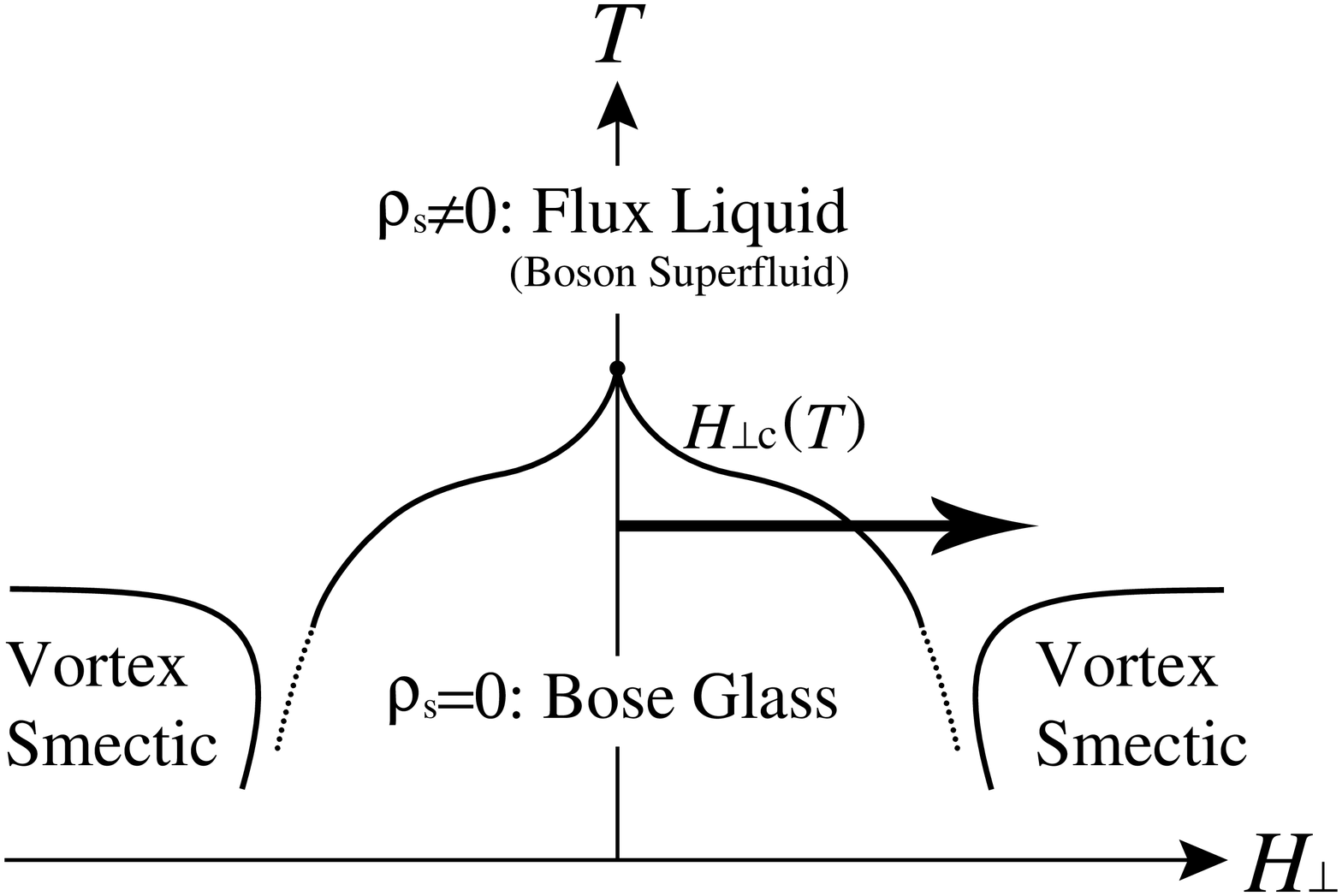}
\begin{figure}
\figcaption{
A schematic phase diagram of high-temperature superconductors
with columnar defects.\protect\cite{Nelson93a}
The abscissa indicates the transverse component of the external 
magnetic field.
}
\label{phase-diagram}
\end{figure}
temperatures.
For a periodic potential, thermally activated kinks always lead to 
{\em extended} eigenfunctions in the equivalent quantum problem (Bloch's 
theorem!), and the effect vanishes.
See Ref.~\cite{Balents95} for a discussion.

Nelson and Vinokur \cite{Nelson93a} have described
flux-line pinning by means of the path-integral scheme.
They mapped a phenomenological Hamiltonian for flux lines
with extended defects onto interacting bosons in $2+1$ dimensions with
point impurities.
A low-temperature phase with flux lines localized on extended defects 
was related to 
the Bose-glass phase\cite{MPAFisher89} of disordered boson systems.
When the applied magnetic field is tilted away from the extended 
defects, the corresponding boson problem acquires the 
non-Hermitian field $\mbox{\boldmath $g$}$.\cite{Nelson90,Hatano96}
The flux-line depinning transition due to the magnetic-field tilt
corresponds to a delocalization transition of the non-Hermitian 
quantum problem.
The expected $T$-$H_{\perp}$ phase diagram of the flux-line system 
(with $\mbox{\protect\boldmath $H$}_{\parallel}$, the field component
parallel to the columns, held fixed)
is shown in Fig.~\ref{phase-diagram}.\cite{Nelson93a}
The transition to a flux-liquid or \lq\lq superfluid'' phase 
with increasing temperature for 
$\mbox{\boldmath $H$}_{\perp}=\mbox{\bf 0}$ is of the usual Bose-glass 
type.\cite{MPAFisher89}
In the present paper, we shall focus on the transition curves for 
$\mbox{\boldmath $H$}_{\perp}\neq\mbox{\bf 0}$.
Hence we fix the temperature and follow the trajectory of increasing
$\mbox{\boldmath $H$}_{\perp}$,
indicated in Fig.~\ref{phase-diagram} as a thick arrow.
We first describe the delocalization transition of the 
non-Hermitian quantum problem, and next translate the results into the 
language of flux-line depinning.
We mainly discuss the one-body (or one-flux-line) problem of the 
non-Hermitian quantum mechanics.
Some of our results were summarized in a previous publication.\cite{Hatano96}

One advantage of discussing the flux-line depinning in terms of an 
equivalent \lq\lq quantum'' system comes from the fact that many ideas are already 
available concerning localization in Hermitian random systems, 
such as the Anderson transition and Mott's variable-range hopping.
Our problem is perhaps the simplest non-Hermitian generalization of 
these phenomena.
The inverse localization length of a wave function, when it 
delocalizes, is proportional to the critical transverse field $H_{\perp c}$.
As $\mbox{\boldmath $H$}_{\perp}$ increases, 
a precursor of the flux-line depinning appears first near the surface
of the superconductor.
Slightly below the critical point, the ends of the pinned flux line 
begin to tear free from the pinning center.
Thus the transverse Meissner effect first breaks down near the surfaces,
and the typical depth of the region where this bending takes place is a 
``penetration depth'' ({\em different} from the penetration depth 
associated with the underlying superconducting material) which 
accompanies the transverse 
Meissner effect.
A Mott variable-range-hopping description of this depinning will lead 
us to the conclusion that
the penetration depth 
$\tau^\ast$ diverges at the transition point $H_{\perp c}$ as 
\begin{equation}\label{0010}
\tau^\ast\sim(H_{\perp c}-H_{\perp})^{-2}
\end{equation}
for the random columnar defects, and
\begin{equation}\label{0020}
\tau^\ast\sim(H_{\perp c}-H_{\perp})^{-1}
\end{equation}
for the random twin boundaries.
The pinning becomes ineffective above the critical field.
The direction of the bulk transverse magnetization jumps at the critical point,
while the surface transverse magnetization diverges in both cases according to
\begin{equation}\label{0030}
M_{\perp s}\sim(H_{\perp c}-H_{\perp})^{-1}.
\end{equation}
We find {\em stretched-exponential} relaxation of the perpendicular 
magnetization to zero as one goes deep into the bulk of the sample 
from the surface.

The penetration depth $\tau^\ast$ may be observed experimentally  by 
measuring the response to an AC magnetic field superimposed
on the DC field 
$\mbox{\boldmath $H$}_{\perp}$;
only the dangling ends of the flux lines will respond to the AC field,
and the length of these line segments diverges at the transition.
It would be interesting to check our prediction of a diverging 
penetration depth near the Bose-glass transition with such an 
experiment.

A preliminary account of this work appeared in Ref.~\cite{Hatano96}.
For a related problem which arises from the physics of charge density 
waves, with some results applicable to vortex lines, see Ref.~\cite{Chen96}.
See Ref.~\cite{Miller96} for a related problem in fluid mechanics.

The plan of this paper is as follows.
Section~\ref{sec-corres} describes basic relations between the flux-line 
system and non-Hermitian quantum mechanics.
In Sec.~\ref{sec-mechanism}, we explain how the delocalization 
transition occurs in the non-Hermitian random system. 
We present solutions of non-Hermitian systems in
Secs.~\ref{sec-1D1imp}--\ref{sec-2D}.
The predictions for the transverse Meissner effect are
derived in Sec.~\ref{sec-meissner}.
Section~\ref{sec-interaction} discusses the effect of interactions in the 
delocalized phase in both $1+1$ and $2+1$ dimensions.

\section{Flux-lines and non-Hermitian quantum mechanics}
\label{sec-corres}

In this section, we review the basic correspondence between vortex 
lines in superconductors and the non-Hermitian quantum system~(\ref{12020}). 
We then describe what the current and the wave functions of the 
non-Hermitian system mean for vortex trajectories.

\subsection{Path-integral mapping}

We start with the energy of a flux line in 
$(d+1)$-dimensional type-II superconductor:\cite{Nelson93a}
\begin{equation}\label{11010}
E_{\rm flux}
=\int_0^{L_{\tau}}d\tau
\left(
\frac{\tilde{\varepsilon}_1}{2}
\left|\frac{d\mbox{\boldmath $x$}}{d\tau}\right|^2
+V(\mbox{\boldmath $x$})
-\frac{\phi_0\mbox{\boldmath $H$}_\perp}{4\pi}\cdot
\frac{d\mbox{\boldmath $x$}}{d\tau}
\right).
\end{equation}
This is a phenomenological Hamiltonian for the flux-line system where all 
columnar defects or twin boundaries are parallel.
Here $\tau$ denotes the coordinate parallel to the 
extended defects, while 
$\mbox{\boldmath $x$}$ denotes the $d$-dimensional coordinates perpendicular 
to the defects.
We describe the flux line as a single-valued function
$\mbox{\boldmath $x$}(\tau)$ by neglecting overhangs of the flux line.
The thickness of the superconductor in the $\tau$ direction is denoted by
$L_{\tau}$.

The first kinetic-energy-like term of the integrand in~(\ref{11010}) 
comes from the harmonic 
approximation to the energy increase due to local tilt of the flux line.
The \lq\lq mass,'' or tilt modulus is denoted by $\tilde{\varepsilon}_1$.
When using quantum language to describe this problem, we shall make the 
correspondence
\begin{equation}\label{mass-tilt}
\tilde{\varepsilon}_{1} \longleftrightarrow m.
\end{equation}

The potential $V$ is generated by columnar defects or twin boundaries.
Each columnar defect is specified by a point $\mbox{\boldmath $X$}_k$
in the $\mbox{\boldmath $x$}$ plane.
We assume that the columnar defects are distributed randomly:
\begin{equation}
V(\mbox{\boldmath $x$})=\sum_{k=1}^M
V_1(\mbox{\boldmath $x$}-\mbox{\boldmath $X$}_k),
\end{equation}
where $M$ denotes the number of the columnar defects, $V_1$ is the potential
of an individual defect, and $\mbox{\boldmath $X$}_k$ is the random
position of the $k$th columnar defect.
In the case of twin boundaries, we consider only the 
situation where all twins are parallel to each other.
By projecting out the degree of freedom perpendicular to $\tau$ and 
parallel to the twin boundaries,\cite{Nelson93a,Marchetti95} 
we can reduce the dimensionality of the $\mbox{\boldmath $x$}$-space by one.
(We neglect effects due to an enhanced concentration of point disorder 
which may occur in twins.)
Thus each twin boundary acts like a line defect in $1+1$ dimensions.
In short, we can set $d=2$ for columnar defects and $d=1$ for twin 
boundaries.

The third term of the integrand in~(\ref{11010}) is due to the
transverse component of the magnetic field $\mbox{\boldmath $H$}_\perp$.
The flux quantum $2\pi\hbar c/(2e)$ is denoted by $\phi_0$.
In the absence of\linebreak 

\epsfxsize=3.375in
\epsfbox{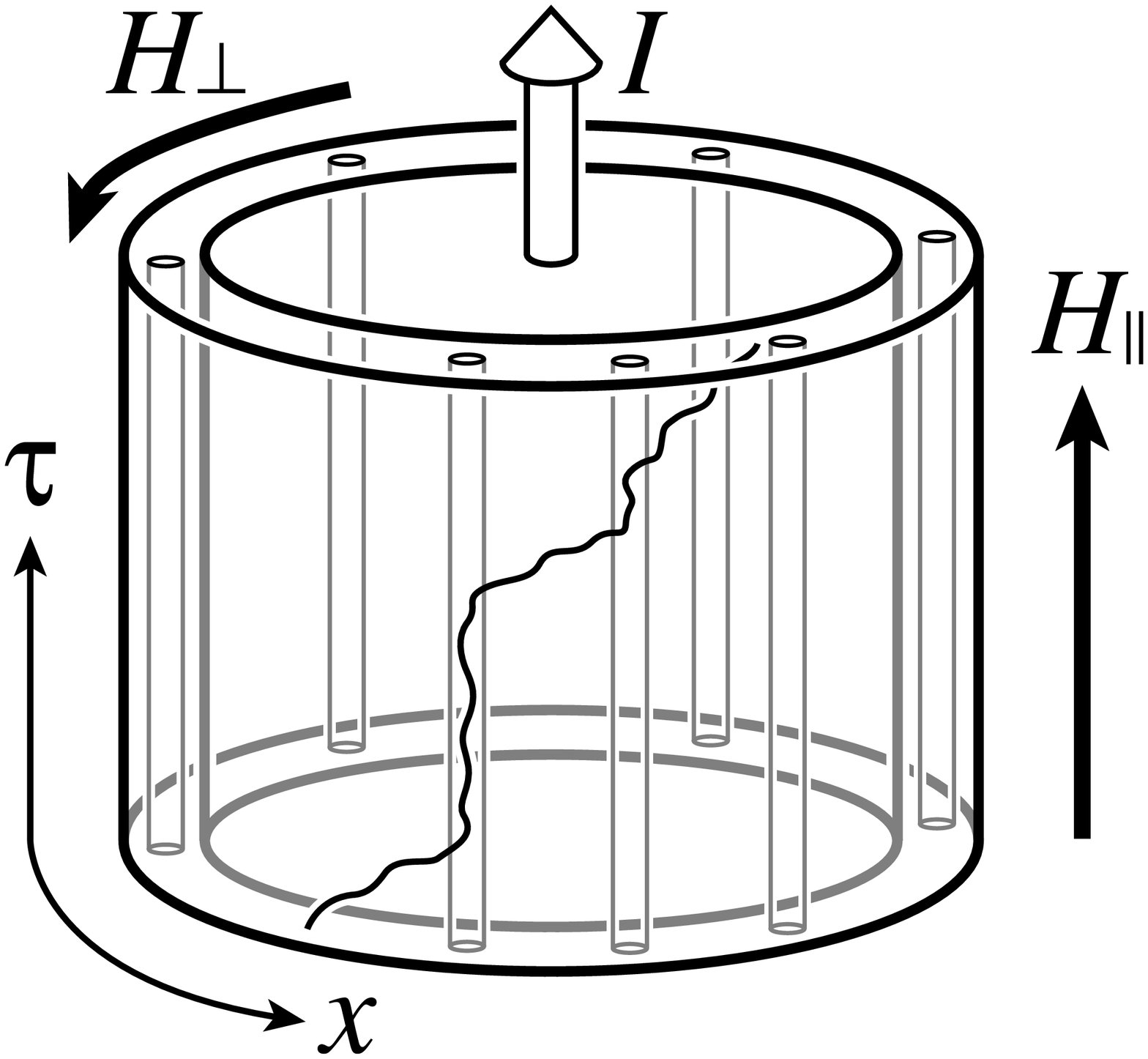}
\begin{figure}
\figcaption{
One flux line (wavy curve) induced by the field 
$\mbox{\protect\mibsmall H}_{\parallel}$ 
and interacting with columnar defects in a cylindrical superconducting shell.
The transverse component 
$\mbox{\protect\mibsmall H}_\perp$ 
is generated by the current 
$\mbox{\protect\mibsmall I}$ 
threading the ring.
}
\label{1Dpbc}
\end{figure}
a pinning potential, this term is optimized 
energetically when the flux line tilts to follow the external field.

Figure~\ref{1Dpbc} shows a flux-line system described by the above
phenomenology in $(1+1)$ dimensions with columnar pinning.
We assume periodic boundary conditions in the $\mbox{\boldmath $x$}$ 
direction, although these could be replaced by reservoirs of flux lines 
at the edges of the sample.
In this geometry, the transverse component $\mbox{\boldmath $H$}_\perp$ 
is generated by the current threading the cylinder of a thin superconductor.
As we shall see, the physics resembles the Aharonov-Bohm experiment 
carried out on a mesoscopic quantum ring with, however, an {\em 
imaginary} \lq\lq flux'' through the ring.

The partition function of the flux-line system~(\ref{11010}) is given by
\begin{equation}\label{11020}
{\cal Z}=\int{\cal D}\mbox{\boldmath $x$}
\exp\left(-\frac{E_{\rm flux}}{k_BT}\right),
\end{equation}
where $\int{\cal D}\mbox{\boldmath $x$}$ denotes integration over all possible
configurations $\mbox{\boldmath $x$}(\tau)$ of the flux line.
Upon identifying Eq.~(\ref{11020}) with Eq.~(\ref{11020-1}), we are led to
a correspondence between the flux-line system~(\ref{11010}) and 
the non-Hermitian quantum system~(\ref{12020}) summarized in 
Table~\ref{corres}.
In particular, the thermal fluctuations of the flux-line system are
equivalent to zero-point motion of the quantum system, and
\begin{equation}
k_BT\longleftrightarrow\hbar.
\end{equation}
The tilt term in~(\ref{11010}) is related to the 
non-Hermitian field in~(\ref{12020}) by
\begin{equation}
\mbox{\boldmath $g$}\equiv\frac{\phi_0\mbox{\boldmath $H$}_\perp}{4\pi}.
\end{equation}
\end{multicols}
\begin{table}
\caption{
The correspondence between the $(d+1)$-dimensional flux-line system 
and the $d$-dimensional quantum system.
}
\label{corres}
\begin{tabular}{ll}
Partition function (\protect\ref{11020}) & 
Partition function (\protect\ref{11030}) \\
Thickness $L_{\tau}$ & 
Imaginary time $\tau$ \\
Temperature $k_{B}T$ & 
Planck parameter $\hbar$ \\
Tilt modulus $\tilde{\varepsilon}_1$ & 
Mass $m$ \\
Tilting field 
$\phi_0\mbox{\protect\mibsmall H}_\perp/(4\pi)$ &
Non-hermitian field 
$\mbox{\protect\mibsmall g}$ \\
Tilt slope 
$d\mbox{\protect\mibsmall x}/d\tau$ 
($\propto$ Transverse magnetization) &
${\rm Im}\; J$, where $J=\mbox{current}$ \\
Probability distribution at $\tau=0$ ($L_{\tau}\to\infty$) &
Left-eigenvector 
$\psi^L(\mbox{\protect\mibsmall x})$ \\
\phantom{Probability distribution} at $\tau=L_{\tau}$ &
Right-eigenvector 
$\psi^R(\mbox{\protect\mibsmall x})$ \\
\phantom{Probability distribution} at $\tau=L_{\tau}/2$ &
$\psi^L\psi^R$ \\
\end{tabular}
\end{table}
\begin{multicols}{2}
\noindent
Boundary conditions at the bottom and top surfaces of the 
superconductor are represented by the initial and final vectors of
Eq.~(\ref{11030}) in the forms
\begin{equation}
\left\langle\psi^f\right|
\equiv\int d\mbox{\boldmath $x$}\left\langle\mbox{\boldmath $x$}\right|
\qquad\mbox{and}\qquad
\left|\psi^i\right\rangle
\equiv\int d\mbox{\boldmath $x$}\left|\mbox{\boldmath $x$}\right\rangle.
\end{equation}
These boundary conditions in the imaginary-time direction mean that 
we integrate freely over the points where vortex lines enter and exit 
the sample.

Although the derivation of the quantum problem~(\ref{12020}) from the 
classical elastic string described by Eq.~(\ref{11010})
is straightforward in the path-integral 
scheme, it is also useful to think in terms of Galilean 
boosts.\cite{Nelson90,Hwa93b}
We can eliminate the additional term in~(\ref{11010}) by the Galilean
transformation
\begin{equation}\label{12040}
\left\{\begin{array}{ll}
\mbox{\boldmath $x$}'&=\mbox{\boldmath $x$}-\mbox{\boldmath $v$}t,\\
t'&=t,
\end{array}\right.
\end{equation}
where the velocity of the moving frame is
\begin{equation}\label{12050}
\mbox{\boldmath $v$}
=\frac{\phi_0\mbox{\boldmath $H$}_\perp}{4\pi i\tilde{\varepsilon}_1}
\equiv\frac{\mbox{\boldmath $g$}}{im}.
\end{equation}
Note that the velocity is {\em imaginary}, 
because of the imaginary time $t=i\tau$ in the equivalent quantum problem.
The above Galilean transformation changes the kinetic term of the
Hermitian Hamiltonian to the form
\begin{equation}\label{12060}
\frac{\mbox{\boldmath $p$}^2}{2m}
\longrightarrow
\frac{(\mbox{\boldmath $p$}-m\mbox{\boldmath $v$})^2}{2m},
\end{equation}
which results in the kinetic term in~(\ref{12020}).

We do not explicitly include interactions between flux lines in the 
Hamiltonian~(\ref{11010}) nor in the quantum 
Hamiltonians~(\ref{12020}) and~(\ref{41010}).
However, strong short-range repulsive potential may be treated approximately
by forbidding multiple occupancy of localized levels of the quantum 
Hamiltonian.
Since the particle mapped from the flux line follows the Bose 
statistics,\cite{Nelson93a} we may treat many-flux-line problems as 
hard-core bosons in a random potential.
For $d=1$, hard-core bosons are equivalent to fermions,
and hence we can use the Pauli principle directly to treat many-body 
problems in this case.
See Sec.~\ref{sec-interaction} for a discussion of interaction 
effects using a boson formalism in $d$ dimensions.

\subsection{Imaginary current and the transverse magnetization}

We show in the following that the imaginary part of
the current of the non-Hermitian system
yields the average slope of a vortex trajectory as it crosses the 
sample.

Since the non-Hermitian field $\mbox{\boldmath $g$}$ acts like an imaginary
vector potential, we define the current operator as
\begin{equation}\label{12070}
\mbox{\boldmath $J$}\equiv -i
\frac{\partial {\cal H}}{\partial \mbox{\boldmath $g$}}
=\frac{\mbox{\boldmath $p$}+i\mbox{\boldmath $g$}}{m}
=\mbox{\boldmath $J$}_{\rm para}+\mbox{\boldmath $J$}_{\rm dia},
\end{equation}
where
\begin{equation}
\mbox{\boldmath $J$}_{\rm para}\equiv\frac{\mbox{\boldmath $p$}}{m}
\qquad\mbox{and}\qquad
\mbox{\boldmath $J$}_{\rm dia}\equiv i\frac{\mbox{\boldmath $g$}}{m}.
\end{equation}

Note first that the expectation value of the position of the flux line
at the distance $\tau$ from the bottom surface of the superconductor is given
by
\begin{equation}\label{12080}
\left\langle \mbox{\boldmath $x$} \right\rangle_{\tau}
\equiv\frac{1}{\cal Z}
\left\langle\psi^f\left|
e^{-(L_{\tau}-\tau){\cal H}/\hbar}
\mbox{\boldmath $x$}^{\rm op}
e^{-\tau{\cal H}/\hbar}
\right|\psi^i\right\rangle,
\end{equation}
where the partition function ${\cal Z}$ is defined by~(\ref{11030})
and $\mbox{\boldmath $x$}^{\rm op}$ is the position operator acting on 
position eigenstates $|\mbox{\boldmath $x$}\rangle$ such that
$\mbox{\boldmath $x$}^{\rm op}|\mbox{\boldmath $x$}\rangle
=\mbox{\boldmath $x$}|\mbox{\boldmath $x$}\rangle$.
Note also the useful commutation relation
\begin{equation}\label{12090}
\left[ {\cal H},\mbox{\boldmath $x$}^{\rm op}\right]
=-i\hbar\mbox{\boldmath $J$},
\end{equation}
which leads immediately to
\begin{eqnarray}\label{12100}
\frac{\partial}{\partial \tau}
\left\langle \mbox{\boldmath $x$} \right\rangle_{\tau}
&=&\frac{1}{\hbar{\cal Z}}
\left\langle\psi^f\left|
e^{-(L_{\tau}-\tau){\cal H}/\hbar}
[{\cal H},\mbox{\boldmath $x$}^{\rm op}]
e^{-\tau{\cal H}/\hbar}
\right|
\psi^i\right\rangle
\nonumber\\
&=&-i\left\langle \mbox{\boldmath $J$} \right\rangle_{\tau}
={\rm Im}\;\left\langle \mbox{\boldmath $J$} \right\rangle_{\tau}.
\end{eqnarray}
In the last line of Eq.~(\ref{12100}), 
we used the fact that the expectation
value of the current is pure imaginary in the present problem.
The current carried by an individual eigenstate is {\em not} necessarily 
pure imaginary,
where the current of a state $\varepsilon_n$ is defined by
\begin{equation}\label{12120}
\mbox{\boldmath $J$}_n
\equiv-i\frac{\partial \varepsilon_n}{\partial \mbox{\boldmath $g$}}.
\end{equation}
However, a purely imaginary expectation value of the current arises owing to
a pairing property of eigenvalues and eigenfunctions: if there is a complex 
eigenvalue $\varepsilon$ with a right-eigenvector $\psi^R$ for this 
non-Hermitian problem, there is also the 
eigenvalue $\varepsilon^\ast$ with the right-eigenvector $(\psi^R)^\ast$.
When we sum Eq.~(\ref{12120}) 
over all the eigenstates to calculate the expectation value,
the real part of the current of a state cancels out in pairwise fashion.

The transverse component of the magnetization per vortex line
at depth $\tau$ below the surface of the sample is
proportional to the tilt slope,
\begin{equation}\label{12102}
\mbox{\boldmath $m$}_{\perp}(\tau)\equiv\frac{\partial}{\partial \tau}
\left\langle \mbox{\boldmath $x$} \right\rangle_{\tau}
={\rm Im}\;\left\langle \mbox{\boldmath $J$} \right\rangle_{\tau}.
\end{equation}
The total transverse magnetization is proportional to the net
displacement of the flux line per unit length
between the bottom surface ($\tau=0$) and the top surface ($\tau=L_{\tau}$):
\begin{eqnarray}\label{12107}
\mbox{\boldmath $M$}_{\perp}&\equiv& n_{v}\frac{\phi_{0}}{L_{\tau}}
\int_{0}^{L_{\tau}}\mbox{\boldmath $m$}_{\perp}(\tau)d\tau
=\frac{\phi_{0}n_{v}}{L_{\tau}}
{\rm Im}\;\int_0^{L_{\tau}}d\tau
\left\langle \mbox{\boldmath $J$}\right\rangle_{\tau}
\nonumber\\
&=&\frac{\phi_{0}n_{v}}{L_{\tau}}
\left(\left\langle \mbox{\boldmath $x$}\right\rangle_{L_{\tau}}
-\left\langle \mbox{\boldmath $x$}\right\rangle_{0}\right),
\end{eqnarray}
where $n_{v}$ is the density of vortices.
As we show below,
the last factor in the brackets is given by
\begin{equation}\label{12105}
\left\langle \mbox{\boldmath $x$}\right\rangle_{L_{\tau}}
-\left\langle \mbox{\boldmath $x$}\right\rangle_{0}
=\hbar\frac{\partial}{\partial \mbox{\boldmath $g$}}\ln {\cal Z}.
\end{equation}
Equation~(\ref{12107}) is an indicator of the delocalization transition;
the transverse magnetization must remain finite in the limit 
$L_{\tau}\to\infty$
in a depinned phase, while it must vanish macroscopically
in a pinned glassy phase, which is the transverse Meissner effect.
In fact, we show later that the vortex tilt 
$\mbox{\boldmath $m$}_{\perp}(\tau)$ 
appears only close to surfaces in the pinning regime, 
and thus $\mbox{\boldmath $M$}_{\perp}$ vanishes as $L_{\tau}\to\infty$.
In the limit $L_{\tau}\to\infty$, Eq.~(\ref{12105}) becomes
\begin{equation}\label{12110}
\lim_{L_{\tau}\to\infty}
\left(\left\langle \mbox{\boldmath $x$}\right\rangle_{L_{\tau}}
-\left\langle \mbox{\boldmath $x$}\right\rangle_{0}\right)
=-L_{\tau}\frac{\partial \varepsilon_{\rm gs}}%
{\partial \mbox{\boldmath $g$}},
\end{equation}
where $\varepsilon_{\rm gs}$ denotes the ground-state energy of 
${\cal H}(\mbox{\boldmath $g$})$.
(The ground state is defined by the lowest {\em real} part of the 
eigenenergy.\cite{footnote1})
Hence, in the pinning regime, the ground-state energy must be independent of
$\mbox{\boldmath $g$}$.
In the depinning regime, 
$-\partial\varepsilon_{\rm gs}/\partial\mbox{\boldmath $g$}$ is 
nonzero and equals the mean tilt slope of the corresponding flux line.

To prove Eq.~(\ref{12105}), we proceed as follows.
As is stated above, the non-Hermitian field $\mbox{\boldmath $g$}$ acts like
an imaginary vector potential.
Because of this, we can gauge away the non-Hermitian field by
applying the imaginary gauge transformation to the non-Hermitian 
Hamiltonian~(\ref{12020}):
\begin{equation}\label{12170}
{\cal U}^{-1}{\cal H}(\mbox{\boldmath $g$}){\cal U}=
{\cal H}(\mbox{\bf 0}),
\end{equation}
where
\begin{equation}\label{12180}
{\cal U}\equiv\exp\left(
\frac{\mbox{\boldmath $g$}\cdot
\mbox{\boldmath $x$}^{\rm op}}{\hbar}\right).
\end{equation}
(This is just another expression of the Galilean
transformation~(\ref{12060}).)
Therefore, we have
\begin{equation}
e^{-L_{\tau}{\cal H}(\mbox{\mibscriptsize g})/\hbar}
={\cal U}e^{-L_{\tau}{\cal H}(\mbox{\bfscriptsize 0})/\hbar}{\cal U}^{-1},
\end{equation}
and
\begin{equation}
\hbar\frac{\partial}{\partial\mbox{\boldmath $g$}}
e^{-L_{\tau}{\cal H}(\mbox{\mibscriptsize g})/\hbar}
=\mbox{\boldmath $x$}^{\rm op}e^{-L_{\tau}{\cal H}(\mbox{\mibscriptsize g})/\hbar}
-e^{-L_{\tau}{\cal H}(\mbox{\mibscriptsize g})/\hbar}\mbox{\boldmath $x$}^{\rm op}.
\end{equation}
Upon taking thermodynamic averages (expectation values in quantum 
language), we obtain Eq.~(\ref{12105}).

A conductivity-like transport coefficient may be defined by
\begin{equation}\label{12190}
\sigma_{\mu\nu}\equiv
-i\frac{\partial}{\partial g_{\nu}}\langle J_{\mu} \rangle.
\end{equation}
The quantity defined for each eigenstate 
\begin{equation}\label{12200}
\sigma^{(n)}_{\mu\nu}\equiv
-\frac{\partial^2 \varepsilon_{n}}{\partial g_{\mu} \partial g_{\nu}}
\end{equation}
is a measure of the stiffness of the eigenfunction.
In many-body problems, $\sigma_{\mu\nu}$ is related 
to the superfluid density of the relevant boson system.\cite{Uwe97}

\subsection{Probability distribution of a flux line and 
eigenvectors of the non-Hermitian Hamiltonian}
\label{sec-probdist}

We now outline the correspondence between the probability distribution of
a flux line and the eigenvectors of the quantum system.
The flux line fluctuates because of thermal excitations.
This thermal agitation is described by quantum fluctuations of 
the corresponding fictitious non-Hermitian quantum particle.
The probability distribution of the flux line due to thermal fluctuation
is thus related to wave functions of the quantum system.

The probability distribution of the flux line at the distance $\tau$ from
the bottom surface is given by \cite{Nelson93a}
\begin{equation}\label{13010}
P(\mbox{\boldmath $x$};\tau)
\equiv\frac{1}{\cal Z}
\left\langle\psi^f\right|
e^{-(L_{\tau}-\tau){\cal H}/\hbar}
\left|
\mbox{\boldmath $x$}\right\rangle
\left\langle\mbox{\boldmath $x$}\right|
e^{-\tau{\cal H}/\hbar}
\left|\psi^i\right\rangle,
\end{equation}
so that we can rewrite Eq.~(\ref{12080}) as
\begin{equation}\label{13020}
\left\langle \mbox{\boldmath $x$} \right\rangle_{\tau}
=\int\mbox{\boldmath $x$}P(\mbox{\boldmath $x$};\tau)d^d\mbox{\boldmath $x$}.
\end{equation}
Let us calculate $P(\mbox{\boldmath $x$};\tau)$ in the limit
$L_{\tau}\to\infty$.
In this limit the exponential operator $\exp(-L_{\tau}{\cal H}/\hbar)$ 
can be approximated by
\begin{equation}\label{13030}
\lim_{L_{\tau}\to\infty}e^{-L_{\tau}{\cal H}/\hbar}\simeq
\left|\psi_{\rm gs}\right\rangle
e^{-L_{\tau}\varepsilon_{\rm gs}/\hbar}
\left\langle\psi_{\rm gs}\right|.
\end{equation}
Note here that left and right eigenvectors may be expressed as
\begin{equation}
\psi^R(\mbox{\boldmath $x$})=
\left\langle\mbox{\boldmath $x$}\bigm|\psi\right\rangle
\quad\mbox{and}\quad
\psi^L(\mbox{\boldmath $x$})=
\left\langle\psi\bigm|\mbox{\boldmath $x$}\right\rangle.
\end{equation}
Upon combining Eq.~(\ref{13010}) and  Eq.~(\ref{13030}), we find
\begin{eqnarray}\label{13050}
\lim_{L_{\tau}\to\infty}P(\mbox{\boldmath $x$};0)
&=&\frac{\psi_{\rm gs}^L(\mbox{\boldmath $x$})}%
{\int\psi_{\rm gs}^L(\mbox{\boldmath $x$})
d^d\mbox{\boldmath $x$}},
\\ \label{13060}
\lim_{L_{\tau}\to\infty}P(\mbox{\boldmath $x$};L_{\tau})
&=&\frac{\psi_{\rm gs}^R(\mbox{\boldmath $x$})}%
{\int\psi_{\rm gs}^R(\mbox{\boldmath $x$})
d^d\mbox{\boldmath $x$}},
\end{eqnarray}
and
\begin{equation}
\lim_{L_{\tau}\to\infty}P(\mbox{\boldmath $x$};L_{\tau}/2)
=\psi_{\rm gs}^L(\mbox{\boldmath $x$})\psi_{\rm gs}^R(\mbox{\boldmath $x$}).
\end{equation}
Since the ground state wave function is positive definite,
the distributions~(\ref{13050}) and~(\ref{13060}) are well-defined.

We remark on the nature of the limit in Eq.~(\ref{13030}). 
For one-impurity systems (see Sec.~\ref{sec-1D1imp}), corrections to the 
right-hand side of Eq.~(\ref{13030}) are of order 
$\exp(-L_{\tau}\Delta\varepsilon)$, where $\Delta\varepsilon$ is the 
energy gap above the ground state.
For random systems, the argument in Sec.~\ref{sec-meissner} yields
corrections which vanish as a stretched exponential function of 
$L_{\tau}$.

For a finite density of flux lines, we can treat a
strong short-range repulsive potential approximately by forbidding multiple 
occupancy of localized levels of the quantum Hamiltonian.\cite{Nelson93a}
We then fill up the localized levels in order of increasing energy up to
a certain level $\varepsilon=\mu$. 
The system is now characterized by an average chemical potential
$\mu=\mu(H_{\parallel})$, where $H_{\parallel}$ is the external magnetic field 
along the $\tau$ axis. 
This chemical potential controls the flux-line density and separates
occupied and unoccupied levels.
Under this assumption, we can estimate the probability distribution of
the most weakly bound flux line by using the state at the chemical potential,
$\psi_\mu$, instead of $\psi_{\rm gs}$ in the above analysis.

\section{Delocalization transition in non-Hermitian quantum mechanics}
\label{sec-mechanism}

We now sketch the mechanism of the delocalization in the non-Hermitian
quantum system described above.
Consider first localized states in a small transverse field 
$\mbox{\boldmath $g$}$.
Assume that we know the eigenvalues $\varepsilon_n$ and the eigenfunctions 
$\psi_n(\mbox{\boldmath $x$})$ for $\mbox{\boldmath $g$}=\mbox{\bf 0}$.
We can use the imaginary gauge transformation~(\ref{12180}) to 
determine the eigenfunctions for small $\mbox{\boldmath $g$}$.
The right-eigenfunctions and the left-eigenfunctions are given by
\begin{eqnarray}\label{21010}
\psi_n^R(\mbox{\boldmath $x$};\mbox{\boldmath $g$})
&=&{\cal U}\psi_n(\mbox{\boldmath $x$};\mbox{\boldmath $g$}=\mbox{\bf 0}),
\\ \label{21020}
\psi_n^L(\mbox{\boldmath $x$};\mbox{\boldmath $g$})
&=&\psi_n(\mbox{\boldmath $x$};\mbox{\boldmath $g$}=\mbox{\bf 0}){\cal U}^{-1},
\end{eqnarray}
where

\noindent
\hskip 0.4375in
\begin{minipage}{2.5in}
\epsfxsize=2.5in
\epsfbox{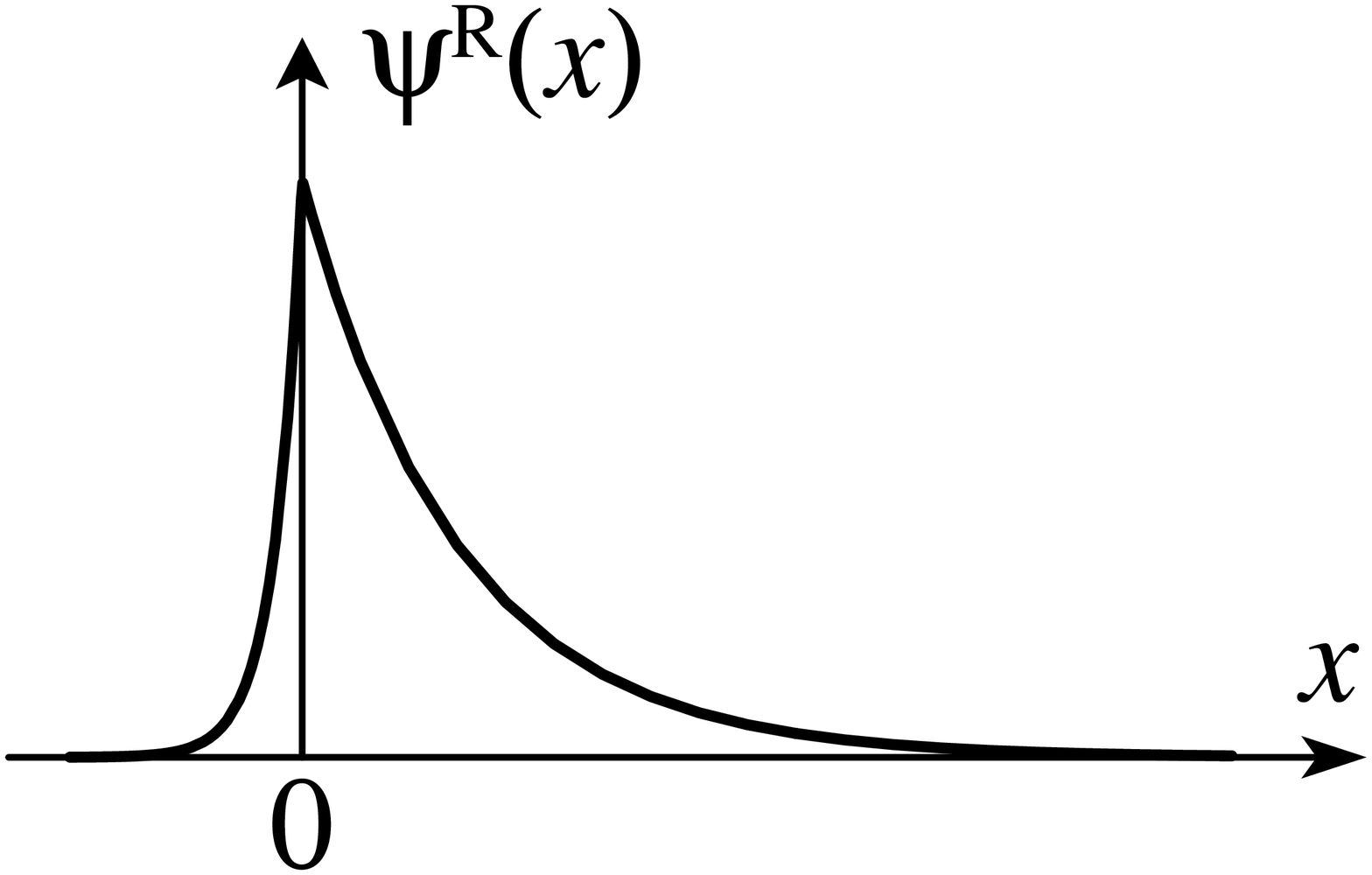}
\end{minipage}
\begin{figure}
\figcaption{
The wave function~(\protect\ref{21050}) or~(\protect\ref{31121}), 
the ground state of a single impurity with $0<g<g_c$ in one dimension.
}
\label{fig1D1pin-WF}
\end{figure}
\begin{equation}\label{21030}
{\cal U}\equiv\exp\left(
\frac{\mbox{\boldmath $g$}\cdot\mbox{\boldmath $x$}^{\rm op}}{\hbar}\right)
\quad\mbox{and}\quad
{\cal U}^{-1}\equiv\exp\left(-
\frac{\mbox{\boldmath $g$}\cdot\mbox{\boldmath $x$}^{\rm op}}{\hbar}\right).
\end{equation}
The corresponding energy eigenvalue $\varepsilon_n$ is then unchanged 
provided that this imaginary gauge transformation is applicable.
However, the above wave functions $\psi_n^R$ and $\psi_n^L$ may diverge
as $|\mbox{\boldmath $x$}|\to\infty$ and become unnormalizable.\cite{Hwa93b}
If we assume an asymptotic form of the wave function for 
$\mbox{\boldmath $g$}=\mbox{\bf 0}$ of the form
\begin{equation}\label{21040}
\psi_n(\mbox{\boldmath $x$};\mbox{\boldmath $g$}=\mbox{\bf 0})
\mathop\sim_{x\to\infty}\exp\left(-\kappa_n|\mbox{\boldmath $x$}
-\mbox{\boldmath $x$}_n|\right), 
\end{equation}
the condition for non-divergence of $\psi_n^R$ and $\psi_n^L$ 
in $|\mbox{\boldmath $x$}|\to\infty$ 
is $|\mbox{\boldmath $g$}|<\hbar\kappa_n$.
Then the normalized wave functions are approximately given by (see 
Fig.~\ref{fig1D1pin-WF})
\begin{eqnarray}\label{21050}
\psi_n^R&\simeq&\sqrt{\frac{(2\kappa_n)^d}{\Gamma(d)\Omega_d}}
\exp\left[
\frac{\mbox{\boldmath $g$}\cdot(\mbox{\boldmath $x$}-\mbox{\boldmath $x$}_n)}{\hbar}
-\kappa_n|\mbox{\boldmath $x$}-\mbox{\boldmath $x$}_n|
\right],
\\ \label{21060}
\psi_n^L&\simeq&\sqrt{\frac{(2\kappa_n)^d}{\Gamma(d)\Omega_d}}
\exp\left[
-\frac{\mbox{\boldmath $g$}\cdot(\mbox{\boldmath $x$}-\mbox{\boldmath $x$}_n)}{\hbar}
-\kappa_n|\mbox{\boldmath $x$}-\mbox{\boldmath $x$}_n|
\right],
\end{eqnarray}
where $\mbox{\boldmath $x$}_n$ is the localization center for 
$\mbox{\boldmath $g$}=\mbox{\bf 0}$, $\Omega_d$ is the total solid angle of the 
$d$-dimensional space, and the normalization condition is
\begin{equation}\label{21070}
\int\psi_n^L\psi_n^Rd^d\mbox{\boldmath $x$}=1.
\end{equation}

We naturally regard the point $|\mbox{\boldmath $g$}|=\hbar\kappa_n$
as the delocalization point of the $n$th state.
Since the eigenfunctions are proportional to the probability distribution
of a flux line at the surfaces of the superconductor
(see Eqs.~(\ref{13050}) and~(\ref{13060})), we interpret
the divergence in $|\mbox{\boldmath $x$}|\to\infty$ as depinning of the flux 
line.
Because localized states with finite 
localization length are stable against a small $\mbox{\boldmath $g$}$ field,
flux-line pinning is {\em robust} against small 
transverse field: this is the transverse Meissner 
effect.\cite{Nelson93a}

The diverging tilt slope in Eq.~(\ref{0030}) follows, because
the localization length of the right-eigenfunction is 
$(\kappa_{n}-|\mbox{\boldmath $g$}|/\hbar)^{-1}$ in the direction of 
$\mbox{\boldmath $g$}$, and 
$(\kappa_{n}+|\mbox{\boldmath $g$}|/\hbar)^{-1}$
in the opposite direction.
The former diverges as $|\mbox{\boldmath $g$}|\to{\hbar\kappa_n}^{-}$,
hence $H_{\perp c}=4\pi\hbar\kappa_{n}/\phi_{0}$ in Eq.~(\ref{0030}).
According to the relation~(\ref{12107}), 
the surface displacement or \lq\lq localization length'' 
of the flux line is 
proportional to the surface transverse magnetization $M_{\perp s}$.

We need to specify boundary conditions in order to obtain a well-defined
wave function in the thermodynamic limit.
Consider for simplicity the periodic boundary conditions
\begin{equation}\label{21080}
\psi_n^R(L_x/2,y,\ldots)=\psi_n^R(-L_x/2,y,\ldots),
\end{equation}
where we put the $x$ axis parallel to $\mbox{\boldmath $g$}$.
(The one-dimensional periodic system is realized in the setup shown in
Fig.~\ref{1Dpbc}.)
A wave function of the form~(\ref{21050}) localized at the origin
has a mismatch of the factor
$\exp[-(\kappa_n-|\mbox{\boldmath $g$}|/\hbar)L_x]$ at the boundaries 
$x=\pm L_x/2$.
This mismatch is exponentially small
in the pinning regime $|\mbox{\boldmath $g$}|<\hbar\kappa_n$.
Hence the wave functions~(\ref{21050}) and~(\ref{21060}) are 
excellent approximations
when we take the thermodynamic limit, imposing the periodic boundary 
conditions~(\ref{21080}).

In the depinning regime $|\mbox{\boldmath $g$}|\geq\hbar\kappa_n$, however,
we need to change the wave functions drastically to meet the boundary 
conditions.
To see what happens in this regime, consider the
limit in which the random potential
$V(\mbox{\boldmath $x$})$ may be neglected:
\begin{equation}\label{21090}
{\cal H}_0=\frac{(\mbox{\boldmath $p$}+i\mbox{\boldmath $g$})^2}{2m}.
\end{equation}
(We argue later that the potential in fact has only
perturbative effects for large $\mbox{\boldmath $g$}$ when $d=1$.)
The periodic boundary conditions~(\ref{21080}) are then
satisfied only by the extended Bloch wave 
\begin{equation}\label{21100}
\psi^R(\mbox{\boldmath $x$})
=\exp(i\mbox{\boldmath $k$}\cdot\mbox{\boldmath $x$})
\end{equation}
with $k_\nu=2n_\nu\pi/L_\nu$, where $n_\nu$ is an integer and $L_\nu$ is the 
system size in the $x_\nu$ direction.
The eigenvalue of this wave function is complex:
\begin{equation}\label{21110}
\varepsilon(\mbox{\boldmath $k$})=
\frac{(\hbar\mbox{\boldmath $k$}+i\mbox{\boldmath $g$})^2}{2m}.
\end{equation}

We can interpret
the appearance of the imaginary part of the energy in the following way:
A depinned flux line in a periodic system has a spiral trajectory,
and hence periodicity in the imaginary time direction.
(See Fig.~\ref{spiral}.)
Because the single-line partition function at \lq\lq time'' $\tau$ 
may be written 
${\cal Z}(\tau)=\sum_{n}c_{n}\psi_{n}^R(\mbox{\boldmath $x$})
e^{-\varepsilon_{n}\tau/\hbar}$,
where the constants $c_{n}$ are coefficients depending on \lq\lq initial 
conditions'' at, say, the bottom surface of the sample,
the period of motion in the imaginary-time direction associated with 
the $n$th eigenstate is given by
$\hbar/{\rm Im}\;\varepsilon_n$, where $n$ denotes the wave function 
describing the depinned flux line.
Thus a complex energy, as well as an imaginary part of the 
current~(\ref{12105}), is an indicator of the depinning
transition in periodic systems.

\epsfxsize=3.375in
\epsfbox{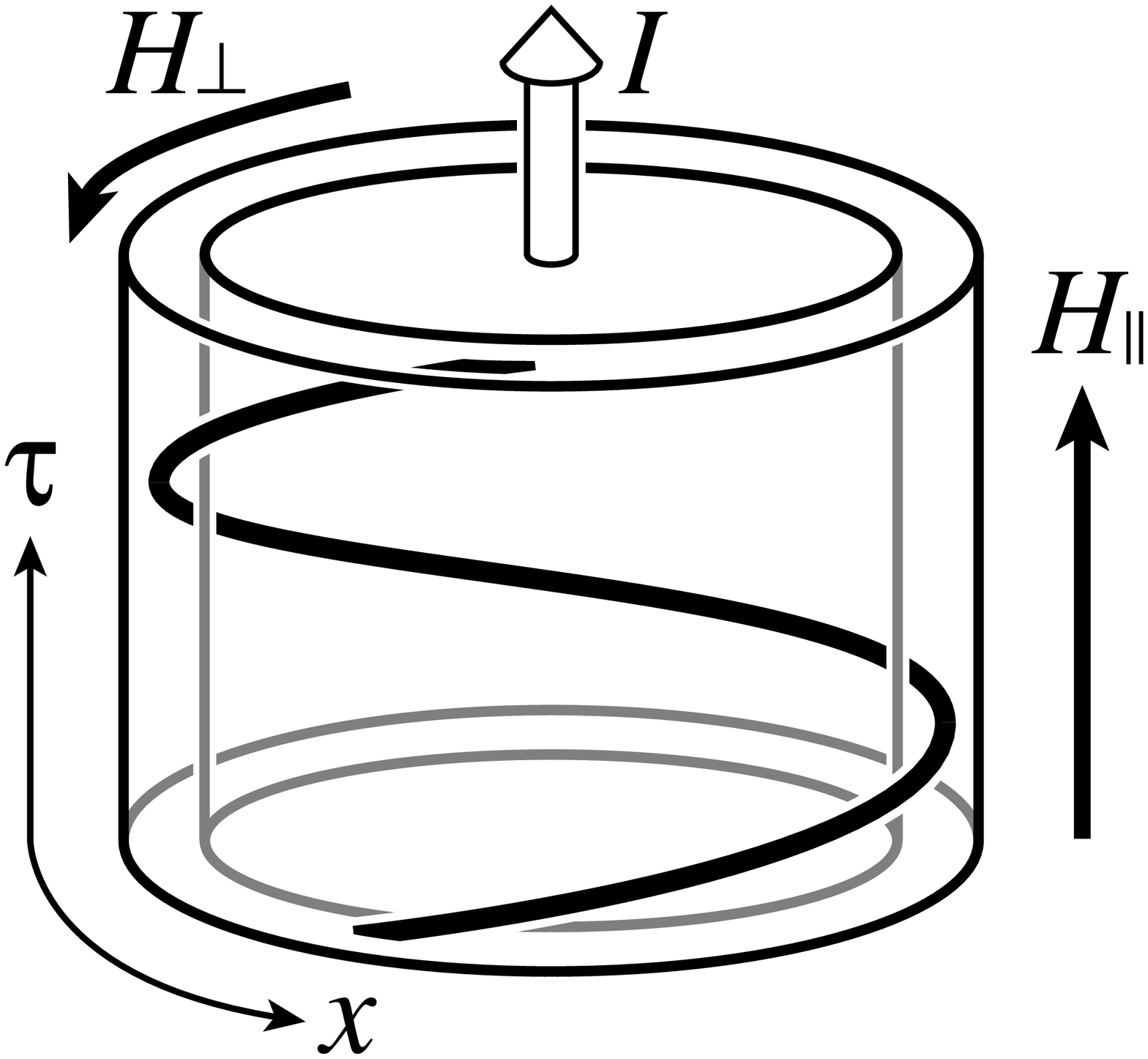}
\begin{figure}
\figcaption{
A spiral trajectory of a flux line in the system of 
Fig.~\protect\ref{1Dpbc}, reflecting the imaginary eigenvalues, 
which appear in non-Hermitian systems when boundary conditions allow
nonzero tilt of vortex lines across the system.
}
\label{spiral}
\end{figure}

We use periodic boundary conditions in all directions perpendicular 
to the imaginary-time axis throughout this paper.
The effect of \lq\lq free'' boundary conditions in the time-like 
direction is discussed in Sec.~\ref{sec-meissner}.
For superconducting flux lines, unusual geometries such as 
Fig.~\ref{1Dpbc} would be required to exactly implement periodic 
boundary conditions in the direction parallel to the tilt field 
$\mbox{\boldmath $g$}$.
For conventional slab-like superconducting samples, tilt causes extra 
vortices to enter and exit the sample on the sides perpendicular to 
$\mbox{\boldmath $g$}$.
These boundary conditions correspond to reservoirs and sinks of 
particles at opposite ends of the sample in the quantum analogy.
Because periodic boundary conditions mimic the effects of such 
sources and sinks, we expect that the basic results of this paper 
apply to slab-like geometries as well.

\section{One-dimensional one-impurity problem}
\label{sec-1D1imp}
In the present section, we analyze the non-Hermitian 
Hamiltonian~(\ref{12020}) analytically in one dimension with 
a single impurity.
The calculation is instructive as well as of physical relevance for 
vortex lines in the case of a twin boundary.
Throughout this section, we assume without loss of generality
that the non-Hermitian field is 
applied in the positive $x$ direction, {\it i.e.}\
$g>0$.

\subsection{Exact solution: a point impurity}
\label{sec-1D1imp-exact}

To determine the effect of an isolated impurity,
we calculate the right-eigenvectors of 
the Schr\"odinger equation 
${\cal H}\psi^R(x)=\varepsilon\psi^R(x)$ with the Hamiltonian defined by
\begin{equation}\label{31010}
{\cal H}
\equiv
\frac{(p+ig)^2}{2m}-V_0\delta(x)
=-\frac{(\hbar\nabla-g)^2}{2m}-V_0\delta(x)
\end{equation}
with $V_{0}>0$ and  the periodic boundary 
conditions,
\begin{equation}\label{31015}
\psi^R(L_x)=\psi^R(0).
\end{equation}

The details of the solution are given in Appendix~\ref{app-1D}.
There is a unique localized ground state for $g<g_{c}$, where
the critical field is 
\begin{equation}\label{31111}
g_{c}\equiv\hbar\kappa_{\rm gs}=\frac{mV_{0}}{\hbar}.
\end{equation}
(Recall that the ground state is the state with the lowest 
{\em real} part of the eigenenergy.)
This solution is the only localized state in the entire spectrum.
There are extended excited states for $g<g_{c}$ as well as extended 
ground {\em and} excited states for $g>g_{c}$.
For later use in the next subsection, we write down the solutions
of the extended states including finite-size corrections.

First, the ground state for $g<g_{c}$ has the energy
\begin{equation}\label{31101}
\varepsilon_{\rm gs}=-\frac{(\hbar\kappa_{\rm gs})^2}{2m}
=-\frac{mV_{0}^2}{2\hbar^2}
\end{equation}
with
\begin{equation}
\kappa_{\rm gs}\equiv\frac{mV_{0}}{\hbar^2}.
\end{equation}
Note that the ground-state energy is independent of $g$ and equal, 
in particular, to the ground-state energy in the Hermitian case $g=0$.
This observation is consistent with the pinning criterion given below 
Eq.~(\ref{12110}).
The current carried by the ground state therefore vanishes;
\begin{equation}\label{32010}
{\rm Im}\; J_{\rm gs}\equiv-\frac{\partial \varepsilon_{\rm gs}}{\partial g}
=0.
\end{equation}
The corresponding eigenvector is a localized function written in
the form
\begin{equation}\label{31121}
\psi^R_{\rm gs}(x)\propto\left\{
\begin{array}{ll}
\exp\left[-\left(\kappa_{\rm gs}-{g}/{\hbar}\right)x\right]
& \quad\mbox{for $x\geq0$,}
\\
\exp\left[-\left(\kappa_{\rm gs}+{g}/{\hbar}\right)|x|\right]
& \quad\mbox{for $x<0$,}
\end{array}\right.
\end{equation}
in the limit $L_{x}\to\infty$.
This is depicted in Fig.~\ref{fig1D1pin-WF}.

Next, the excited states for $g<g_{c}$ are given by
\begin{equation}\label{31218}
\varepsilon_{\rm ex}=\frac{(\hbar K_{\rm ex})^2}{2m},
\end{equation}
where
\begin{equation}\label{31221}
K_{\rm ex}=k_{\rm ex}+i\kappa_{\rm ex}
\end{equation}

\noindent
\hskip 0.4375in
\epsfxsize=2.5in
\epsfbox{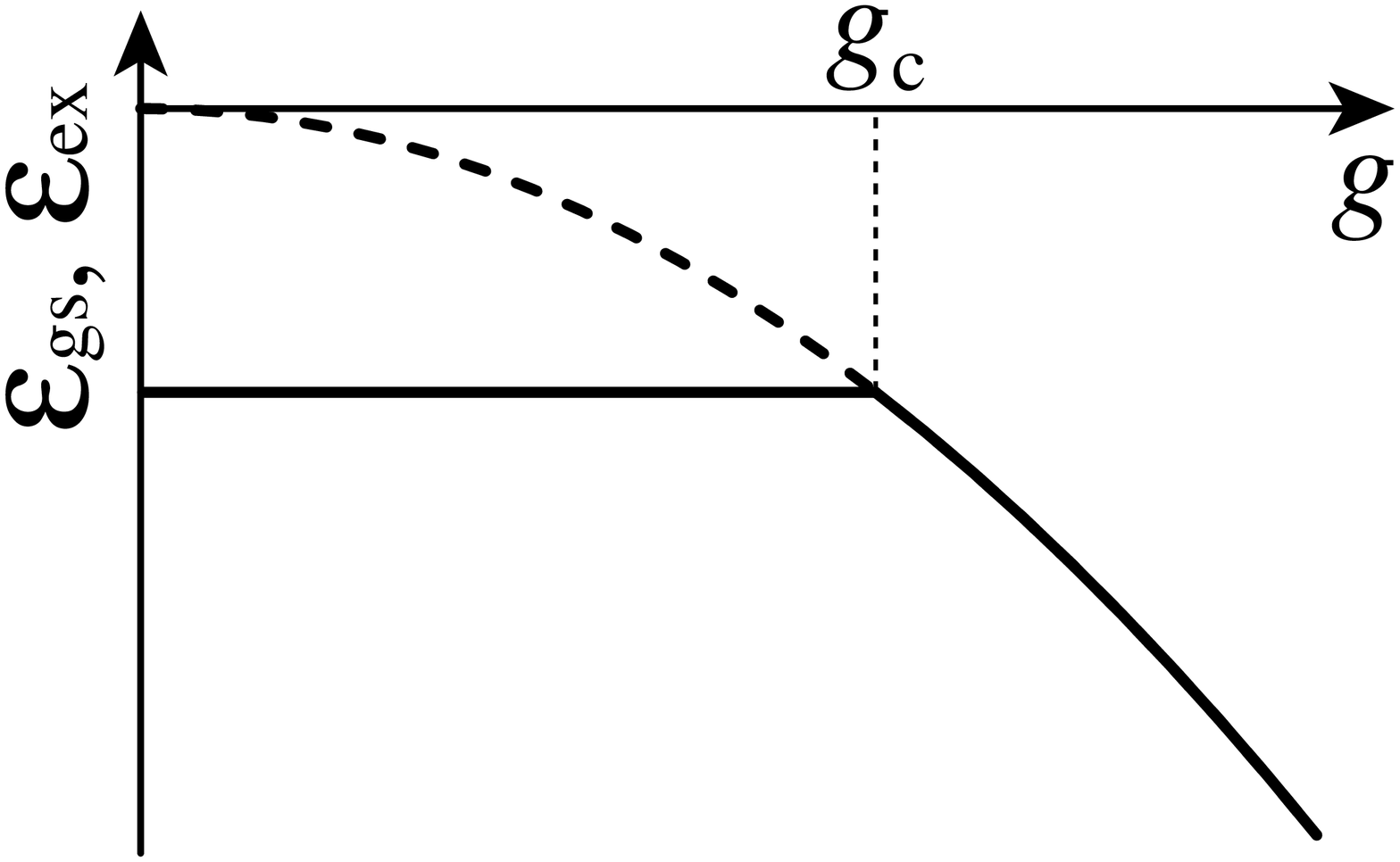}
\begin{figure}
\figcaption{
The $g$ dependence of the ground-state 
energy (solid line) and the first-excited-state energy (dashed line)
of the one-dimensional 
one-impurity problem.
}
\label{fig1D1pin-energy}
\end{figure}
with
\begin{eqnarray}\label{31231}
k_{\rm ex}&=&\frac{(2n+1)\pi}{L_{x}}+O(L_{x}^{-2}),
\\
\label{31241}
\kappa_{\rm ex}&=&\frac{g}{\hbar}
+\frac{1}{L_{x}}\ln\frac{g}{g_{c}-g}+O(L_{x}^{-2}),
\end{eqnarray}
where $n$ is an integer.

The ground and excited states for $g>g_{c}$ are given by
\begin{equation}\label{31219}
\varepsilon_{\rm gs,ex}=\frac{(\hbar K_{\rm gs,ex})^2}{2m},
\end{equation}
where
\begin{equation}\label{31222}
K_{\rm gs,ex}=k_{\rm gs,ex}+i\kappa_{\rm gs,ex}
\end{equation}
with
\begin{eqnarray}\label{31251}
k_{\rm gs,ex}&=&\frac{2n\pi}{L_{x}}+O(L_{x}^{-2}),
\\
\label{31261}
\kappa_{\rm gs,ex}&=&\frac{g}{\hbar}
+\frac{1}{L_{x}}\ln\frac{g}{g-g_{c}}+O(L_{x}^{-2}).
\end{eqnarray}
The case $n=0$ in 
Eq.~(\ref{31251}) describes a delocalized ground state when $g>g_{c}$.

The leading term in the energy eigenvalues for the extended 
states~(\ref{31218}) and~(\ref{31219}) is
\begin{equation}\label{31175}
\lim_{L_{x}\to\infty}\varepsilon_{\rm gs,ex}=-\frac{g^2}{2m}.
\end{equation}
The energies of the ground state and the first excited state depend on $g$
in the limit $L_{x}\to\infty$ as shown in Fig.~\ref{fig1D1pin-energy}.
According to the pinning criterion given below Eq.~(\ref{12110}),
the appearance of the $g$ dependence indicates that these are extended 
states.
The current carried by each state is thus purely imaginary,
\begin{equation}\label{31176}
\lim_{L_{x}\to\infty}J_{\rm gs,ex}=i\frac{g}{m}.
\end{equation}
The eigenvectors corresponding to these extended states are

\epsfxsize=3.375in
\epsfbox{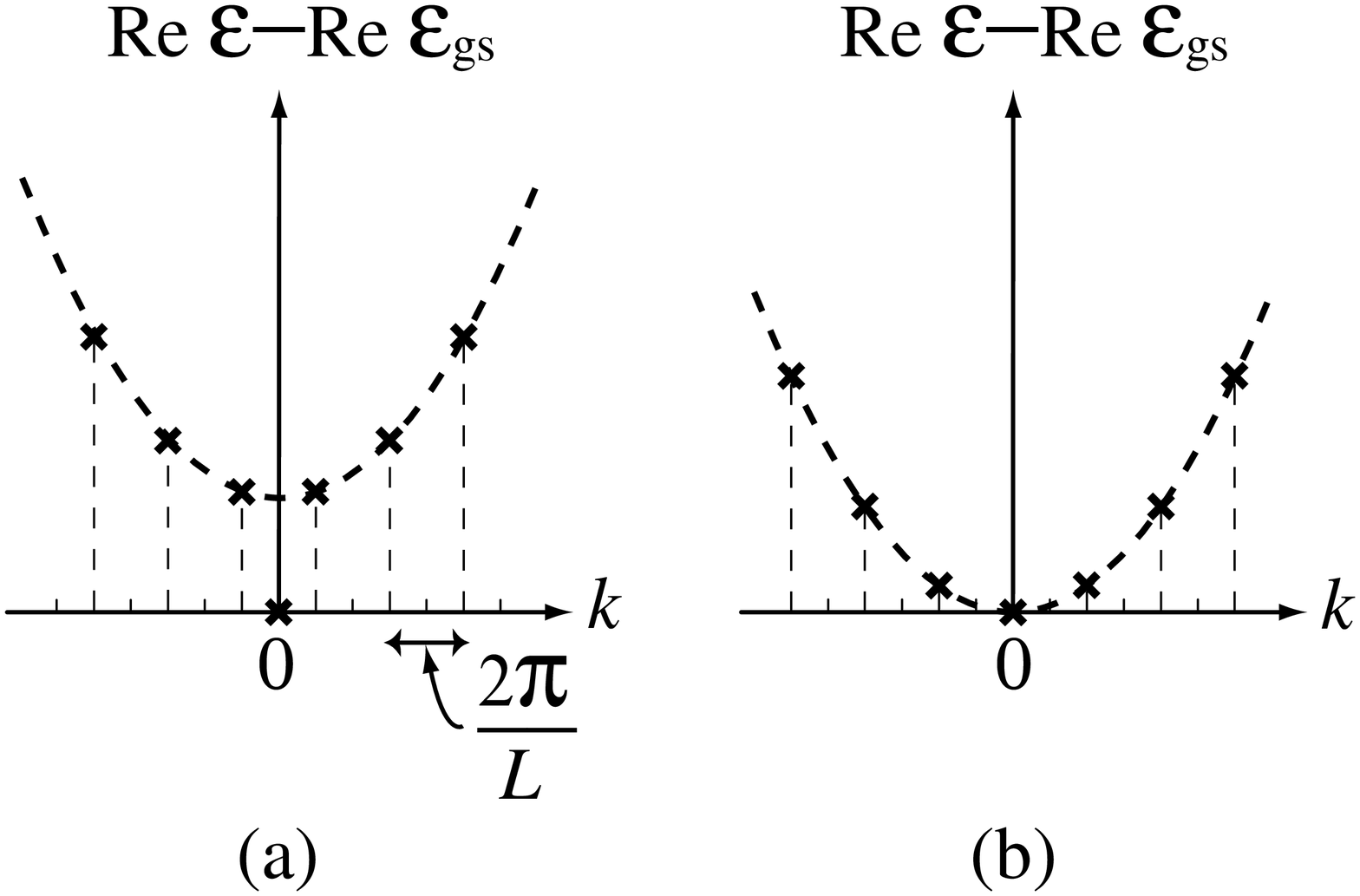}
\begin{figure}
\figcaption{
The dispersion relation ${\rm Re}\;\varepsilon$ 
vs.\ $k(\propto{\rm Im}\;\varepsilon)$ in one-dimensional one-impurity 
case: (a) for $g<g_{c}$ and (b) for $g>g_{c}$.
}
\label{fig-ReEImE1D1imp}
\end{figure}
\begin{equation}\label{31271}
\psi^R_{\rm ex}{(x)}\sim \left\{
\begin{array}{ll}
e^{ik_{\rm ex}x}            & \quad \mbox{for $x>0$,}\\
e^{ik_{\rm ex}x}+ce^{-ik_{\rm ex}x+2gx/\hbar} & \quad \mbox{for $x<0$}
\end{array}\right.
\end{equation}
in the limit $L_{x}\to\infty$, where $c$ is a constant.
The term $e^{ik_{\rm ex}x}$ is a plane wave traveling in the $x$ direction, 
and the second term of the second 
is the reflection due to the delta potential at the origin.

In Fig.~\ref{fig-ReEImE1D1imp}, we plot the dispersion relation,
that is ${\rm Re}\;\varepsilon$ against $k(\propto{\rm Im}\;\varepsilon)$.
Note the level repulsion between the impurity level and the 
extended states;
as we increase $g$, the localized impurity level approaches 
to the series of extended excited states, and must squeeze itself into the 
sequence.

Finally, the solutions at the critical point $g=g_{c}$ are again 
given by the form~(\ref{31219}), but with
\begin{eqnarray}\label{31281}
k_{\rm gs,ex}&=&O(L_{x}^{-1}\ln L_{x}),
\\
\label{31282}
\kappa_{\rm gs,ex}&=&\frac{g}{\hbar}+O(L_{x}^{-1}\ln L_{x}).
\end{eqnarray}

\subsection{Flux-line depinning from a defect}
\label{sec-1D1imp-depin}

We now apply the above solutions to flux-line physics in a thin 
cylindrical superconducting shell with one defect in it.

The flux line undergoes a depinning transition when $g=g_c$.
Its tilt slope, given as $L_{\tau}\to\infty$ by the imaginary current 
of the ground state, vanishes for $g<g_c$, but jumps to a finite value
for $g>g_c$:
\begin{equation}\label{32020}
{\rm Im}\;J_{\rm gs}=
\left\{\begin{array}{ll}
0           &\quad\mbox{for}\quad g<g_c, \\
g/m &\quad\mbox{for}\quad g>g_c.
\end{array}\right.
\end{equation}
This directly shows that the flux line is depinned from the defect at $g=g_c$.
At this depinning point, the contribution of this flux line to the
bulk transverse magnetization jumps by
\begin{equation}\label{32030}
\Delta M_{\perp}=\frac{\phi_{0}g_c}{Am},
\end{equation}
where $A$ is the cross-sectional area,
because of the relations~(\ref{12107}) and~(\ref{12110}).

Although the above jump appears to indicate a first-order 
transition,\cite{Chen96} we shall argue in 
Sec.~\ref{sec-interaction} that the bulk magnetization 
in fact grows continuously at the 
transition when interactions are taken into account.
There are, moreover, diverging precursors of the destruction of the 
transverse Meissner effect as $g\to {g_{c}}^{-}$.
As discussed in Subsec.~\ref{sec-probdist},
the normalized right-eigenvector gives the probability
distribution of the corresponding flux line at the top surface.
The wave function~(\ref{31121}) in the pinning regime ($g<g_c$) 
defines a surface localization length $\xi_\perp$ of the flux line
as $\xi_{\perp}\equiv(\kappa_{\rm gs}-g/\hbar)^{-1}$, 
which diverges as $g\to {g_{c}}^{-}$,
\begin{equation}\label{32040}
\xi_\perp\sim\frac{\hbar}{g_c-g}.
\end{equation}
The transverse magnetization near the surface diverges like 
$\xi_{\perp}$,
\begin{eqnarray}
M_{\perp s}&\propto&\left\langle x\right\rangle_{L_{\tau}}
-\left\langle x\right\rangle_{L_{\tau}/2}
=\frac{(\kappa_{\rm gs}-g/\hbar)^{-2}+(\kappa_{\rm gs}+g/\hbar)^{-2}}%
{(\kappa_{\rm gs}-g/\hbar)^{-1}+(\kappa_{\rm gs}+g/\hbar)^{-1}}
\nonumber\\
&\simeq&
\frac{\hbar}{g_{c}-g}
\end{eqnarray}
(in the limit $L_{\tau},L_{x}\to\infty$), as the vortex line begins to 
tear free.

We show below that the displacement of the flux line from the defect 
depends on the depth from the surface according to
\begin{equation}\label{32050}
\left\langle x \right\rangle_{\tau}
\sim
\exp\left(-\frac{\tau}{\tau^\ast}\right)
\end{equation}
as $\tau\to\infty$, where $\tau$ denotes the distance below the surface.
(See Sec.~\ref{sec-meissner} for the many-defect case.)
The penetration depth $\tau^\ast$ diverges at the depinning point $g=g_c$
as
\begin{equation}\label{32090}
\tau^\ast\simeq\frac{m\hbar/g_c}{g_c-g}
\qquad\mbox{as}\quad
g\to {g_c}^{-}.
\end{equation}
The exponent of the penetration-depth divergence is the same 
as the one of the surface-localization-length divergence~(\ref{32040}):
If we define the exponent $z$ by $\tau^\ast\sim(\xi_\perp)^z$, we have $z=1$.
(The exponent is different in the many-defect case; see 
Sec.~\ref{sec-meissner}.)

The divergence~(\ref{32090}) is derived as follows:
Expand Eq.~(\ref{12080}) with respect to eigenfunctions and
take the limit $L_{\tau}\to\infty$ in~(\ref{12080}) to obtain
\begin{equation}\label{32060}
\left\langle x \right\rangle_{\tau}
\simeq
\sum_n c_n
\int dx \langle\psi_{\rm gs}\bigm|x\rangle
x\langle x\bigm|\psi_n\rangle
e^{-\tau\Delta\varepsilon_n/\hbar},
\end{equation}
where 

\noindent
\hskip 0.6875in
\epsfxsize=2in
\epsfbox{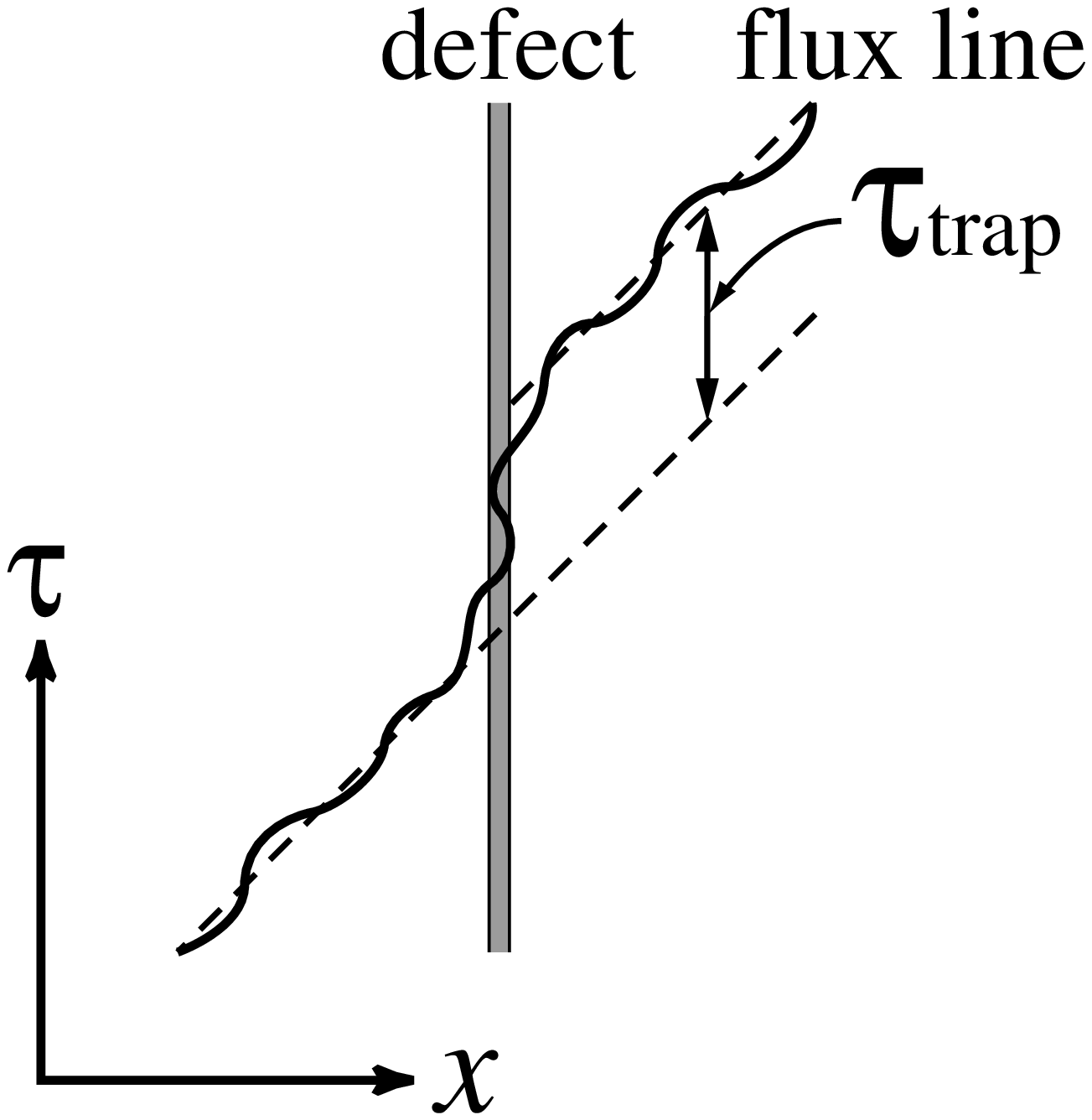}
\begin{figure}
\figcaption{
The definition of the trapping length $\tau_{\rm trap}$.
}
\label{fig1D1pin-traptime}
\end{figure}
\begin{equation}\label{32070}
c_n\equiv
\frac{\int\langle\psi_n\bigm|x'\rangle dx'}%
{\int\langle\psi_{\rm gs}\bigm|x'\rangle dx'}
\end{equation}
and
\begin{equation}\label{32075}
\Delta\varepsilon_n\equiv\varepsilon_n-\varepsilon_{\rm gs}.
\end{equation}
The term $\psi_n=\psi_{\rm gs}$ in the summation over $n$
gives the position of the
localization center of the ground state, namely the origin.
Hence the leading term is due to the first excited state.
Then the penetration depth $\tau^\ast$ in~(\ref{32050})
is given by the inverse energy gap
between the ground state and the first excited state,
\begin{equation}\label{32080}
\tau^\ast\equiv\frac{\hbar}{\Delta \varepsilon}=\frac{2m\hbar}{g_c^2-g^2},
\end{equation}
which leads to Eq.~(\ref{32090}).

There are also diverging precursors on the other side of the transition
($g\to {g_{c}}^{+}$) in the form of a diverging finite-size correction, 
namely the trapping length.
The trapping length is defined as follows (see Fig.~\ref{fig1D1pin-traptime}).
The average slope of the flux line for $x\neq0$ should be given by
the tilt slope of a flux line in the absence of defects, or $g/m$.
Assume that the flux line is trapped by the defect at $x=0$ 
over the distance $\tau_{\rm trap}$.
Then the total imaginary time necessary for a flux line to circle the periodic
system of circumference $L_x$ is $L_xm/g+\tau_{\rm trap}$.
Thus the mean tilt slope (or the imaginary part of the current) should be
given by ${\rm Im}\;J_{\rm gs}=L_x\left(L_xm/g+\tau_{\rm trap}\right)^{-1}$.
This, in turn, gives the definition of the trapping length as follows;
\begin{equation}
\tau_{\rm trap}\equiv
L_x\left(\frac{1}{{\rm Im}\;J_{\rm gs}}-\frac{m}{g}\right).
\end{equation}
For this to be finite in the limit $L_x\to\infty$, 
the imaginary part of the current should have a finite-size correction 
of the form
\begin{equation}
{\rm Im}\;J_{\rm gs}
\simeq\frac{g}{m}\left(1-\frac{\tau_{\rm trap}g/m}{L_x}\right).
\end{equation}

\noindent
\hskip 0.6875in
\epsfxsize=2in
\epsfbox{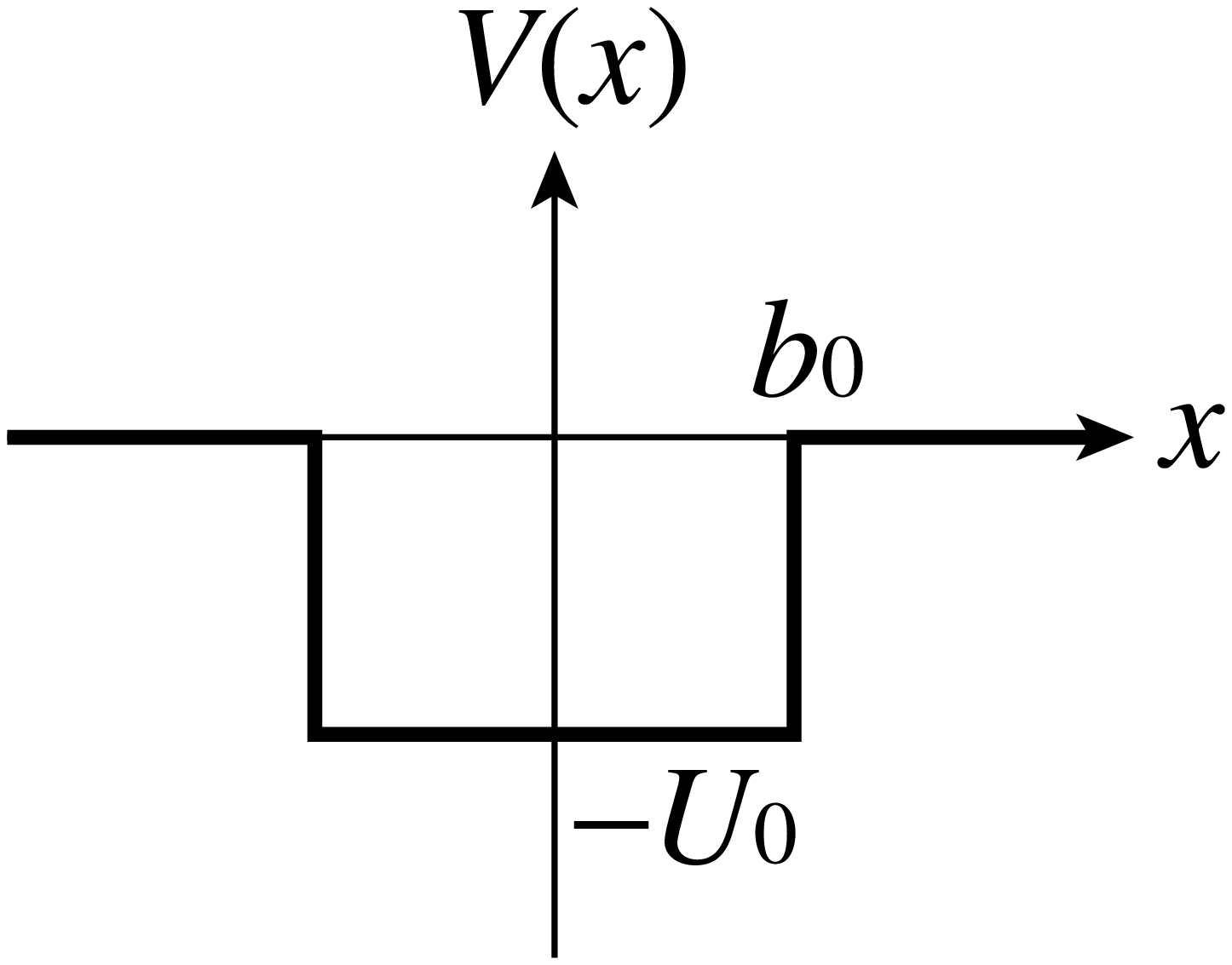}
\begin{figure}
\figcaption{
The square-well 
impurity potential treated in Sec.~\protect\ref{sec-squarewell}.
}
\label{fig1D1imp-square}
\end{figure}
In fact, the correction~(\ref{31261}) has precisely this form:
\begin{eqnarray}
{\rm Im}\;J_{\rm gs}&=&
\frac{\partial}{\partial g}
\frac{\hbar^2(\kappa_{\rm gs}(\infty)+\delta\kappa_{\rm gs}(L_x))}{2m}
\nonumber\\
&\simeq&\frac{g}{m}\left(1+\frac{\hbar\delta\kappa_{\rm gs}}{g}
+\hbar\frac{\partial\delta\kappa_{\rm gs}}{\partial g}\right).
\end{eqnarray}
We thereby obtain
\begin{equation}
\tau_{\rm trap}=\frac{m\hbar}{g^2}
\left(\frac{g_c}{g-g_c}-\ln\frac{g}{g-g_c}\right)
\simeq\frac{m\hbar/g_c}{g-g_c},
\end{equation}
which has a simple pole
as we approach the pinning point $g=g_c$ from the depinning regime.

\subsection{Exact solution: a square-well potential}
\label{sec-squarewell}

We can also solve the one-dimensional problem exactly for a single
square-well
impurity as shown in Fig.~\ref{fig1D1imp-square}.
We here restrict ourselves to localized states.
The equation~(\ref{A31151}) for the $\delta$-potential case
is now replaced by
\begin{eqnarray}\label{31510}
\lefteqn{
2k\kappa\left(\cosh(L_xg/\hbar)-\cos(2kb_0)\cosh[\kappa(L_x-2b_0)]\right)
}\nonumber\\
&&
+(k^2-\kappa^2)\sin(2kb_0)\sinh[\kappa(L_x-2b_0)]=0,
\end{eqnarray}
where
\begin{equation}
k\equiv\frac{\sqrt{2m(U_0-|\varepsilon|)}}{\hbar}
\quad\mbox{and}\quad
\kappa\equiv\frac{\sqrt{2m|\varepsilon|}}{\hbar}.
\end{equation}
The equation for the $\delta$ potential is recovered in the limit
$b_0\to0$ and $U_0\to\infty$ with $2b_0U_0=V_0$.

The results are quite similar to the $\delta$-potential case.
The critical field of the delocalization transition is again given by
\begin{equation}\label{31420}
g_c=\lim_{L_{x}\to\infty}\hbar\kappa(L_{x}),
\end{equation}
where $\kappa(L_{x})$ is a solution of Eq.~(\ref{31510}) with $g=0$.
In the limit $L_{x}\to\infty$ with $g=0$, Eq.~(\ref{31510}) is 
reduced to
\begin{equation}
2k\kappa\cos(2kb_{0})+(\kappa^2-k^2)\sin(2kb_{0})=0,
\end{equation}

\epsfxsize=3.375in
\epsfbox{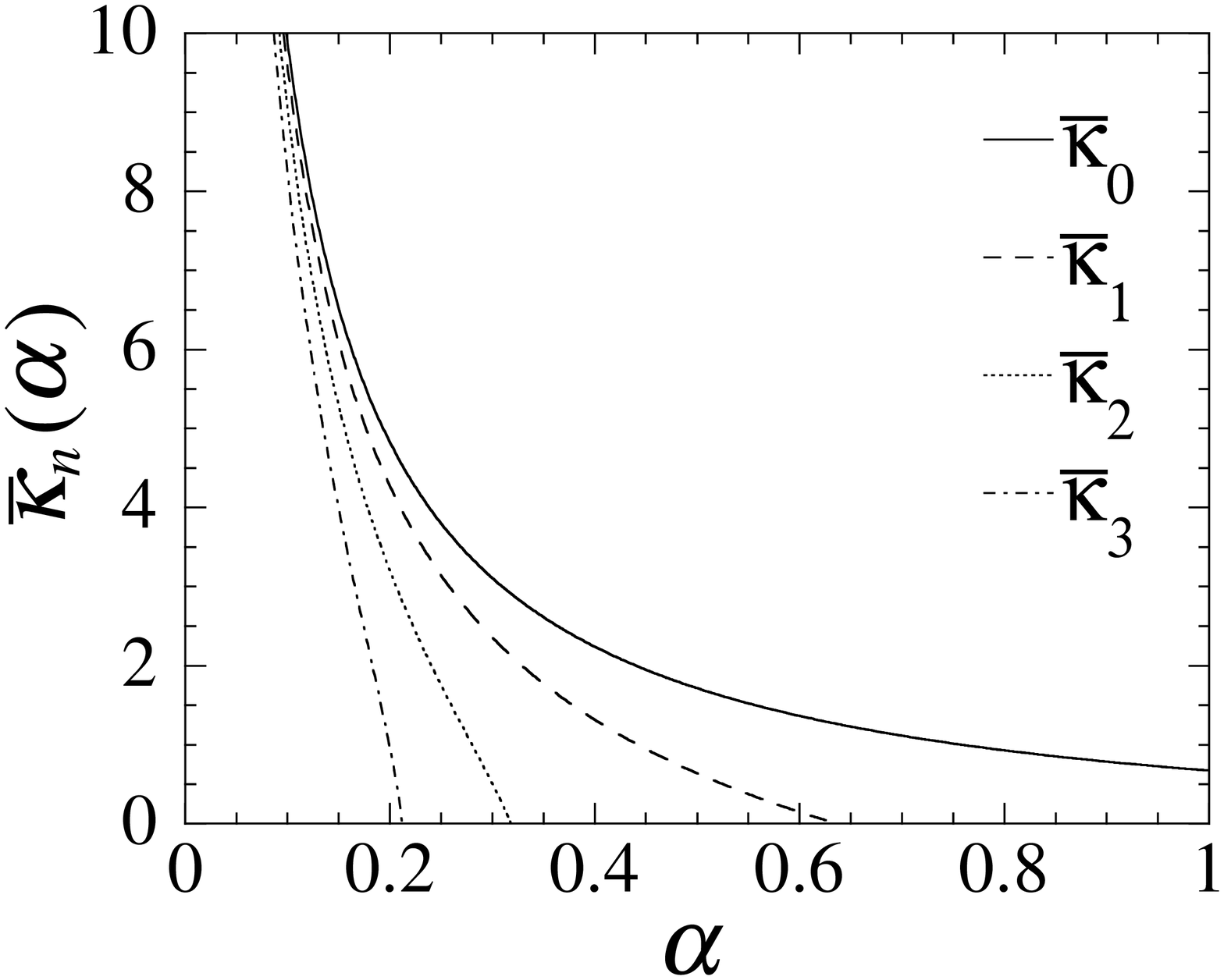}
\begin{figure}
\figcaption{
The solution $\overline{\kappa}_n(\alpha)$ 
of Eqs.~(\protect\ref{31430}) 
and~(\protect\ref{31432}), 
which is proportional to the depinning field $H_{\perp c}$ of a single
square-well impurity.
Both the ground-state function $\overline{\kappa}_0(\alpha)$
as well as the first three excited states are shown.
}
\label{fig-solution}
\end{figure}
or
\begin{equation}
\left[
\kappa\sin(kb_{0})+k\cos(kb_{0})
\right]
\left[
\kappa\cos(kb_{0})-k\sin(kb_{0})
\right]
=0.
\end{equation}
Upon putting $\overline{\kappa}\equiv b_{0}\kappa$ 
and $\eta\equiv b_{0}k$, we are 
led to
\begin{equation}\label{31430}
\overline{\kappa}^2+\eta^2=\alpha^{-2}
\end{equation}
and
\begin{equation}\label{31432}
\overline{\kappa}=\eta\tan\eta
\quad\mbox{or}\quad
\overline{\kappa}=-\eta\cot\eta,
\end{equation}
where
\begin{equation}
\alpha\equiv\frac{\hbar}{b_{0}\sqrt{mU_{0}}}
\end{equation}
is the ``reduced temperature''\cite{Nelson93a}
in the sense that $\hbar$ corresponds 
to the temperature of the flux-line system.
There is always at least one solution
of Eqs.~(\ref{31430}) and~(\ref{31432}), including the ground state.
More localized excited states are available for small $\alpha$, which 
corresponds to the low temperature of the flux-line system.
Each state is localized for $g<g_{c}=\hbar\kappa(\infty)$, where $\kappa$ 
is the inverse localization length of the state at $g=0$.
Figure~\ref{fig-solution} shows the $\alpha$-dependence of 
$\overline{\kappa}_n(\alpha)$ for the ground state and the first few excited states.
The depinning transverse field for the $n$th state is
$H_{\perp c}=(4e/cb_{0})\overline{\kappa}_n(\alpha)$
for the ground state as well as for the excited states.

The divergences discussed in the previous subsection for the 
$\delta$-function potential problem also apply to the present case.

\section{One-dimensional random non-Hermitian tight-binding model}
\label{sec-1Drandom}

In this section, we present numerical results for a one-dimensional random 
system.
To make the numerical calculation tractable, we use lattice
non-Hermitian random tight-binding models~(\ref{41010}), where the sites 
represent column positions with random binding energies 
and/or hopping matrix elements, as in Fig.~\ref{1Dpbc}.

\subsection{Site-random model}

First, we discuss the site-random model.
The second-quantized Hamiltonian in $d$ dimensions is given by 
Eq.~(\ref{41010}).
(Boson notation is used here, 
because flux lines behave like bosons in the delocalized regime,
although statistics are irrelevant when the lines are strongly 
localized.)
The hopping element is approximately \cite{Nelson93a}
\begin{equation}\label{41020}
t\sim V_{\rm bind}\exp(-\sqrt{2mV_{\rm bind}}a/\hbar), 
\end{equation}
where $V_{\rm bind}$ is a typical binding energy of the defect, and
$a$ is the lattice spacing.
We again apply periodic boundary conditions 
\begin{equation}\label{41030}
b_{\mbox{\mibscriptsize x}+N_\nu\mbox{\mibscriptsize e}_\nu}=
b_{\mbox{\mibscriptsize x}}
\qquad \mbox{for} \quad \nu=1,2,\ldots,d, 
\end{equation}
where $N_\nu\equiv L_\nu/a$.

The issues involved in determining the spectrum of Eq.~(\ref{41010}) 
are especially easy to illustrate in one dimension.
In an $N$-site basis with lattice spacing $a$, ${\cal H}$ takes an 
approximately tridiagonal form,
\begin{equation}\label{matrix}
{\cal H}=-\frac{1}{2}t
\left(\begin{array}{ccccccc}
v_{1} & e^{\overline{g}} & & & & & e^{-\overline{g}} \\
e^{-\overline{g}} & v_{2} & e^{\overline{g}} & & & \bigzerol & \\
& e^{-\overline{g}} & v_{3} & e^{\overline{g}} & & & \\
& & e^{-\overline{g}} & v_{4} & \ddots & & \\
& & & \ddots & \ddots & \ddots & \\
& \bigzerou & & & \ddots & v_{N-1} & e^{\overline{g}} \\
e^{\overline{g}} & & & & & e^{-\overline{g}} & v_{N}
\end{array}\right)
\end{equation}
with $v_{n}\equiv-2V_{n}/t$ and $\overline{g}\equiv ga_{0}/\hbar$.
Delocalization of the eigenfunctions is traditionally associated with 
an extreme sensitivity to boundary 
conditions;\cite{Edwards72,Licciardello75}
the periodic boundary conditions used here are reflected in the 
nonzero matrix elements in the upper right and lower left corners.
For the present non-Hermitian problem, delocalization is reflected 
in an extreme sensitivity of the {\em eigenvalues}.
Suppose the (real) eigenvalue spectrum is known exactly for the 
Hermitian site-random problem with $\overline{g}=0$.
One expects all states to be localized for this one-dimensional 
problem.
If the entries in the upper right and lower left corners are 
arbitrarily set to zero, it is easily shown that {\em all} eigenvalues for 
general $\overline{g}$ remain real and strictly equal to their values for 
$\overline{g}=0$.
As will be shown in this section, the eigenvalue spectrum for the 
original periodic problem becomes complex in the middle of the band for 
$\overline{g}$ above a threshold value of $\overline{g}_{c}>0$.
These complex eigenvalues are thus entirely due to the presence of 
nonzero upper right and lower left matrix elements.
For this non-Hermitian random problem, complex eigenvalues therefore indicate 
directly the extreme sensitivity to boundary conditions associated 
with delocalized states.

For columnar defects, an important component of the randomness often
comes from position of the extended defects 
rather than on-site disorder as assumed in this tight-binding model.
However, if we coarse-grain a system with the positional randomness, 
the resulting effective Hamiltonian will contain on-site 
disorder.\cite{Nelson93a}
By varying the energy and type of heavy ions which produce columnar 
defects, one can also generate on-site disorder directly.
We expect that similar energy spectra and delocalization phenomena 
arise for both random-site and random-hopping models, and will present 
results for a random hopping model in 
Subsec.~\ref{subsec-1Drandom-hopping}.

Similar to the discussion below Eq.~(\ref{12120}) for the continuum model, 
the eigenvalues of the non-Hermitian lattice model~(\ref{41010})
also appear in complex conjugate pairs,
thus insuring that the partition function ${\cal Z}$ is real.
Another symmetry is 
\begin{equation}\label{41040}
{\cal H}(\mbox{\boldmath $g$})^T={\cal H}(-\mbox{\boldmath $g$}).
\end{equation}
Because of this symmetry, a right-eigenfunction of
${\cal H}(\mbox{\boldmath $g$})$ equals to the left-eigenfunction
of ${\cal H}(-\mbox{\boldmath $g$})$ with the same eigenvalue.

If $V_{\mbox{\mibscriptsize x}}\equiv0$, eigenstates are Bloch waves and
the eigenvalues for general dimension $d$ are
\begin{equation}\label{41050}
\varepsilon(\mbox{\boldmath $k$})
=-t\sum_{\nu=1}^d\cos[(k_\nu+ig_\nu/\hbar)a],
\end{equation}
or
\begin{eqnarray}\label{41055}
{\rm Re}\;\varepsilon(\mbox{\boldmath $k$})
&=&
-t\sum_{\nu=1}^d\cos (k_{\nu}a) \cosh (g_{\nu}a/\hbar)
\nonumber\\
{\rm Im}\;\varepsilon(\mbox{\boldmath $k$})
&=&
t\sum_{\nu=1}^d\sin (k_{\nu}a) \sinh (g_{\nu}a/\hbar)
\end{eqnarray}
with $k_\nu=2n_\nu\pi/L_\nu$, where $n_\nu$ is an integer.
In one dimension, the eigenvalues lie on an ellipse, given by
\begin{equation}\label{ellipse}
\left[\frac{{\rm Re}\; \varepsilon}{\cosh (ga/\hbar)}\right]^2
+\left[\frac{{\rm Im}\; \varepsilon}{\sinh (ga/\hbar)}\right]^2
=t^2.
\end{equation}
Its low-energy structure is the same as the dispersion 
relation~(\ref{21110}) of the impurity-free continuum model~(\ref{21090}).
The high end of the ellipse~(\ref{ellipse}) is the dispersion relation 
for hole excitations.
The eigenfunctions take the form
$\psi^R(\mbox{\boldmath $x$})\propto
\exp(i\mbox{\boldmath $k$}\cdot\mbox{\boldmath $x$})$ with
$\psi^L(\mbox{\boldmath $x$})\propto
\exp(-i\mbox{\boldmath $k$}\cdot\mbox{\boldmath $x$})$.

Numerical calculations for $V_{\mbox{\mibscriptsize x}}\neq0$
were carried out with a random potential
$V_{\mbox{\mibscriptsize x}}$ uncorrelated in space and\linebreak

\epsfxsize=3.375in
\epsfbox{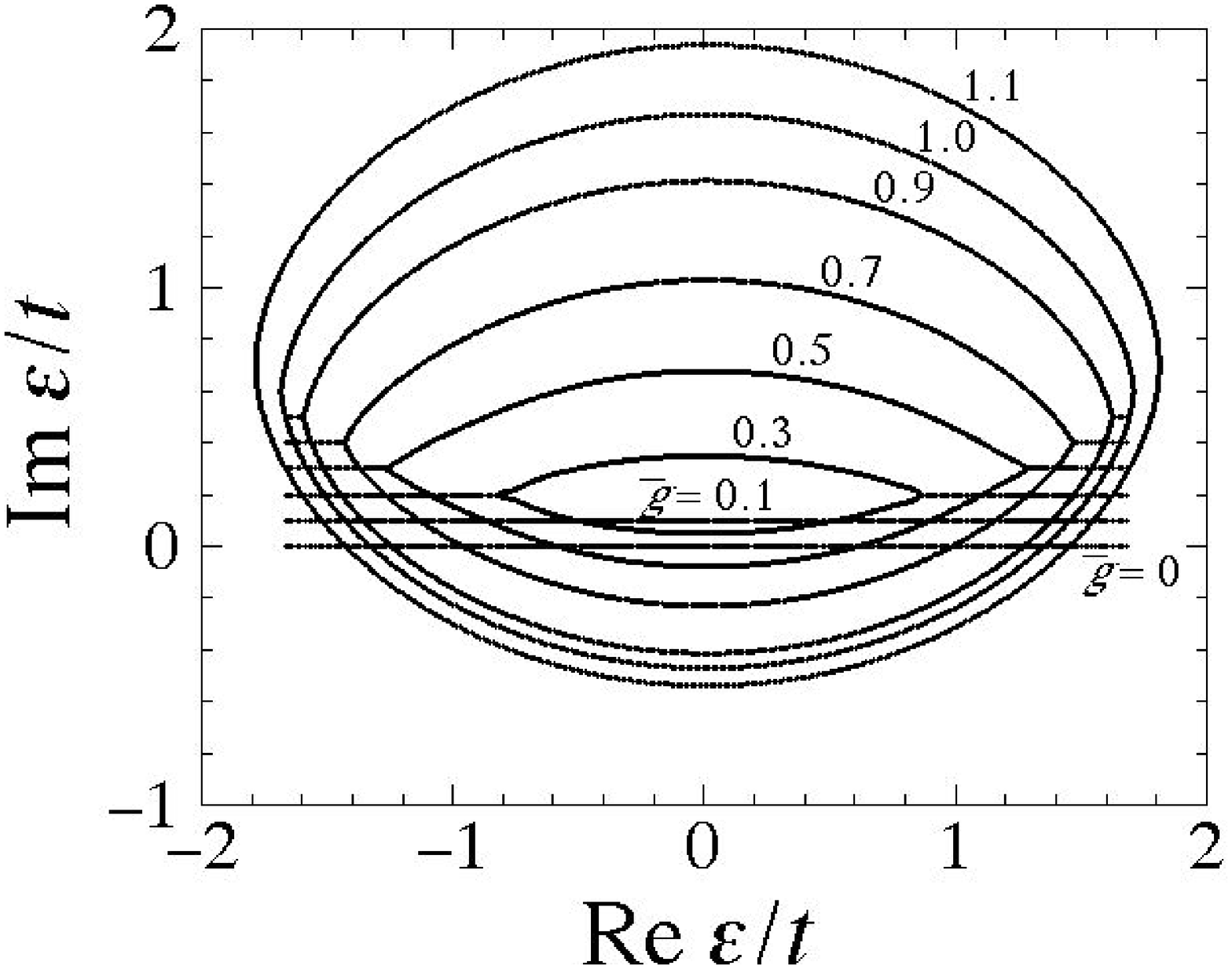}
\begin{flushleft}(a)\end{flushleft}
\vskip 10pt
\epsfxsize=3.375in
\epsfbox{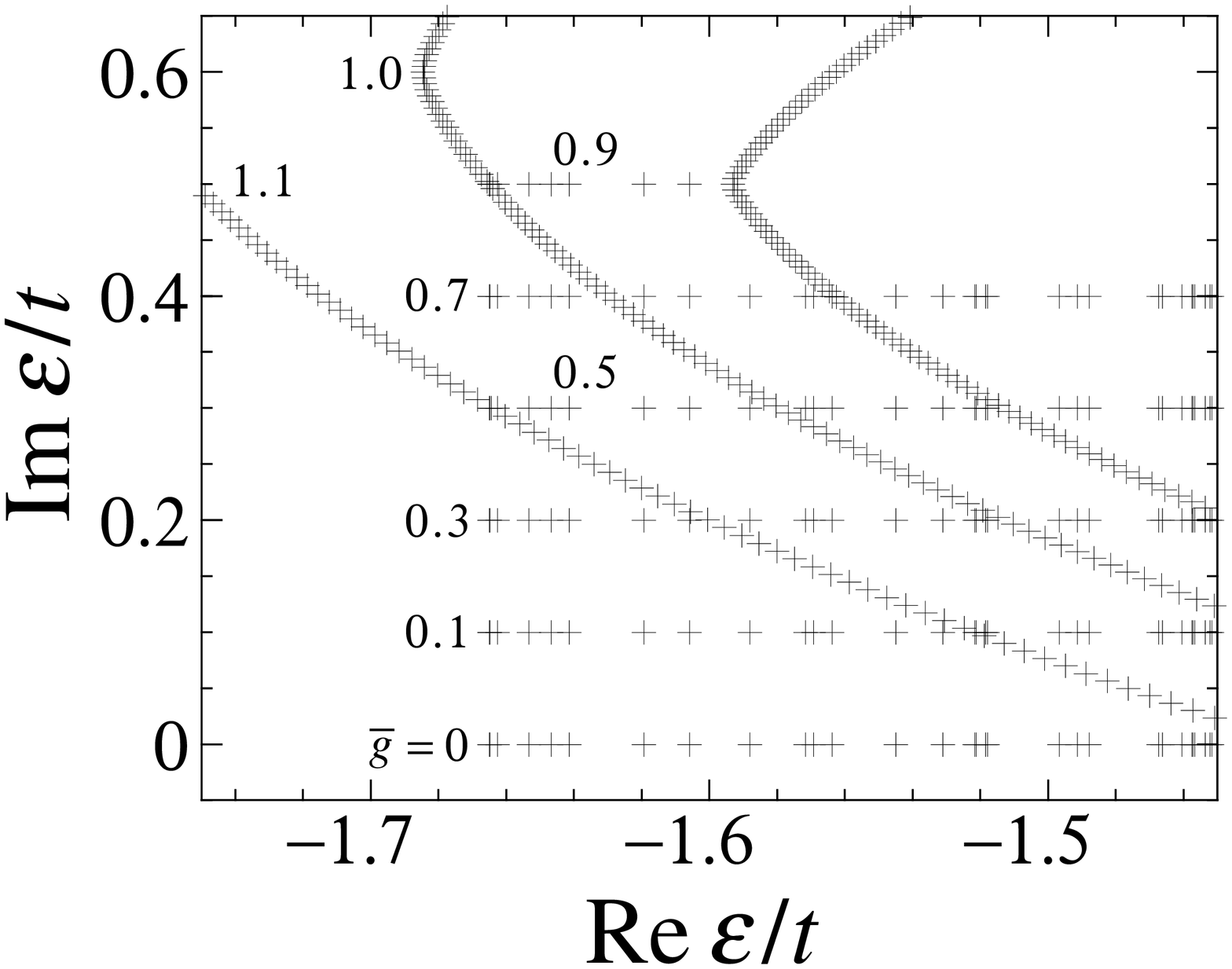}
\begin{flushleft}(b)\end{flushleft}
\begin{figure}
\figcaption{
(a) The energy spectrum of the one-dimensional tight-binding model with
randomness $\Delta/t=1$ and $L_x=1000a$. 
Plots for different values of $\overline{g}\equiv ga/\hbar$ 
are offset for clarity.
The same realization of the random potential was used for all plots here.
Each eigenstate is marked by a cross.
(b) A blowup of a part of (a);
Note that the real eigenvalues, corresponding to localized states,
are independent of $\overline{g}$.
}
\label{fig1Drandom}
\end{figure}
uniformly distributed in the range $[-\Delta,\Delta]$.
For such a symmetric distribution, 
the complex spectrum is statistically symmetric with respect to 
the axis ${\rm Re}\;\varepsilon=0$.

Figure~\ref{fig1Drandom}(a) shows the $d=1$ spectrum with $N=1000$ sites
for various values of $\overline{g}$.
As is discussed above, a complex eigenvalue 
indicates that a flux line occupying\linebreak

\epsfxsize=3.375in
\epsfbox{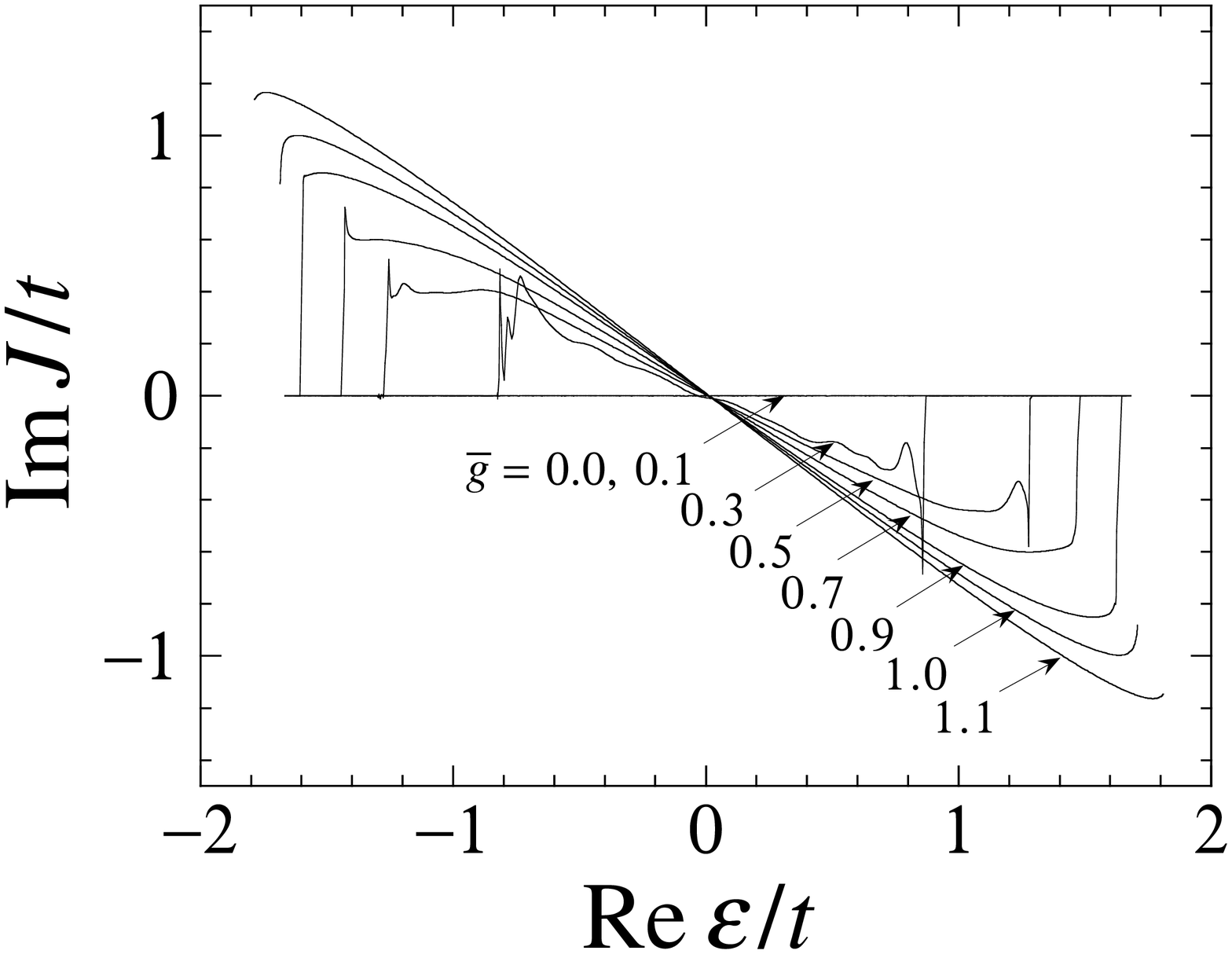}
\begin{figure}
\figcaption{
The imaginary part of the current (or the tilt slope of the corresponding 
flux line)
plotted against ${\rm Re}\;\varepsilon$ for the same sample as in 
Fig.~\protect\ref{fig1Drandom}.
}
\label{fig1Dcurrent}
\end{figure}
that state is depinned.
There is a region of $\overline{g}$ where all the eigenstates are localized.
As we increase $\overline{g}$ further, the first delocalized 
state appears in the band center, and a pair of mobility edges move 
outward toward the band edges.
As long as an eigenvalue is real, or the eigenstate is localized, the 
eigenvalue is independent of $\overline{g}$;
see Fig.~\ref{fig1Drandom}(b) for an expanded version of the bottom of 
the band:
The localized eigenvalues are in perfect registry for different values of 
$\overline{g}$.
The behavior of the delocalized states, on the other hand, is similar to the
impurity-free case Eq.~(\ref{ellipse}) except near the mobility edges.
Close to the mobility edge the imaginary part of the eigenvalue appears to
vanish {\em linearly} with the real part of the eigenvalue.

In Fig.~\ref{fig1Dcurrent}, we show the imaginary part of the current
defined by Eq.~(\ref{12120}), another indicator of the delocalization 
transition.
Upon comparing Fig.~\ref{fig1Dcurrent} with 
Fig.~\ref{fig1Drandom}(a), we note that, for each value of $\overline{g}$,
the states with complex eigenvalues coincide with those
carrying a nonzero imaginary current.
This observation is consistent with the mechanism of the delocalization 
transition presented in Sec.~\ref{sec-mechanism}.
The negative imaginary part of the current in the upper half of the band 
is due to delocalization of hole excitations.

States near the band center get delocalized first, because the inverse 
localization lengths $\kappa$ are smaller near the band center than near 
the band edges.
Figure~\ref{fig1Dkeff} shows the result of a numerical calculation of an
approximate inverse localization length for $\overline{g}=0$,
defined by
\begin{equation}\label{41060}
\kappa_n'\equiv
\left\langle\sqrt{\int x^2 |\psi_n(x)|^2dx-\left(\int x |\psi_n(x)|^2\right)^2}
\right\rangle_{\rm av},
\end{equation}
where we took the random average $\langle\cdots\rangle_{\rm av}$ over
one hundred realizations of the random potential.
According to\linebreak

\epsfxsize=3.375in
\epsfbox{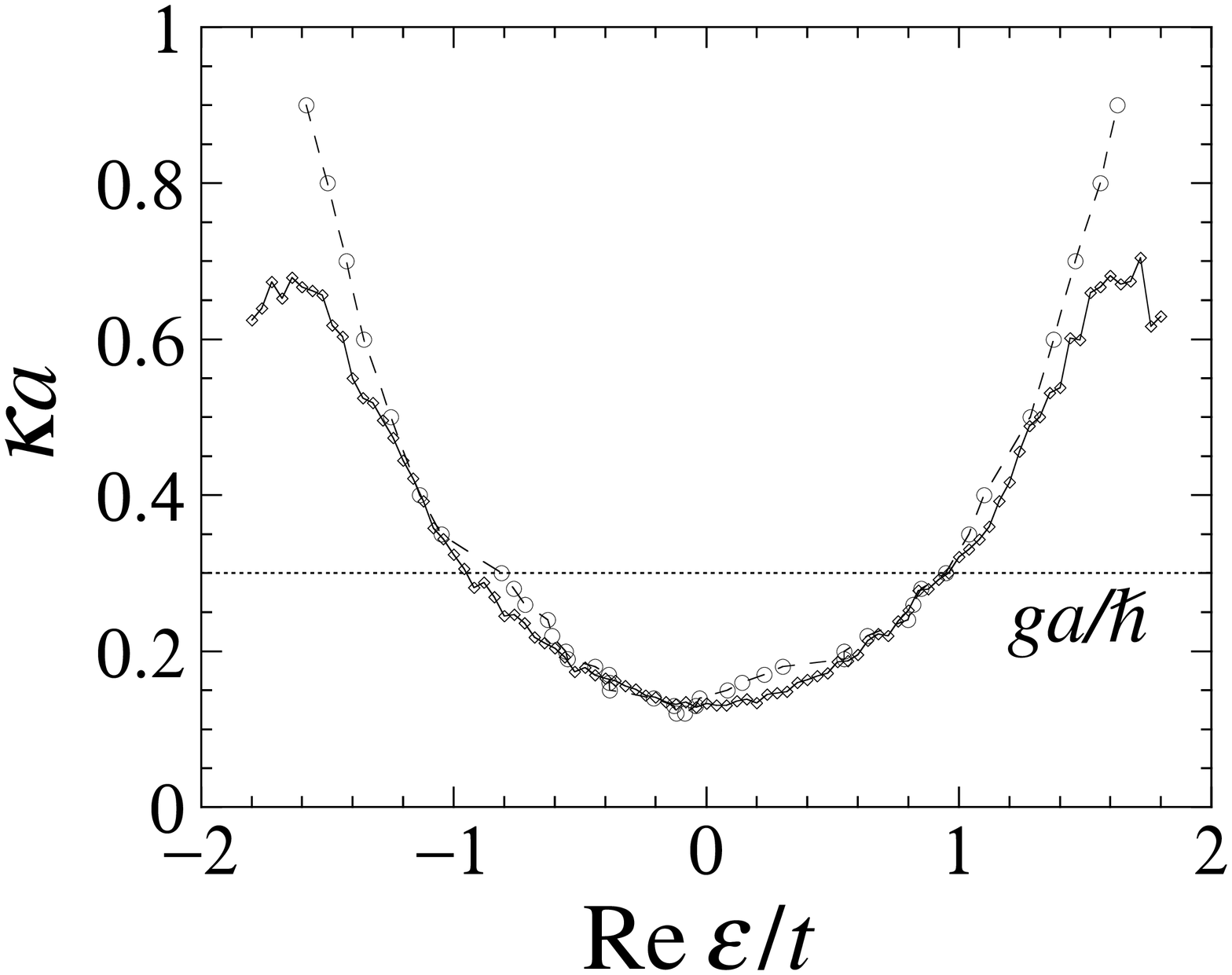}
\begin{figure}
\figcaption{
The solid curve shows an approximate inverse localization length $\kappa'$,
as determined from
Eq.~(\protect\ref{41060}), plotted against ${\rm Re}\;\varepsilon$
for $\Delta/t=1$, $L_x=500a$ and $g=0$.
The value was averaged over one hundred samples binned with energy window 
$0.04t$.
The dashed line shows an estimation of $\kappa$ based on the delocalization 
criterion $g_c=\hbar\kappa$.
For this purpose, we applied increasing $g$ to a sample of size $L_{x}=500a$.
The states at the mobility edges for a value of $g$ 
({\it e.g.}\ the dotted line) have the inverse
localization length $\kappa=g/\hbar$.
}
\label{fig1Dkeff}
\end{figure}
the delocalization criterion $g_{c}=\hbar\kappa$,
delocalized states appear first at the band center and the mobility edges 
move outward, in agreement with our numerical results.

In practice, it is convenient to {\em define} the inverse localization 
length using the delocalization criterion 
$g_{c}=\hbar\kappa$ \cite{Hwa93b} instead of using Eq.~(\ref{41060});
if a state becomes delocalized at a certain value 
$g_{c}$, the inverse localization length of the state is 
then $\kappa=g_{c}/\hbar$, shown as a dotted line
in Fig.~\ref{fig1Dkeff}.
The numerical estimation of $\kappa$ using Eq.~(\ref{41060})
is difficult near the band edges, where the number of data points 
is small, and hence a large statistical error appears.
In addition, the definition~(\ref{41060}) is not tied directly to
the asymptotic behavior $\exp(-\kappa|x-x_n|)$.

\subsection{Probability distribution of a flux line}

Diagonalization of the lattice Hamiltonian~(\ref{41010}) enables us to 
calculate the imaginary-time evolution of the wave function, and more 
importantly the probability distribution of the flux line, Eq.~(\ref{13010}).
The expansion of Eq.~(\ref{13010}) with respect to the energy
eigenstates results in
\begin{eqnarray}\label{41500}
P(\mbox{\boldmath $x$};\tau)
&=&\frac{1}{\cal Z}
\sum_{m,n}
\left\langle\psi^f\bigm|\psi_{m}\right\rangle
\left\langle\psi_{m}\bigm|\mbox{\boldmath $x$}\right\rangle\times
\nonumber\\
&&
\left\langle\mbox{\boldmath $x$}\bigm|\psi_{n}\right\rangle
\left\langle\psi_{n}\bigm|\psi^i\right\rangle
e^{-(L_{\tau}-\tau)\varepsilon_{m})/\hbar-\tau\varepsilon_{n}/\hbar}
\end{eqnarray}
with

\epsfxsize=3.375in
\epsfbox{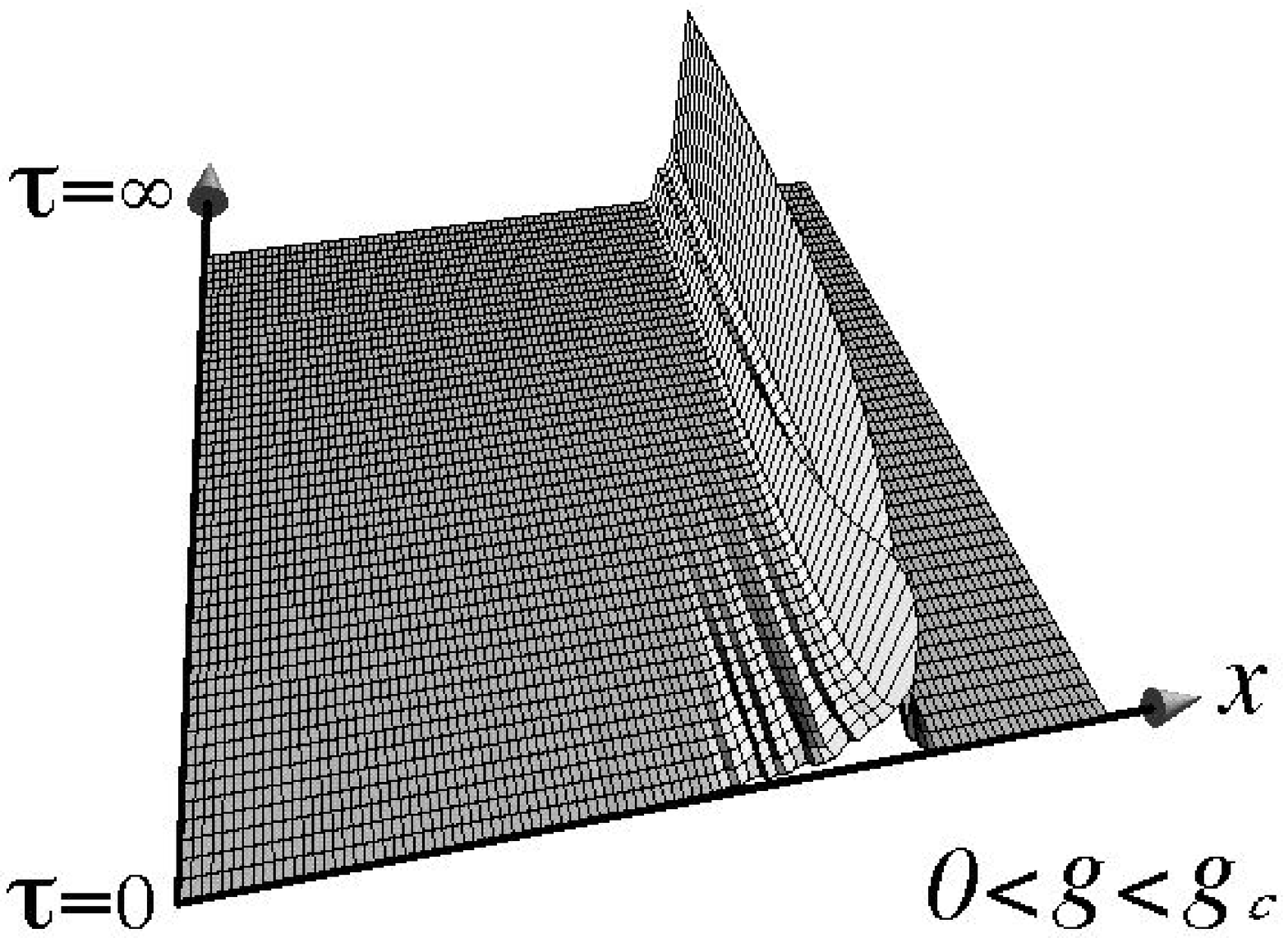}
\begin{flushleft}(a)\end{flushleft}
\vskip 10pt
\epsfxsize=3.375in
\epsfbox{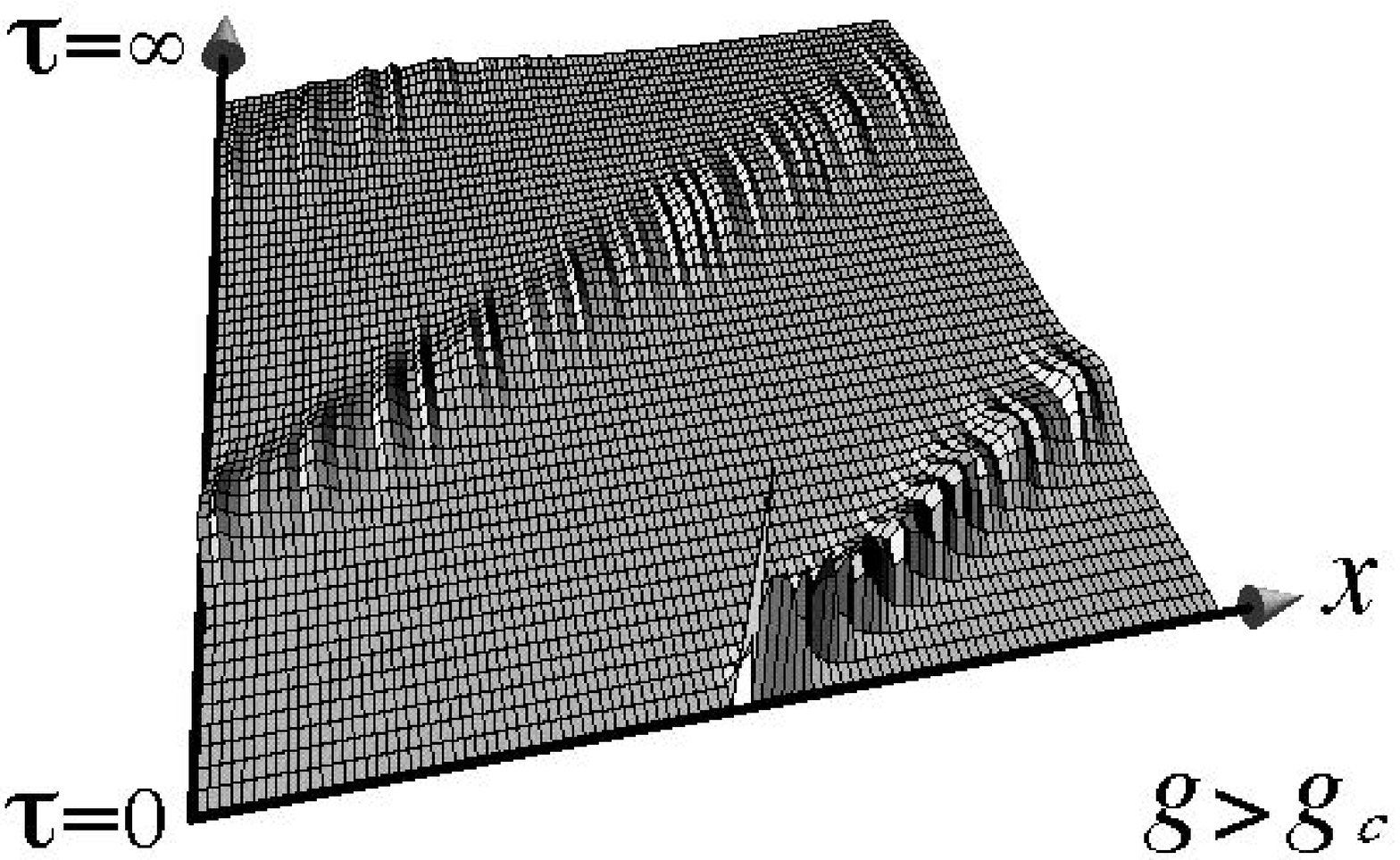}
\begin{flushleft}(b)\end{flushleft}
\begin{figure}
\figcaption{
The probability distribution of a flux line as a function of $\tau$ 
for the one-dimensional random tight-binding 
model~(\protect\ref{41010}) with a specific random potential 
$V(\mbox{\protect\mibsmall x})$ 
with $\Delta/t=1$. 
The spatial size is $L_{x}=100a$.
The front end of the figures corresponds to the bottom surface of the 
superconductor, while the bulk part $\tau>50\hbar/t$ is omitted.
(a) Case $\overline{g}=0.8<\overline{g}_c$ with free boundary conditions, 
(b) Case $\overline{g}=2.0>\overline{g}_c$ with a delta-function initial 
condition.
(The peak at $\tau=0$ in (b) has been reduced in size for visualization 
purposes.)
}
\label{figevol}
\end{figure}
\begin{equation}\label{41505}
{\cal Z}=\sum_{n}e^{-L_{\tau}\varepsilon_{n}/\hbar}
\left\langle\psi^f\bigm|\psi_{n}\right\rangle
\left\langle\psi_{n}\bigm|\psi^i\right\rangle.
\end{equation}
In the limit $L_{\tau}\to\infty$, in particular, we have
\begin{equation}\label{41510}
P(\mbox{\boldmath $x$};\tau)
\simeq
\frac{\left\langle\psi_{\rm gs}\bigm|\mbox{\boldmath $x$}\right\rangle}%
{\left\langle\psi_{\rm gs}\bigm|\psi^i\right\rangle}
\sum_{n}
\left\langle\mbox{\boldmath $x$}\bigm|\psi_{n}\right\rangle
\left\langle\psi_{n}\bigm|\psi^i\right\rangle
e^{-\tau\Delta\varepsilon_{n}/\hbar},
\end{equation}
where 
$\Delta\varepsilon_{n}\equiv\varepsilon_{n}-\varepsilon_{\rm gs}$.

We used this formula to demonstrate numerically,
for a particular realization of the random potential, that a flux line in 
the one-dimensional random system tips over near the surface under 
the influence of the transverse magnetic field.
For $g=g_c$, the flux line is strictly localized for large $\tau$
at the strongest pinning center.
As we increase $g$, kink configurations arise which allow hopping from 
one pin to the next.
Figure~\ref{figevol}(a) shows $P(\mbox{\boldmath $x$};\tau)$ for 
\lq\lq free'' boundary conditions at the bottom ($\tau=0$) of the sample,
corresponding to
\begin{equation}\label{41520}
\left|\psi^i\right\rangle\equiv\int d\mbox{\boldmath $x$}
\left|\mbox{\boldmath $x$}\right\rangle.
\end{equation}
Note that the probability distribution near the surface has been pulled in
the negative $x$-direction by the transverse field.
Once the flux line is depinned, spiral trajectories arise.
To see this more clearly, we show in Fig.~\ref{figevol}(b) the 
probability distribution for large $g$ with the initial vector
\begin{equation}\label{41530}
\left|\tilde{\psi}^i\right\rangle\equiv
\left|\mbox{\boldmath $x$}_{0}\right\rangle,
\end{equation}
so that the end of the flux line is fixed to $x_{0}$ for $\tau=0$.

\subsection{Random-hopping model}
\label{subsec-1Drandom-hopping}

A one-dimensional Hamiltonian with off-diagonal randomness 
may be more suitable for describing the experimental situation 
in Ref.~\cite{Marchetti95} than the present diagonal randomness.
Mutually parallel twin boundaries might represent the pinning potential 
in this situation.
As discussed above, we project out the 
coordinate which is parallel to the twin boundaries and perpendicular 
to the $\tau$ axis.
Randomness arises from the separation of the binding twin boundaries
rather than from the strength of the pinning.
Thus we have
\begin{equation}\label{42010}
{\cal H}\equiv-\frac{1}{2}
\sum_j\left(
t^+_{j}b^\dagger_{j+1}b_{j}+t^-_{j}b^\dagger_{j}b_{j+1}
\right),
\end{equation}
where $t^\pm_{j}\equiv V_{\rm bind}\exp[(-\lambda\pm g)a_{j}/\hbar]$, 
$\lambda\equiv\sqrt{2mV_{\rm bind}}$, and
$a_{j}$ is the separation between the $j$th and $(j+1)$th binding impurities.
If the twin boundaries are located randomly, the random separation follows
the Poisson distribution $P(a_j)=\overline{a}^{-1}e^{-a_j/\overline{a}}$,
where $\overline{a}$ is the average separation.

The delocalization phenomenon discussed above arises in
this off-diagonal case as well;
see Fig.~\ref{fig-randomhop}.
In this example, we neglect the randomness embodied in the factor
$\exp[\pm ga_{j}/\hbar]$ and use a square distribution of the hopping 
elements:
\begin{equation}\label{42011}
t^{\pm}\equiv t e^{\pm g \overline{a}/\hbar},
\end{equation}
where $t$ is a random variable with probability

\epsfxsize=3.375in
\epsfbox{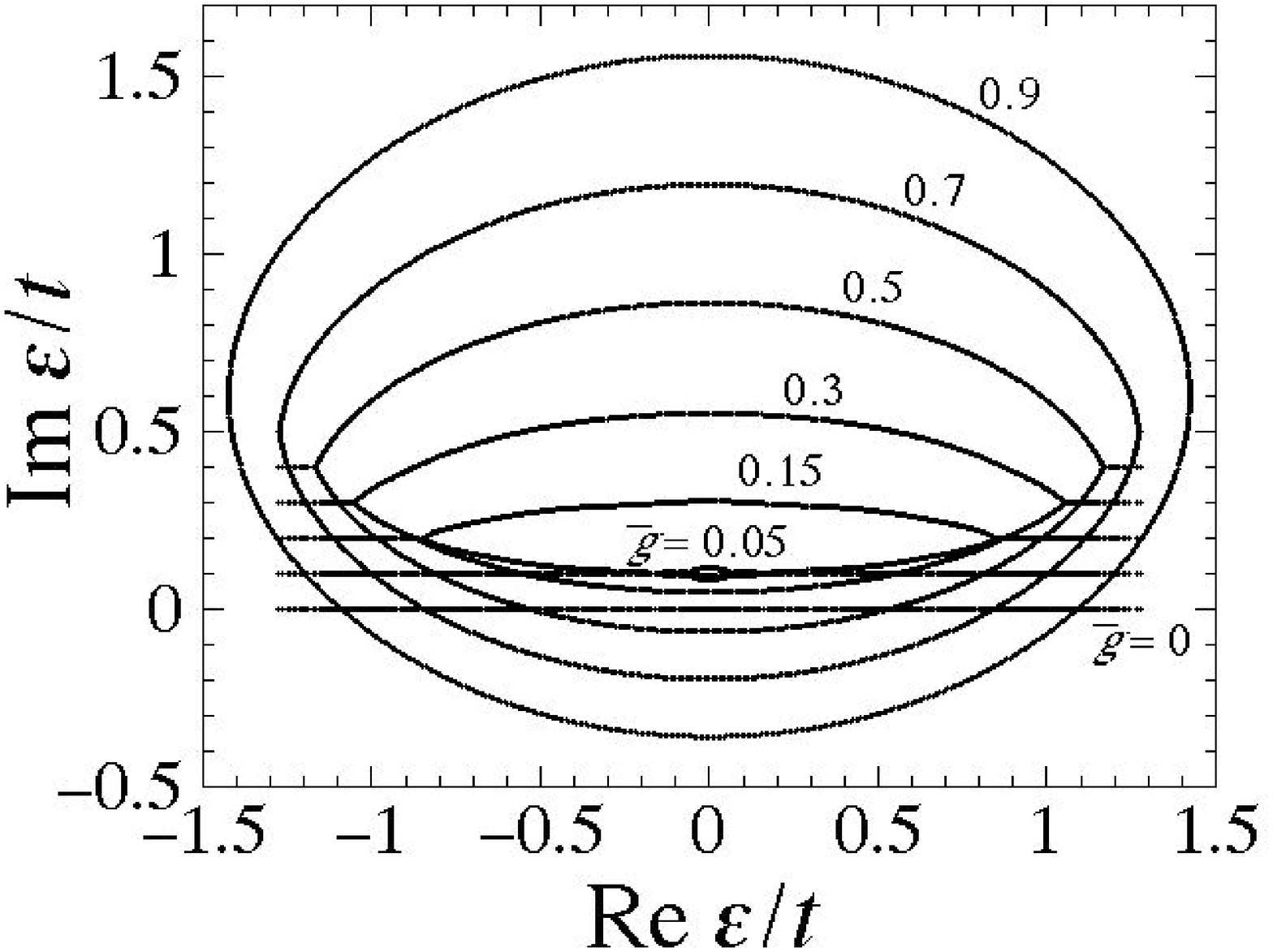}
\begin{figure}
\figcaption{
The energy spectrum of the random-hopping 
model~(\protect\ref{42010}) with the randomness defined by 
Eqs.~(\protect\ref{42011}) and~(\protect\ref{42012}).
The parameter values are $L=1000\overline{a}$ and $\Delta/t_{0}=0.5$.
}
\label{fig-randomhop}
\end{figure}
\begin{equation}\label{42012}
P(t)=\left\{
\begin{array}{ll}
1/(2\Delta) & \quad\mbox{for}\quad t_{0}-\Delta<t<t_{0}+\Delta 
\quad (t_{0}>\Delta), \\
0           & \quad\mbox{otherwise.}
\end{array}\right.
\end{equation}
Figure~\ref{fig-randomhop} is remarkably similar to the one-dimensional
spectrum found earlier for site randomness.\cite{Robbins85}
The currents carried by the extended states resemble those
shown in Fig.~\ref{fig1Dcurrent}.

\section{Two-dimensional non-Hermitian tight-binding model}
\label{sec-2D}

We now discuss numerical results for
the tight-binding model~(\ref{41010}) in two dimensions
for a square lattice with $L_{x}=L_{y}$.
We assume $g_{x}=g_{y}$, {\it i.e.}\ a tilt field along the diagonal, 
in order to reduce artifacts due to lattice periodicity.

\subsection{Impurity-free case and one-impurity case}

First, we describe the impurity-free case of the non-Hermitian 
tight-binding model~(\ref{41010}):
\begin{equation}\label{50010}
V_{\mbox{\mibscriptsize x}}\equiv 0.
\end{equation}
The energy eigenvalues are given by Eq.~(\ref{41055}) with $d=2$.
In the case of $g_{x}=g_{y}\equiv g$, we have
\begin{eqnarray}\label{50020}
{\rm Re}\;\varepsilon&=&-2t\cosh\left(\frac{ga}{\hbar}\right)
\cos\frac{\left(k_{x}+k_{y}\right)a}{2}
\cos\frac{\left(k_{x}-k_{y}\right)a}{2},
\nonumber\\
{\rm Im}\;\varepsilon&=&2t\sinh\left(\frac{ga}{\hbar}\right)
\sin\frac{\left(k_{x}+k_{y}\right)a}{2}
\cos\frac{\left(k_{x}-k_{y}\right)a}{2}
\end{eqnarray}
with

\epsfxsize=3.375in
\epsfbox{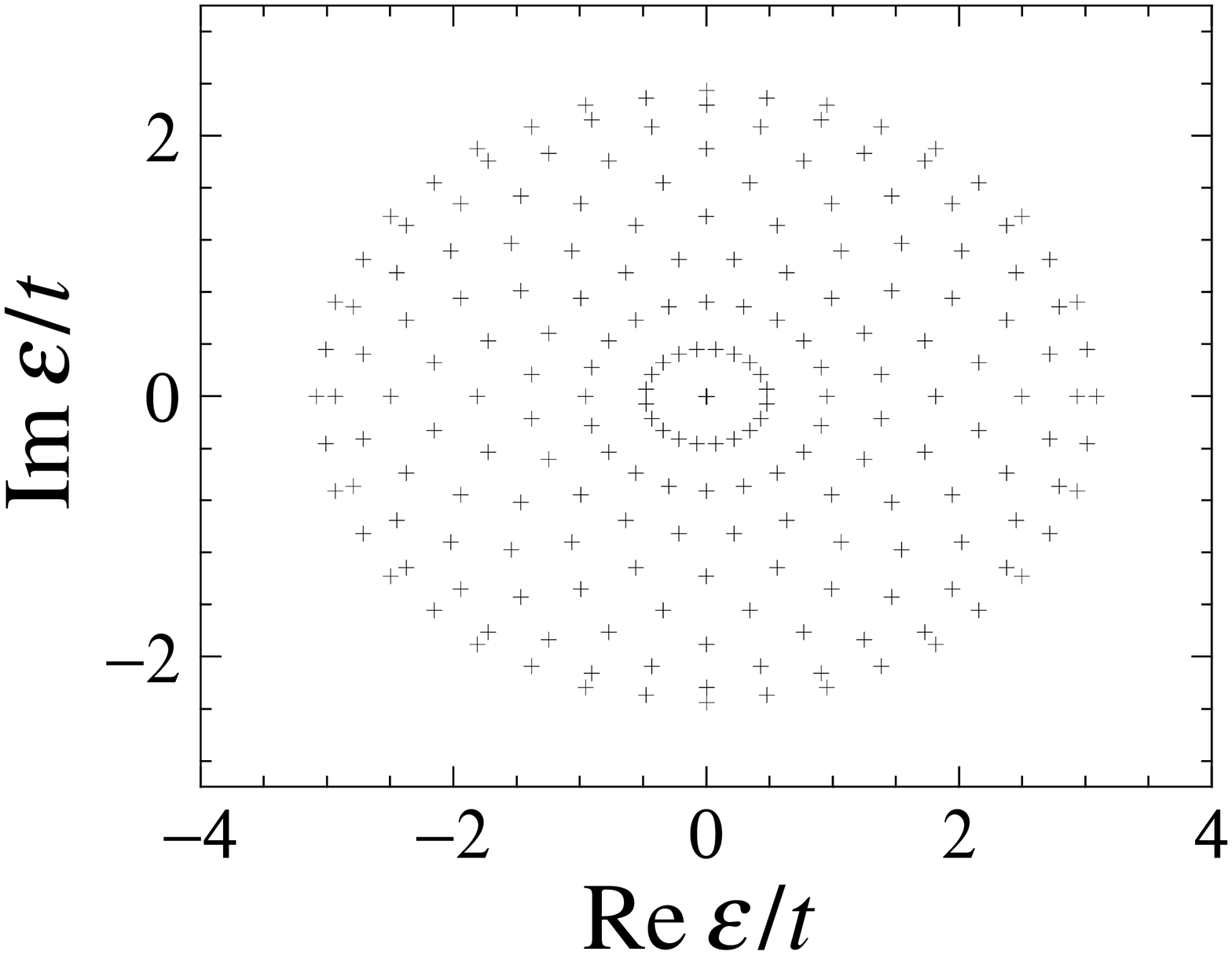}
\begin{figure}
\figcaption{
The energy spectrum 
of the two-dimensional non-Hermitian tight-binding model without 
impurities, with $g_{x}=g_{y}=1.0\times\hbar/a$ 
and $L_{x}=L_{y}=20a$.
}
\label{fig-2DReE-ImE}
\end{figure}
\begin{eqnarray}\label{50025}
k_{x}&=&0,\pm\frac{2\pi}{L_{x}},\pm\frac{4\pi}{L_{x}},\ldots,
\nonumber\\
k_{y}&=&0,\pm\frac{2\pi}{L_{y}},\pm\frac{4\pi}{L_{y}},\ldots,
\end{eqnarray}
or
\begin{equation}\label{50030}
\left[\frac{{\rm Re}\;\varepsilon}{\cosh({ga}/{\hbar})}\right]^2
+\left[\frac{{\rm Im}\;\varepsilon}{\sinh({ga}/{\hbar})}\right]^2
=4t^2\cos^2\left(\frac{k_{x}-k_{y}}{2}a\right).
\end{equation}
Thus the spectrum consists of ellipses with various radii as is shown 
for $N_x=N_y\equiv N=20$ in Fig.~\ref{fig-2DReE-ImE}.
Two levels with $k_{x}$ and $k_{y}$ interchanged 
are degenerate at each cross of the figure except for the point 
$\varepsilon=0$, where $N$ levels with 
$k_{x}-k_{y}=\pi/a$ $({\rm mod}\;2\pi/a)$
are degenerate.
The eigenfunctions have the usual Bloch form:
\begin{equation}\label{50040}
\left|\mbox{\boldmath $k$}\right\rangle
\equiv
b^\dagger_{\mbox{\mibscriptsize k}}|0\rangle
\equiv
\frac{1}{\sqrt{L_{x}L_{y}}}\sum_{\mbox{\mibscriptsize x}}
e^{i\mbox{\mibscriptsize k}\cdot\mbox{\mibscriptsize x}}
b^\dagger_{\mbox{\mibscriptsize x}}|0\rangle.
\end{equation}

The important phenomenon of level repulsion in the complex plane can be 
illustrated with one attractive point impurity:
\begin{equation}\label{51010}
V_{\mbox{\mibscriptsize x}}=
-U_{0}\delta_{\mbox{\mibscriptsize x}\mbox{\mibscriptsize x}_{0}},
\end{equation}
where $\mbox{\boldmath $x$}_{0}$ denotes the position of the impurity.
Figure~\ref{fig2D1imp} shows the results for the $20\times 20$ lattice 
with periodic boundary conditions.
Without the transverse field, we have one impurity level separated 
from a cluster of (delocalized) excited states, 
as is seen in Fig.~\ref{fig2D1imp}(a).
As we introduce $\mbox{\boldmath $g$}$ diagonally ($g_{x}=g_{y}$), 
most excited states acquire imaginary eigenvalues 
(Fig.~\ref{fig2D1imp}(b)), 
showing a spectrum similar to the impurity-free case, 
Fig.~\ref{fig-2DReE-ImE}.
The\linebreak

\epsfxsize=3.375in
\epsfbox{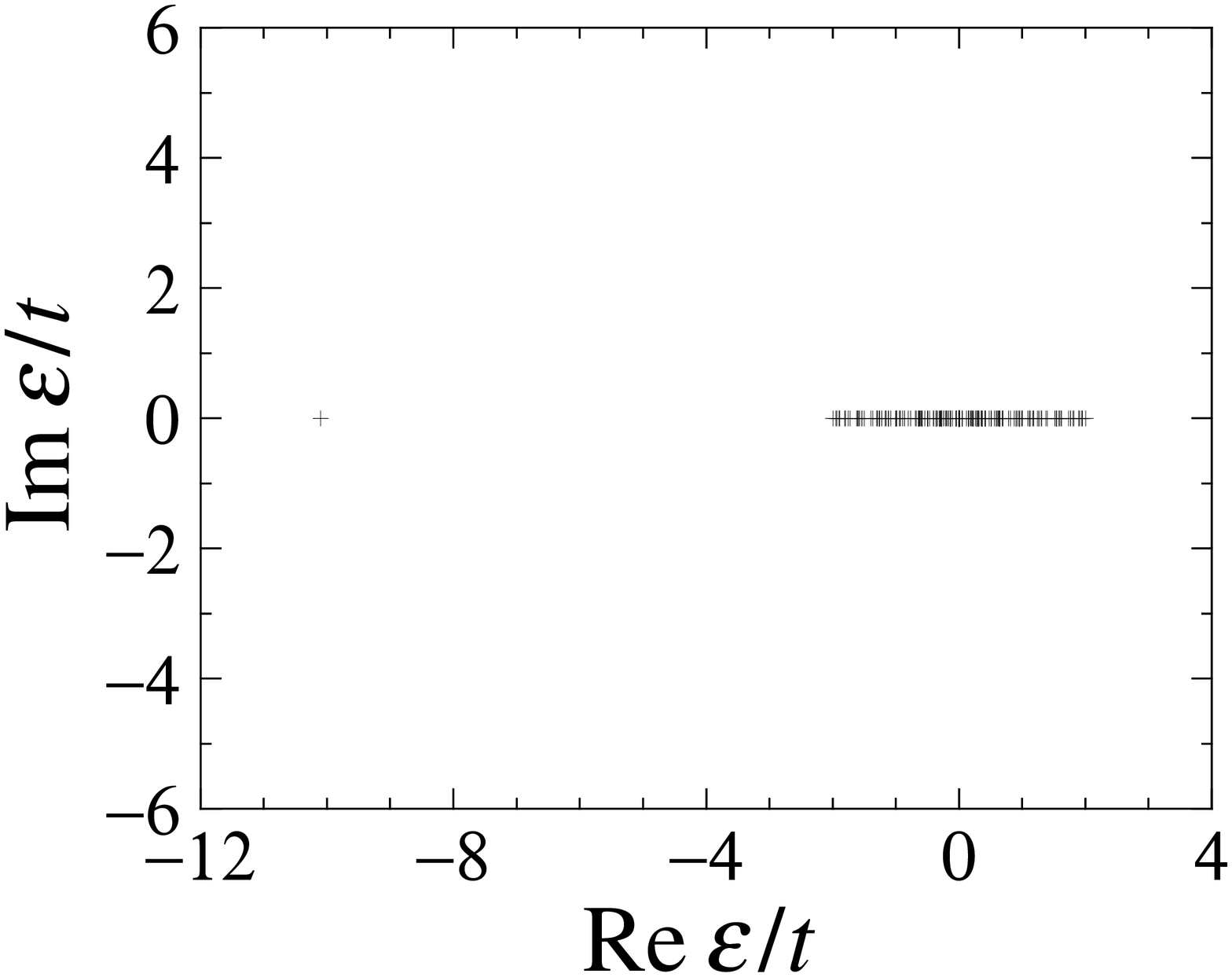}
\vskip -5pt
\begin{flushleft}(a)\end{flushleft}
\epsfxsize=3.375in
\epsfbox{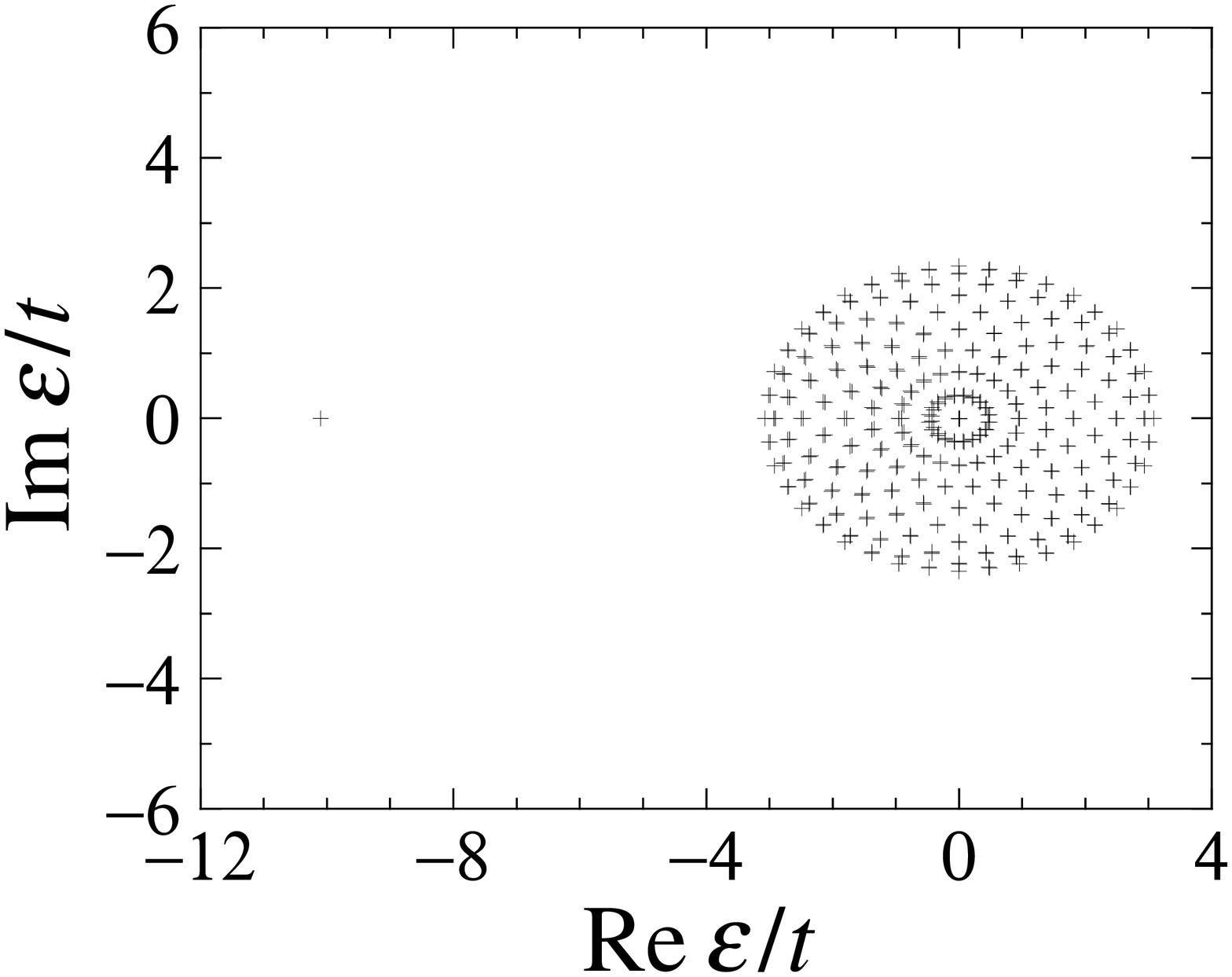}
\vskip -5pt
\begin{flushleft}(b)\end{flushleft}
\epsfxsize=3.375in
\epsfbox{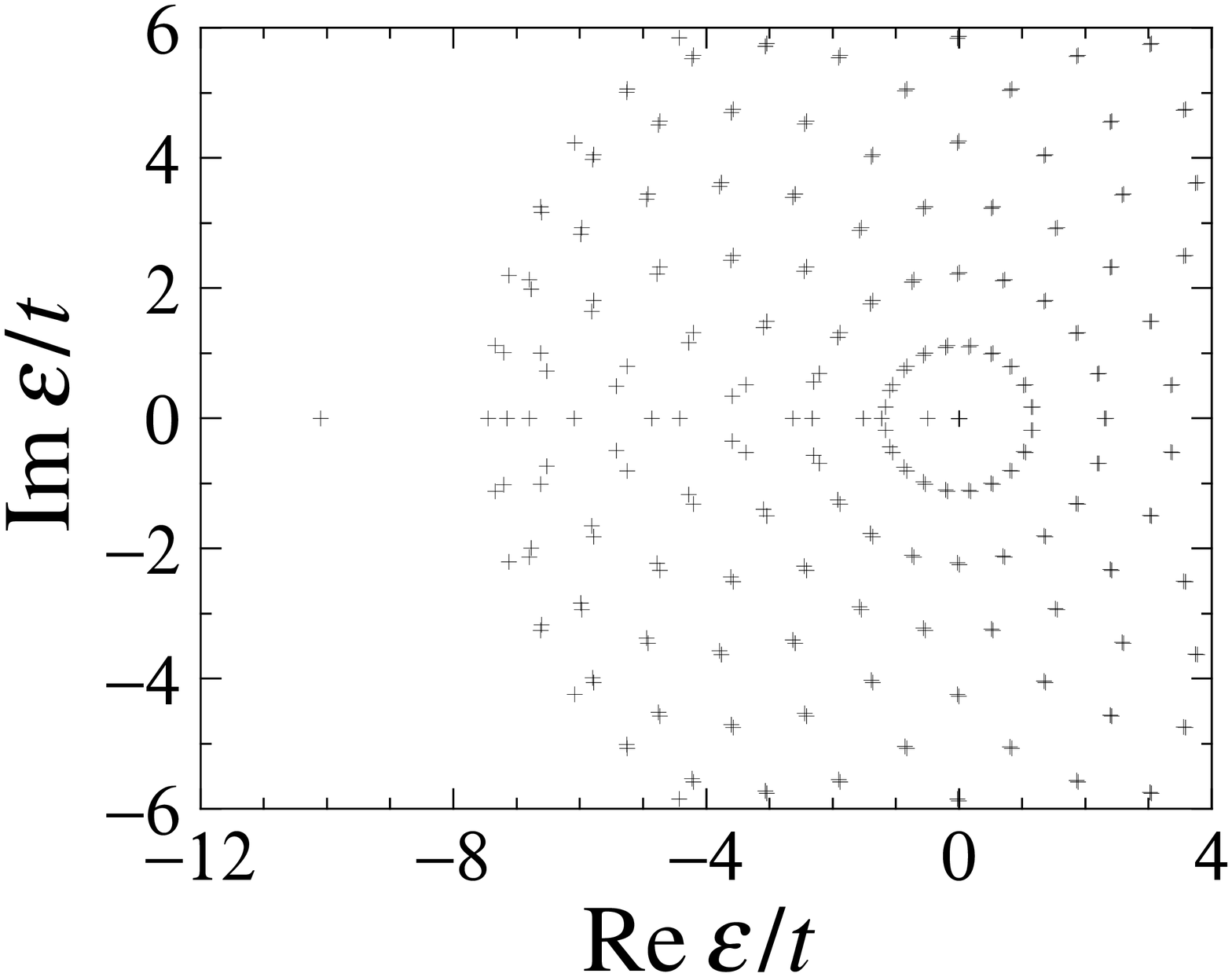}
\vskip -5pt
\begin{flushleft}(c)\end{flushleft}
\begin{figure}
\figcaption{
The energy spectra 
of the two-dimensional non-Hermitian tight-binding model with one 
attractive impurity of well depth $U_0$. 
We show here the case $L_{x}=L_{y}=20a$ and $U_{0}=10t$ with
(a) $g_{x}=g_{y}=0$, (b) $g_{x}=g_{y}=1.0\times\hbar/a$, 
and (c) $g_{x}=g_{y}=2\times\hbar/a$.
}
\label{fig2D1imp}
\end{figure}
localized impurity level does not change its energy as long
as $|\mbox{\boldmath $g$}|<g_{c}$, while the extended excited levels 
expand, following Eq.~(\ref{50030}) of the impurity-free case approximately.
In Fig.~\ref{fig2D1imp}(c), we observe level repulsion between the 
impurity level and the excited levels.
Note that the two-fold degeneracy of the pure system is split on the side 
of the spectrum, where the impurity state is about to enter.
For even larger $|\mbox{\boldmath $g$}|$, the bound state enters the 
region of delocalized levels and the spectrum is again elliptical
with an extra state near the origin.

\subsection{Random case}

In the two-dimensional random tight-binding model, 
we again find energy bands of localized levels bounded by a 
mobility edge.
For certain values of $\mbox{\boldmath $g$}$, however,
extended and localized states are 
{\em mixed} in a complicated way near the band center.
It appears that we have three regimes with respect to the 
field $\mbox{\boldmath $g$}$.
One is the pinning phase for small $\mbox{\boldmath $g$}$.
Another is the Bloch-wave regime for large $\mbox{\boldmath $g$}$.
Finally, there is an exotic intermediate regime with chaotic 
eigenvalue spectra.
We describe these regimes on the basis of our numerical data. 

First,
it is widely accepted for $\mbox{\boldmath $g$}=\mbox{\bf 0}$ 
that all eigenstates of two-dimensional random systems
are localized with finite 
localization lengths (except for a possible exception at the band 
center\cite{Robbins85}).
Hence there is again a region of small $\mbox{\boldmath $g$}$ where 
all states remain localized.
As discussed earlier, these localized states retain the real 
eigenvalues which they had for $\mbox{\boldmath $g$}=\mbox{\bf 0}$;
see the energy spectrum in Fig.~\ref{fig2Drnd}(a) for an example
with $N_x=N_y=40$.

As we increase $\mbox{\boldmath $g$}$, delocalized levels with complex 
eigenvalues
appear as is shown in the energy spectrum in Fig.~\ref{fig2Drnd}(b).
A remarkable difference of the energy spectrum from the 
one-dimensional case is that localized levels and delocalized levels 
coexist in the same range of ${\rm Re}\;\varepsilon$ 
in the two-dimensional case.
This phenomenon can be understood in terms of {\em anisotropic} localization
of the Hermitian system (in the case $\mbox{\boldmath $g$}=\mbox{\bf 0}$).
To take advantage of anisotropic fluctuations in the site potential,
there can be states with different 
localization lengths in specific directions.
As shown schematically in Fig.~\ref{figanisotropy}, in a large system 
there can be another state with nearly the same energy, but whose 
contours are rotated by $90^\circ$ from the first one.
According to the argument developed in Sec.~\ref{sec-mechanism}, 
the delocalization point of each state is 
$g_{c}=\hbar\kappa$,\linebreak 

\epsfxsize=3.375in
\epsfbox{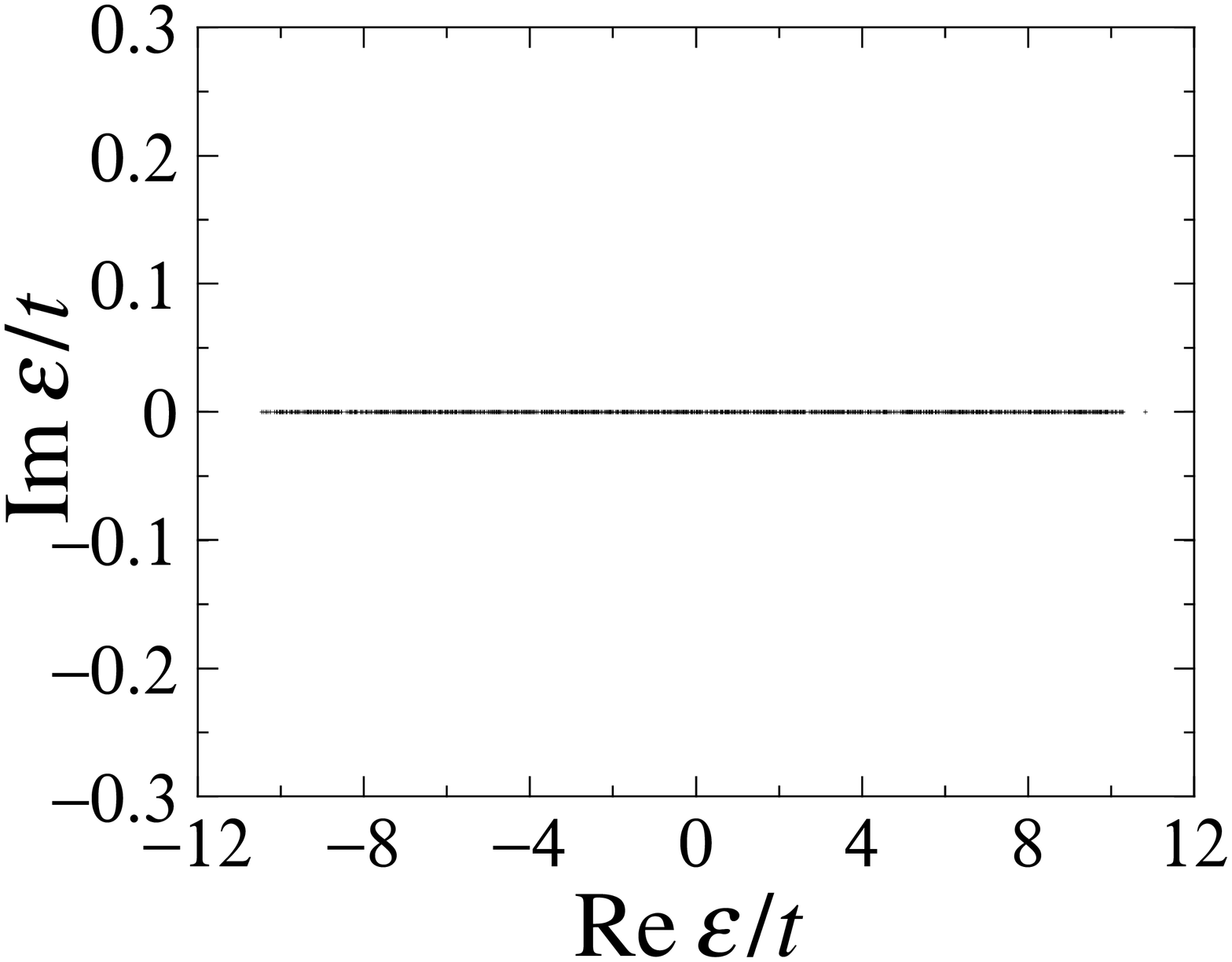}
\vskip -5pt
\begin{flushleft}(a)\end{flushleft}
\epsfxsize=3.375in
\epsfbox{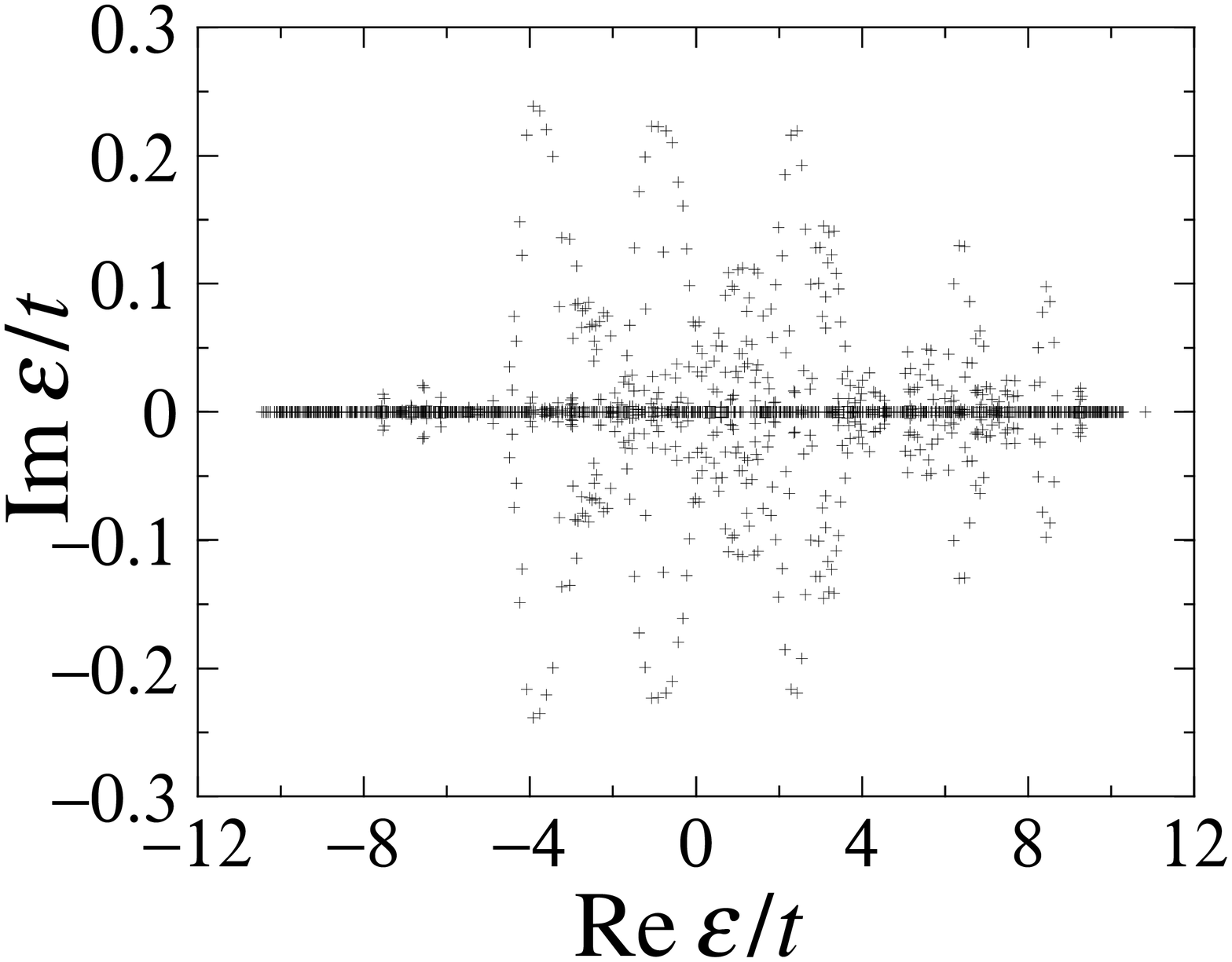}
\vskip -5pt
\begin{flushleft}(b)\end{flushleft}
\epsfxsize=3.375in
\epsfbox{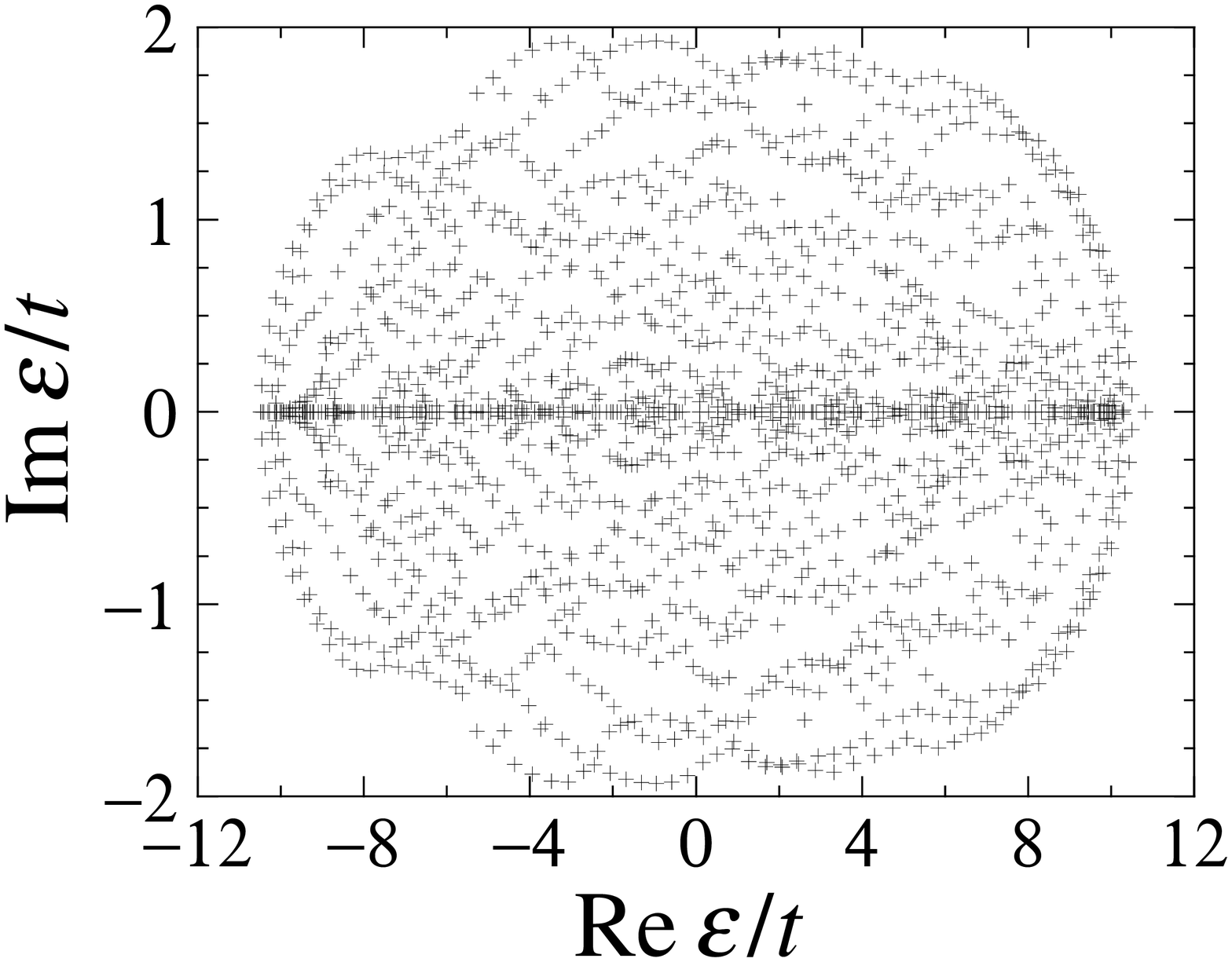}
\vskip -5pt
\begin{flushleft}(c)\end{flushleft}
\epsfxsize=3.375in
\epsfbox{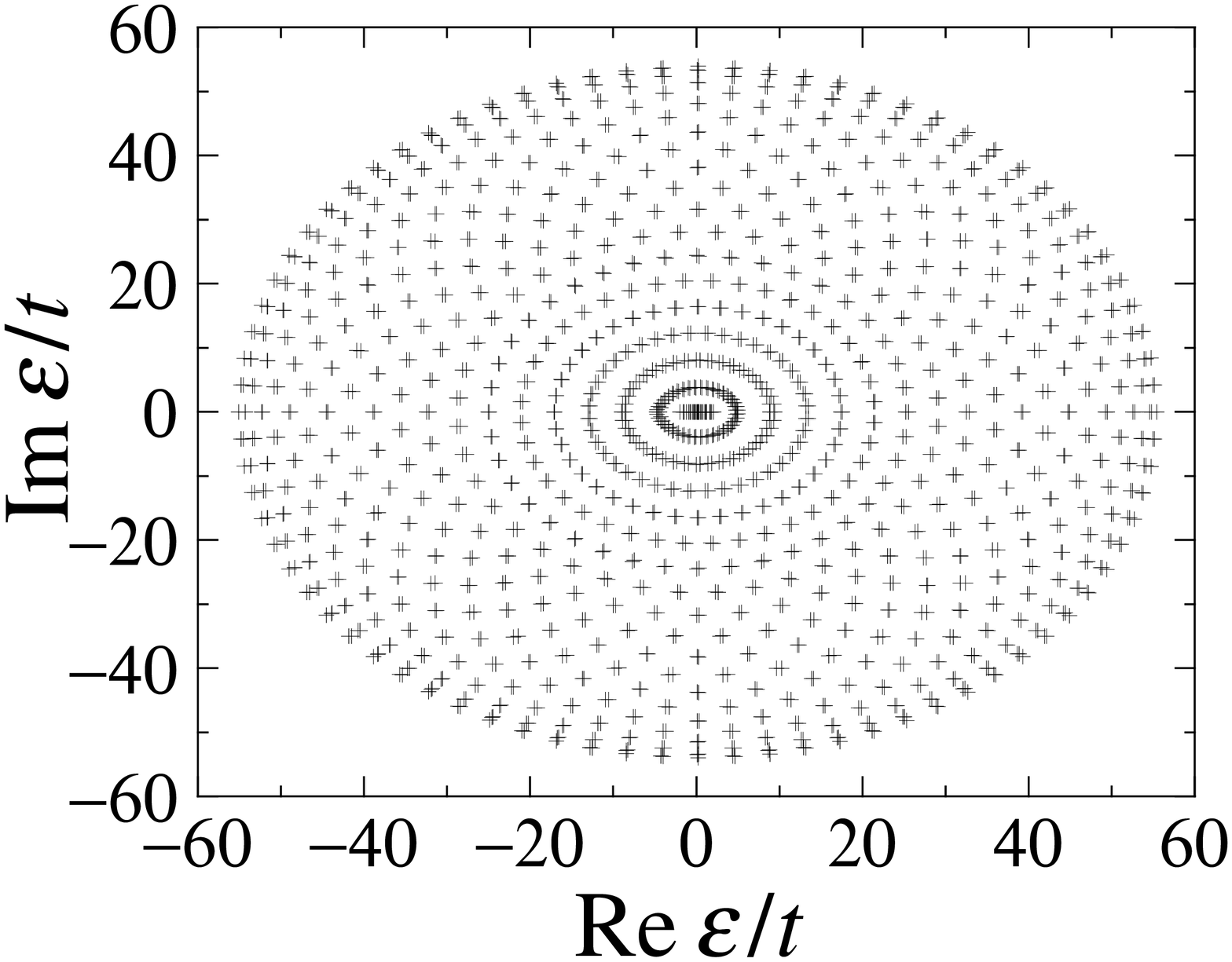}
\vskip -5pt
\begin{flushleft}(d)\end{flushleft}
\begin{figure}
\figcaption{
The energy spectra
of the two-dimensional non-Hermitian tight-binding model with site 
randomness.
We show here the case $L_{x}=L_{y}=40a$ and $\Delta=10t$ with
(a) $g_{x}=g_{y}=0.7\times\hbar/a$, (b) $g_{x}=g_{y}=1.2\times\hbar/a$, 
(c) $g_{x}=g_{y}=1.5\times\hbar/a$, and 
(d) $g_{x}=g_{y}=4.0\times\hbar/a$.
}
\label{fig2Drnd}
\end{figure}
where $\kappa$ is now the inverse localization 
length of the state in the direction of $\mbox{\boldmath $g$}$.
The state with the largest localization length
in the direction of $\mbox{\boldmath $g$}$ gets delocalized first.
Up to the delocalization point, however, both states have nearly 
identical real energies.

After passing through an intermediate region whose energy spectrum is
exemplified in Fig.~\ref{fig2Drnd}(c), 
we move onto the region of large $\mbox{\boldmath $g$}$.
The energy-spectrum structure shown in Fig.~\ref{fig2Drnd}(d) is 
similar to the impurity-free case shown in Fig.~\ref{fig-2DReE-ImE}.

It is tempting to conclude from Fig.~\ref{fig2Drnd}(d) that the 
large-$\mbox{\boldmath $g$}$ limit is well described by the extended 
Bloch wave functions, just as in one dimension.
However, unlike $d=1$, the Bloch approximation breaks down in $d=2$ 
for sufficiently large systems for {\em any} value of $\mbox{\boldmath $g$}$.
To see this, recall that all Bloch states 
$| k_{x},k_{y}\rangle$ with $k_{x}\neq k_{y}$ are two-fold degenerate 
in the impurity-free case:
If $\mbox{\boldmath $g$}$ points along the lattice diagonal as in 
Eq.~(\ref{50020}), then $| k_{x},k_{y}\rangle$ and 
$| k_{y},k_{x}\rangle$ have the same complex energy.
More generally, states related by a reflection across the 
$\mbox{\boldmath $g$}$ axis are degenerate.
(In $d=3$, the degenerate states lie on a 
circle in $(k_x, k_y, k_z)$-space centered on a line
passing through $\mbox{\boldmath $g$}$.)
This degeneracy is split by the randomness.
Degenerate perturbation theory requires diagonalization of the set 
of $2\times2$ matrices,
\begin{equation}\label{david-6a}
\left(\begin{array}{cc}
\varepsilon_{0}(k_{x},k_{y}) & \tilde{V}(k_{x}-k_{y},k_{y}-k_{x}) \\
\tilde{V}(k_{y}-k_{x},k_{x}-k_{y}) & \varepsilon_{0}(k_{x},k_{y}) 
\end{array}\right),
\end{equation}
where

\noindent
\hskip 0.6875in
\epsfxsize=2in
\epsfbox{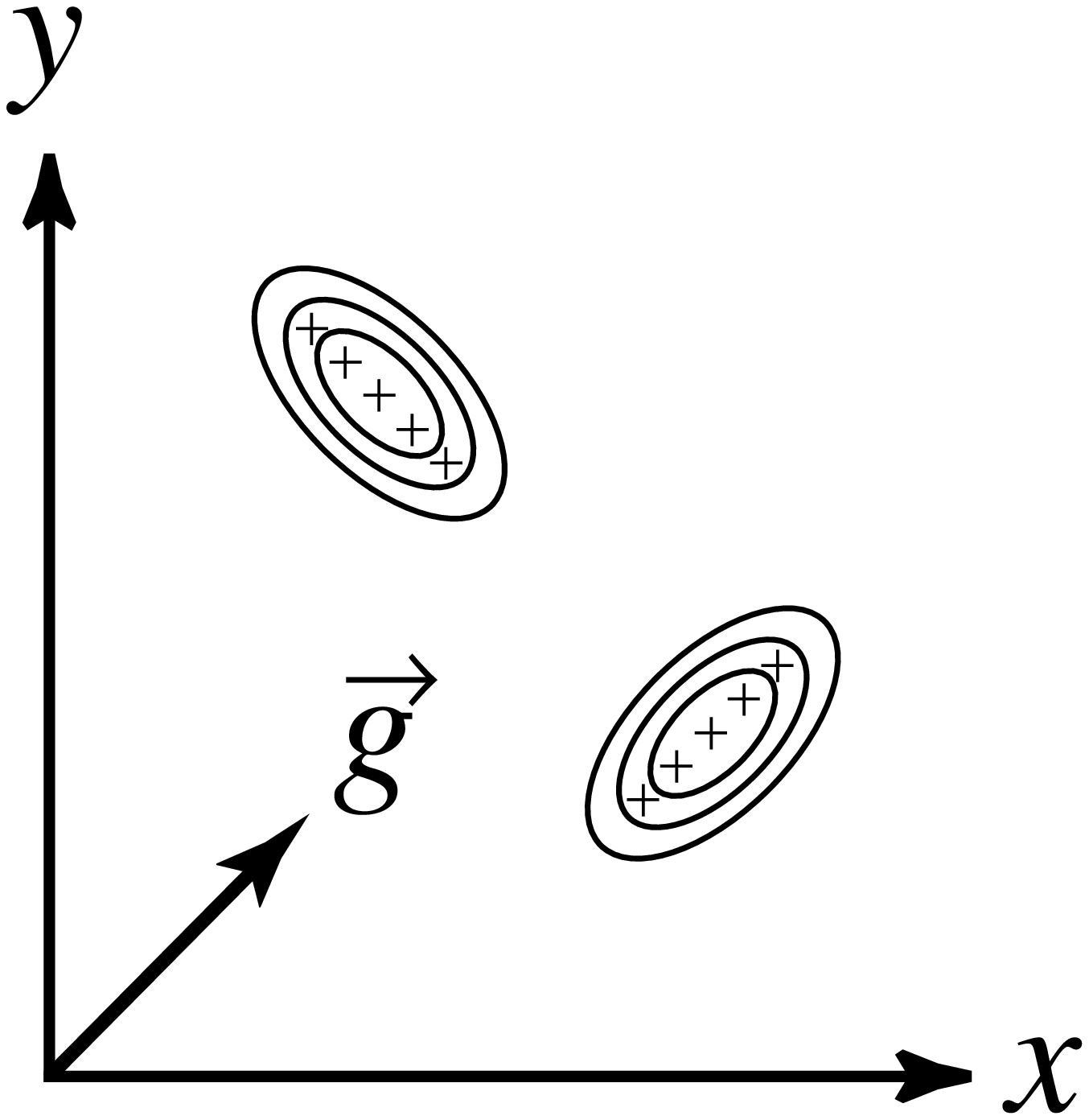}
\begin{figure}
\figcaption{
A schematic view of two localized wave functions 
of the Hermitian system in two dimensions.
The solid curves indicate contours on which the localized wave function takes 
the same value.
The anisotropy arises because of anisotropic fluctuations in the impurity
distribution.
We show an impurity fluctuation which leads to two nearly degenerate
wave functions related by a $90^{\circ}$ rotation.
}
\label{figanisotropy}
\end{figure}
\begin{equation}\label{david-6b}
\tilde{V}(k_{a},k_{b})\equiv\frac{1}{L_{x}L_{y}}
\sum_{x,y}e^{ik_{a}x+ik_{b}y}V_{x,y}
\end{equation}
is the Fourier-transformed disorder potential
and $\varepsilon_{0}(k_{x},k_{y})$ is given by Eq.~(\ref{50020}).
The eigenvalues of Eq.~(\ref{david-6a}) are 
\begin{equation}\label{david-6c}
\varepsilon_{\pm}(k_{x},k_{y})=
\varepsilon_{0}(k_{x},k_{y})\pm\Delta,
\end{equation}
where
\begin{equation}\label{david-6d}
\Delta\equiv \left|\tilde{V}(k_{x}-k_{y},k_{y}-k_{x})\right|.
\end{equation}
The real splitting of the degenerate doublets predicted by 
Eq.~(\ref{david-6c}) is clearly visible in Fig.~\ref{fig2Drnd}(d).

The Bloch spectrum is only a good approximation provided these corrections 
are small compared to the spacing between the doublets.
However, a typical spacing between the real parts of the Bloch levels 
for $L_{x}=L_{y}=L$ given by Eq.~(\ref{50020}) is
\begin{equation}\label{david-6e}
\Delta\varepsilon
\simeq\left|\frac{d{\rm Re}\;\varepsilon_{0}}{dk}\right|\Delta k
\simeq 2ta\cosh\left(\frac{ga}{\hbar}\right)\frac{2\pi}{L}.
\end{equation}
The splitting $\Delta$ due to the randomness is thus only a small 
correction to the Bloch levels whenever
\begin{equation}\label{david-6f}
\frac{4\pi}{L}ta\cosh\left(\frac{ga}{\hbar}\right)>\Delta,
\end{equation}
a condition which is {\em always} violated in sufficiently large systems
as $L\to\infty$.
When the inequality~(\ref{david-6f}) is not satisfied, level 
repulsion interacts with the randomness in a complex way to produce 
chaotic spectra like that in Fig.~\ref{fig2Drnd}(c).
The Bloch states are nondegenerate in $d=1$, so this problem does not arise.

Some insight into the meaning of the chaotic eigenvalue spectra 
follows from tracking the ground-state wave function as a function 
of $\mbox{\boldmath $g$}$.
As shown in Fig.~\ref{david-figZ}, the ground state first streaks out 
in the direction of $\mbox{\boldmath $g$}$.
(For very large $\mbox{\boldmath $g$}$, 
the wave function eventually broadens to cover 
the entire lattice. However, as discussed above, we believe that this 
Bloch-wave behavior is an artifact of the small system size.)
Arguments given in Ref.~\cite{Hatano96} show that the ground state 
can never simply delocalize in the $\mbox{\boldmath $g$}$ direction 
while remaining localized in the perpendicular direction when 
$L\to\infty$.
In Ref.~\cite{Shnerb97}, it is argued that the long-wavelength, 
low-frequency behavior of the non-Hermitian Schr\"{o}dinger equation 
in $2+1$ dimensions for large $\mbox{\boldmath $g$}$ is described by 
a $1+1$ dimensional Burgers' equation.
The tilted flux lines described in this way wander away from the 
$\mbox{\boldmath $g$}$ direction to take advantage of exceptionally 
deep minima in the disorder potential.
This mapping predicts a streaked-out but roughened ground-state wave 
functions with roughness exponent $2/3$.\cite{Shnerb97}
At present, our system sizes are too small to provide a good 
quantitative check of this hypothesis.

For system small enough so that Bloch states are a good first 
approximation for large $\mbox{\boldmath $g$}$, most states are 
approximately two-fold degenerate as discussed above.
However, for an $L\times L$ square lattice, there is a very large 
$L$-fold degeneracy for the state at the origin of the complex energy 
plane.
Perturbation theory explains the removal of the degeneracy in the 
following way.
We show that the first-order perturbation theory reduces to a diagonal
one-dimensional random model, which is readily solved.
We take the hopping term of Eq.~(\ref{41010}) as the 
non-perturbative part and the random-potential term as the 
perturbation so that
the zeroth-order spectrum may be given by Eq.~(\ref{50020}), or
Fig.~\ref{fig-2DReE-ImE}.
The zeroth-order wave functions $\left|k_{x},k_{y}\right\rangle$
of the degenerate levels in question
satisfy the relation $k_{x}-k_{y}=\pi/a$ $({\rm mod}\;2\pi/a)$.
The secular equation for the first-order perturbation 
of the degenerate levels consists of the matrix elements
\begin{eqnarray}\label{52000}
\lefteqn{
\left\langle k_{x},k_{y}\Biggm|
\sum_{\mbox{\mibscriptsize x}}V_{\mbox{\mibscriptsize x}}
b^\dagger_{\mbox{\mibscriptsize x}}b_{\mbox{\mibscriptsize x}}
\Biggm| k_{x}',k_{y}'\right\rangle
}\nonumber\\
&&=
\frac{1}{L^2}\sum_{x,y}e^{i(k_{x}-k_{x}')(x+y)}V_{x,y},
\end{eqnarray}
where we used $k_{y}-k_{y}'=k_{x}-k_{x}'$ $({\rm mod}\;2\pi/a)$.
The above matrix elements are, in fact, equivalent to
the momentum representation of the one-dimensional Hamiltonian
\begin{equation}\label{52010}
{\cal H}_{\rm eff}=\sum_{\xi}V_{\rm eff}(\xi)b^\dagger_{\xi}b_{\xi},
\end{equation}
where
\begin{equation}\label{52020}
V_{\rm eff}(\xi)\equiv\frac{1}{L}\sum_{\eta}V_{x,y}
\end{equation}

\epsfxsize=3.375in
\epsfbox{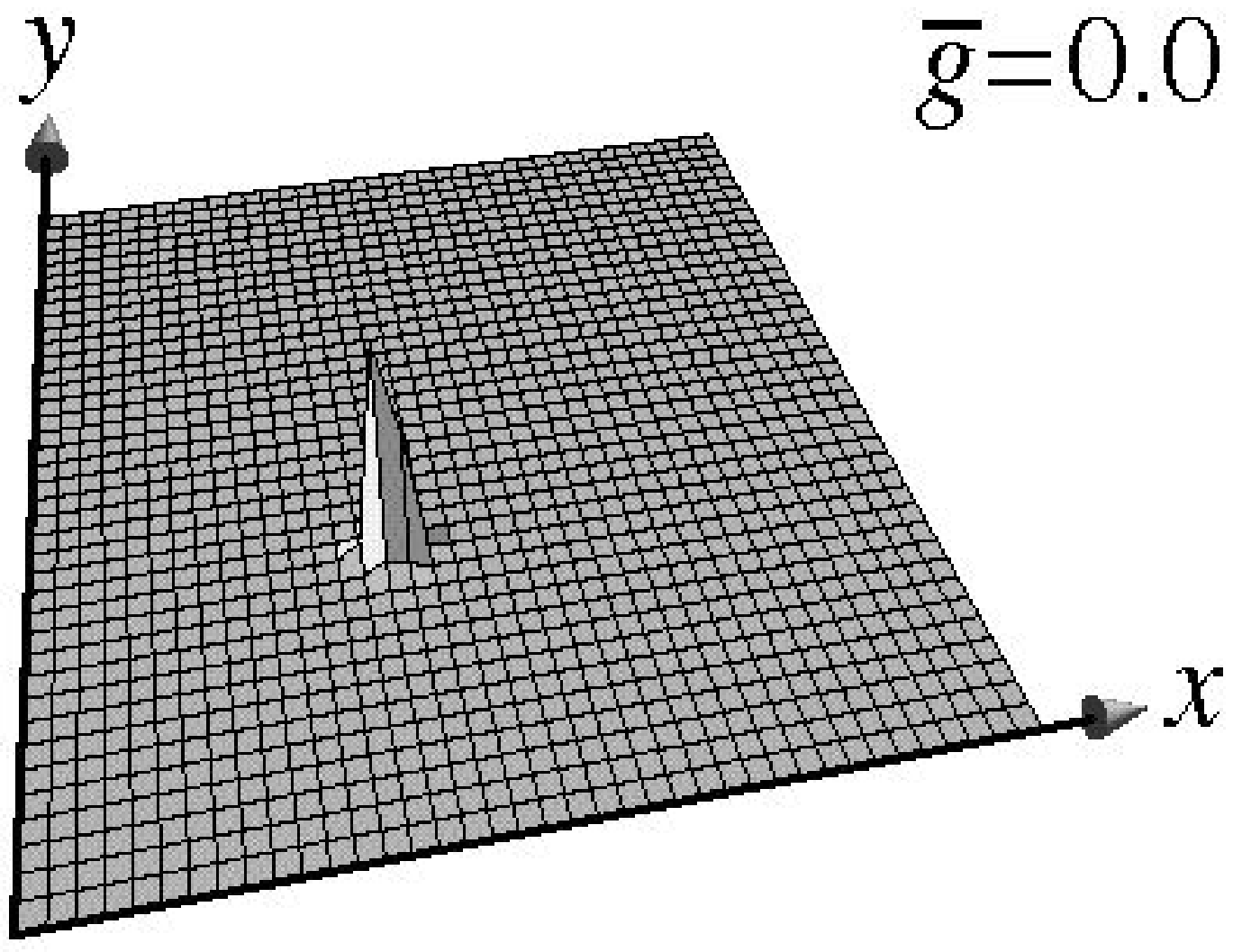}
\vskip -10pt
\begin{flushleft}(a)\end{flushleft}
\vskip -10pt
\epsfxsize=3.375in
\epsfbox{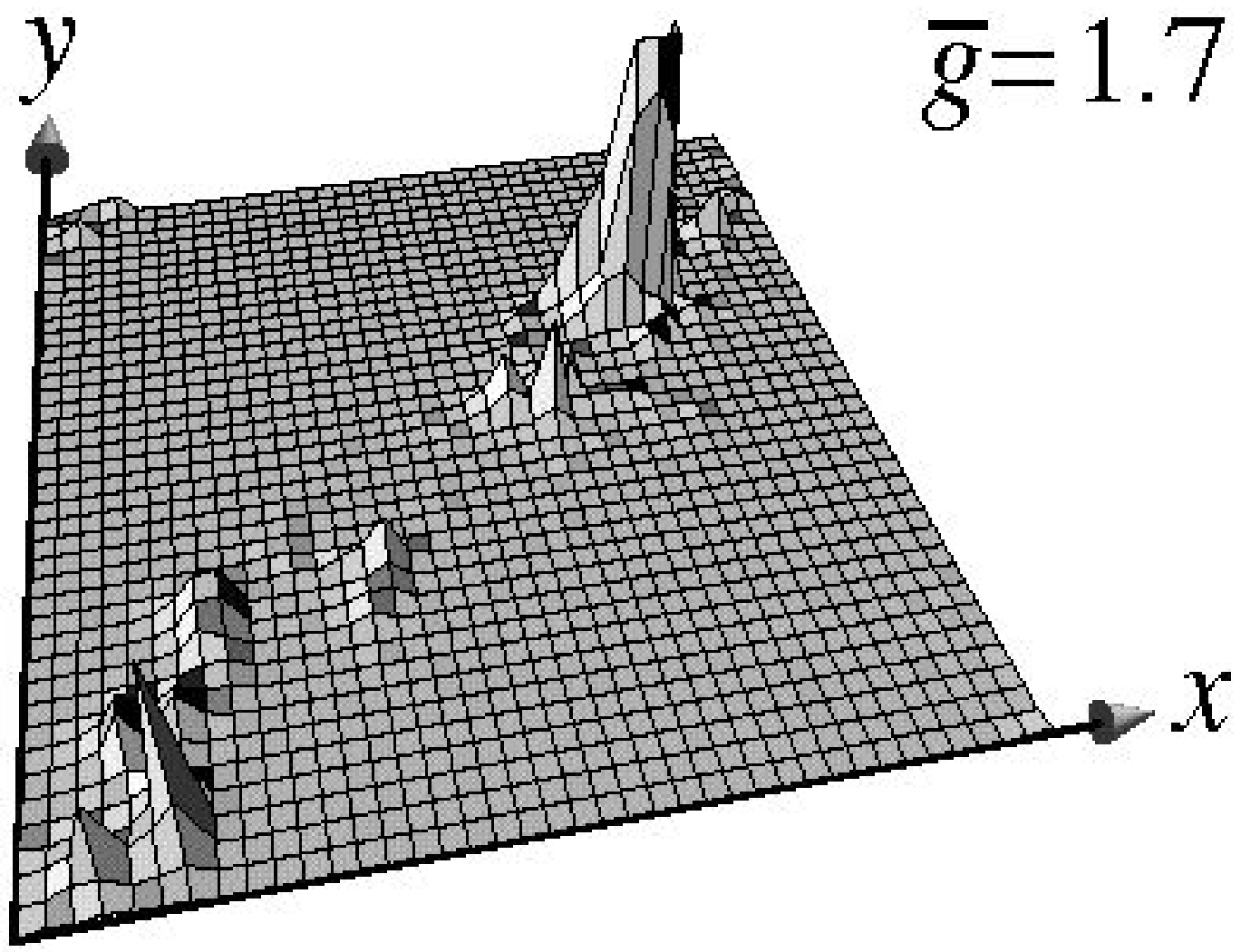}
\vskip -10pt
\begin{flushleft}(b)\end{flushleft}
\vskip -10pt
\epsfxsize=3.375in
\epsfbox{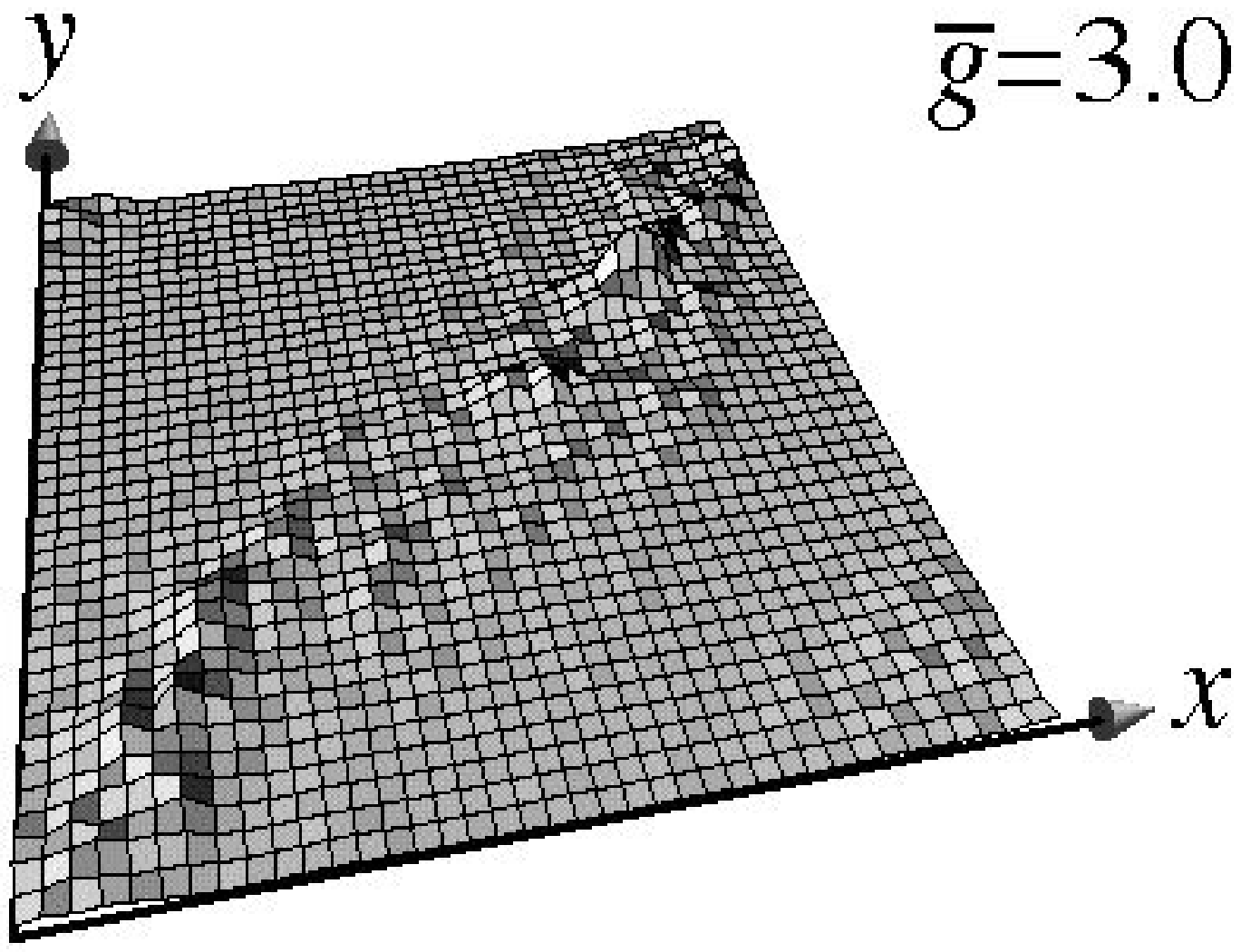}
\vskip -10pt
\begin{flushleft}(c)\end{flushleft}
\epsfxsize=3.375in
\epsfbox{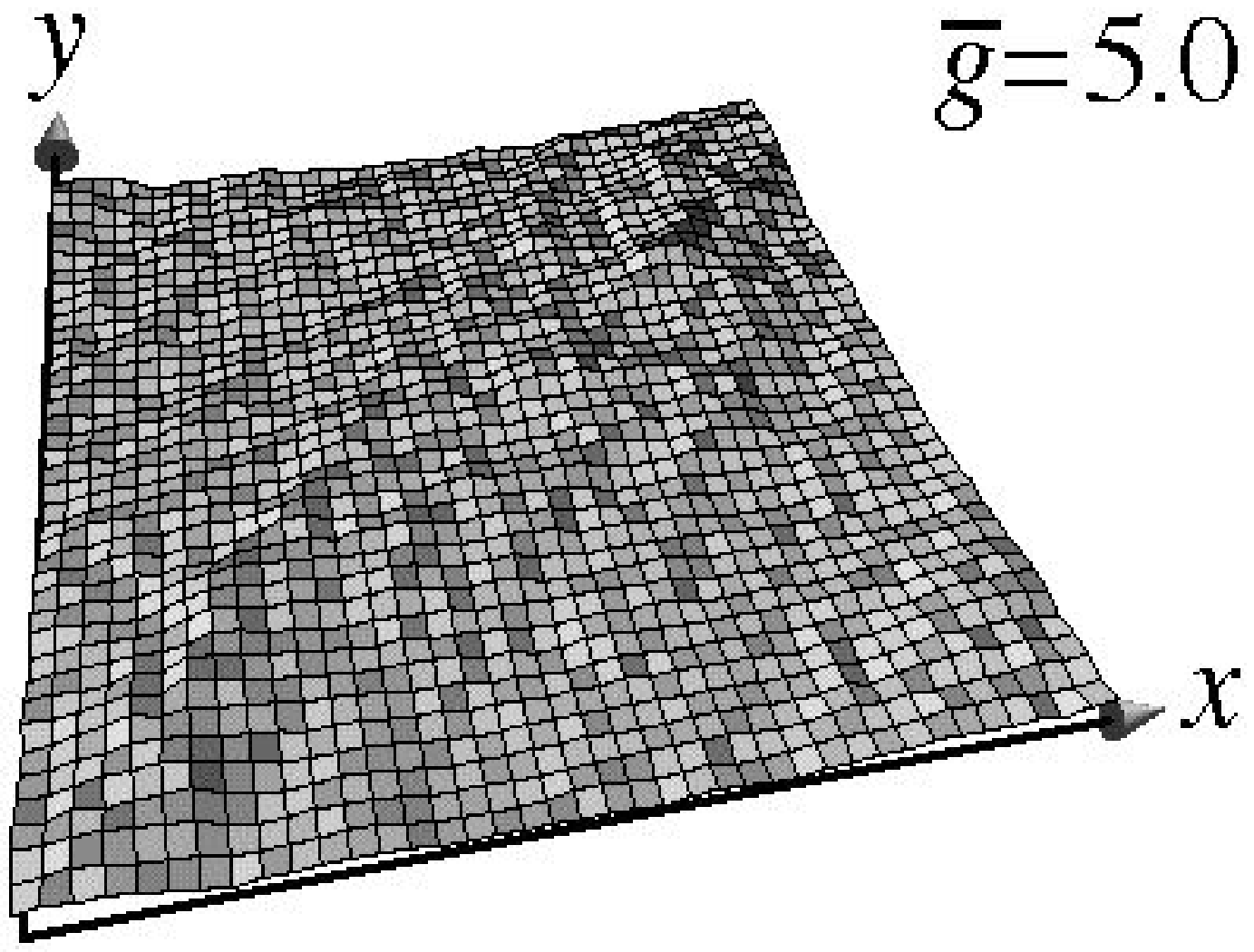}
\vskip -10pt
\begin{flushleft}(d)\end{flushleft}
\begin{figure}
\figcaption{
The ground-state right-eigenfunctions in two dimensions.
We show the case $L_{x}=L_{y}=40a$ and $\Delta=10t$ with
(a) $g_{x}=g_{y}=0$, (b) $g_{x}=g_{y}=1.7\times\hbar/a$, 
(c) $g_{x}=g_{y}=3.0\times\hbar/a$, and 
(d) $g_{x}=g_{y}=5.0\times\hbar/a$.
The vertical scale was rescaled by $1/10$ and $10$ in (b) and (d),
respectively.
The highest peak in (b) was reduced in size for the visualization purposes.)
}
\label{david-figZ}
\end{figure}
with $\xi\equiv x+y$ $({\rm mod}\;L)$ and $\eta\equiv x-y$ $({\rm mod}\;L)$.
The effective potential~(\ref{52020}) is thus obtained by integrating
the original potential $V_{x,y}$ along a line parallel to 
$\mbox{\boldmath $g$}$.
The first-order perturbation spectrum is readily solved,
since the effective Hamiltonian~(\ref{52010}) is already diagonalized.
Thus we arrive at the first-perturbation energy
\begin{equation}\label{1stpert}
\varepsilon^{(1)}_{\xi}=V_{\rm eff}(\xi)
\end{equation}
for $\xi=0,1,\ldots,L-1$, which accounts for the horizontal spread of
eigenvalues in Fig.~\ref{fig2Drnd}(d).

\section{Transverse Meissner effect and its penetration depth}
\label{sec-meissner}

In this section, we discuss the penetration depth associated with 
the transverse Meissner effect for superconductors with columnar 
defects in the localized regime.
The vortex density is assumed to be finite.
The surface deflection which defines the penetration depth for a {\em single}
vortex line is shown in Fig.~\ref{figevol}(a).
We discuss only field penetration in directions parallel to the 
columns.
For a more general discussion of various penetration depths, frequency 
dependences, {\it etc}., see Ref.~\cite{Nelson96}.

As shown in Subsec.~\ref{sec-1D1imp-depin} for one-dimensional case, 
the displacement of the pinned flux line near the surface of
the superconductor
with one defect has, in general, an exponential $\tau$ dependence.
The penetration depth $\tau^\ast$ has the singularity 
$\tau^\ast\sim\xi_\perp\sim(g-g_c)^{-1}$.
We show below that the mean displacement at depth $\tau$ 
in the {\em many}-defect case in $d$ dimensions is given by
a stretched exponential form
\begin{equation}\label{61010}
\left\langle(\langle x \rangle_{\tau}
-\langle x \rangle_\infty)\right\rangle_{\rm av}
\mathop{\sim}_{\tau\to\infty}
\exp[-a\left(\tau/\tau^\ast\right)^{1/(d+1)}],
\end{equation}
where we now assume for simplicity that the field
$\mbox{\boldmath $g$}$ is parallel to the $x$ axis. 
The average displacement in directions perpendicular to $\mbox{\boldmath $g$}$ 
vanishes because of statistical symmetry.
The quantity $\langle x \rangle_\infty$ is the center of the 
localized state in the bulk, which is identical to the average 
position of the corresponding state for 
$\mbox{\boldmath $g$}=\mbox{\bf 0}$.
Moreover, we show that the penetration depth has the singularity
$\tau^\ast\sim\xi_\perp^{z}\sim(g_{c}-g)^{-z}$ with the dynamical 
exponent $z=d$, thus justifying Eqs.~(\ref{0010}) 
and~(\ref{0020}).
Note that $d$ is the dimensionality of the quantum system 
corresponding to flexible lines in $d+1$ dimensions. 

Consider a low but finite density of interacting flux lines.
As discussed in Sec.~\ref{sec-corres}, 
we fill up the localized states in order of increasing energy up to the
average chemical potential $\varepsilon=\mu$.
Consider the deflection of the most unstable (pinned) flux line near the
surface.
Since we forbid double occupancy of localized states,
we can approximate
\begin{equation}\label{61020}
\left\langle x \right\rangle_{\tau}
\simeq
\sum_n\nolimits' c_n
\int d^d\mbox{\boldmath $x$}\langle\psi_\mu\bigm|\mbox{\boldmath $x$}\rangle
x \langle \mbox{\boldmath $x$}\bigm|\psi_n\rangle
e^{-\tau\Delta\varepsilon_n/\hbar},
\end{equation}
where
\begin{equation}\label{61030}
c_n\equiv
\frac%
{\int\langle\psi_n\bigm|\mbox{\boldmath $x$}'\rangle d^d\mbox{\boldmath $x$}'}%
{\int\langle\psi_\mu\bigm|\mbox{\boldmath $x$}'\rangle 
d^d\mbox{\boldmath $x$}'},
\end{equation}
and
\begin{equation}\label{61040}
\Delta\varepsilon_n\equiv\varepsilon_n-\varepsilon_\mu
\end{equation}
in contrast to Eqs.~(\ref{32060}),~(\ref{32070}), and~(\ref{32075})
appropriate to a single line.
Here $\psi_\mu$ denotes an eigenstate at the chemical potential,
and the summation $\sum_n'$ is restricted to states with the energies 
$\varepsilon\geq\mu$.
Note that the coefficient $c_n$ for an extended state will be quite 
small owing to oscillatory factors in the integration over the space.
Because of the exponential factor in~(\ref{61020}), 
the main contributions come from localized states near and above the
chemical potential.
We use the asymptotic form~(\ref{21050}) for localized states to
estimate Eq.~(\ref{61020}).
The term with $\psi_n=\psi_\mu$ gives $\left\langle x \right\rangle_\infty$.
The other leading terms after the integration are approximately
\begin{equation}\label{61050}
\left|\left\langle x \right\rangle_{\tau}
-\left\langle x \right\rangle_\infty\right|
\simeq
\sum_n\nolimits'' 
\tilde{c}_ne^{-f_n(\tau)},
\end{equation}
where 
\begin{equation}\label{61060}
f_n(\tau)\equiv(\hbar\kappa(\mu)-g\cos\theta_n)r_n
+\tau\Delta\varepsilon_n/\hbar,
\end{equation}

\noindent
\hskip 0.6875in
\epsfxsize=2in
\epsfbox{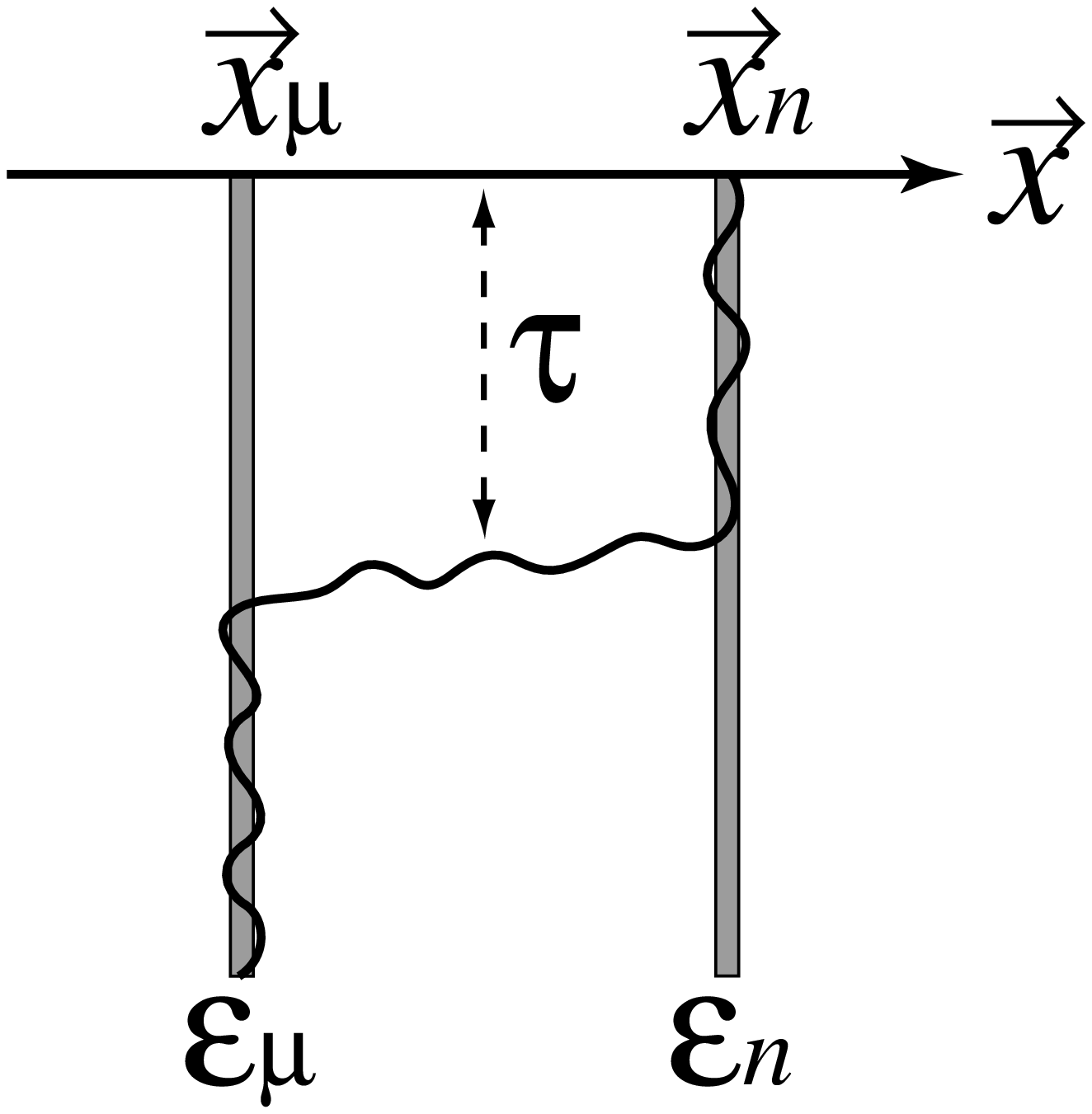}
\begin{figure}
\figcaption{
A kink in a vortex configuration near the top surface of a superconducting
sample.
}
\label{fig-kink}
\end{figure}
$r_n\equiv|\mbox{\boldmath $x$}_n-\mbox{\boldmath $x$}_\mu|$,
$\cos\theta_n\equiv\mbox{\boldmath $g$}\cdot
(\mbox{\boldmath $x$}_\mu-\mbox{\boldmath $x$}_n)
/(|\mbox{\boldmath $g$}|r_n)$,
and $\sum_n''$ is the summation over localized excited states with
the energies $\varepsilon>\mu$.
The inverse localization length $\kappa_n$ was approximated by $\kappa(\mu)$.
The new coefficient $\tilde{c}_n$ is
$\tilde{c}_n\sim c_nr_n\cos\theta_n$.
As in conventional Mott variable-range hopping in 
semiconductors,\cite{Shklovskii84}
we estimate the energy difference $\Delta\varepsilon_{n}$ 
from \cite{Nelson93a}
\begin{equation}\label{61070}
g(\mu)r_n^d\Delta\varepsilon_n\sim1,
\end{equation}
where $g(\mu)$ is the density of states at the chemical potential.

The quantity $f_n(\tau)$ gives the energy of 
a tilted flux-line configuration; see Fig.~\ref{fig-kink}.
The first term is the energy due to a kink joining columns at
$\mbox{\boldmath $x$}_\mu$ and $\mbox{\boldmath $x$}_n$.
Hence the energy is proportional to the width $r_{n}$ of the kink.
The second term is the energy loss arising because the
flux line stays at $\mbox{\boldmath $x$}_n$ over the distance $\tau$
rather than at the most stable position $\mbox{\boldmath $x$}_\mu$.
This energy loss is proportional to $\tau$.
Because of Eq.~(\ref{61070}), the first and the second terms compete:
The further the flux line hops from $\mbox{\boldmath $x$}_\mu$, the more the
kink energy costs, but the lower the binding energy at
$\mbox{\boldmath $x$}_n$.
In variable-range hopping of electrons in semiconductors,\cite{Shklovskii84}
$\Delta\varepsilon_n$ is the energy difference between localized electronic
states, and the kink energy is replaced by a WKB tunnelling matrix element.

Figure~\ref{fig-twodefect} shows the relaxation function
$\left\langle x \right\rangle_{\tau}
-\left\langle x \right\rangle_\infty$ 
for the one-dimensional
tight-binding model with a particular realization of randomness.
In this example,
only two states contribute significantly to the summation $\sum_n''$ 
in Eq.~(\ref{61050}).
Upon averaging over many realizations of randomness
(or close to the depinning transition), we expect contributions
from many states as depicted schematically in Fig.~\ref{fig-manydefect}.
For a fixed $\tau$, the largest contribution to the summation
in~(\ref{61050}) comes from the state at
\begin{equation}
r_n=\left(\frac{\tau d}{\hbar g(\mu)[\hbar\kappa(\mu)-g]}\right)^{1/(d+1)},
\end{equation}
which yields

\epsfxsize=3.375in
\epsfbox{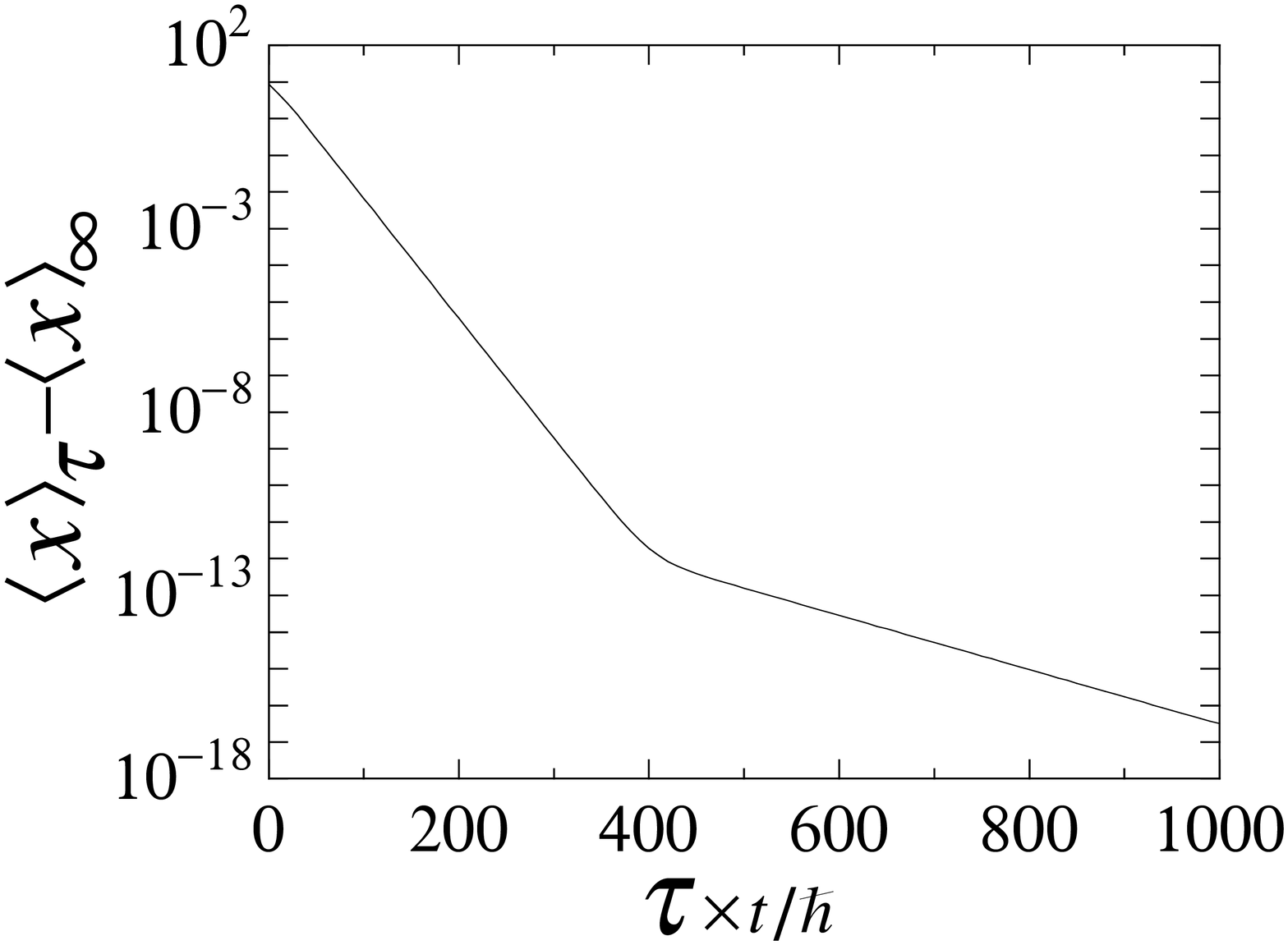}
\begin{figure}
\figcaption{
The $\tau$ dependence of the displacement of a flux line
for a realization of randomness in the one-dimensional tight-binding model.
Since only two terms of $f_n$ contribute to the summation $\sum_n''$ 
in~(\protect\ref{61050}) in this example, the whole curve approximately
consists of two lines. 
The slope of each line is $-\Delta\varepsilon_n/\hbar$, while its
$y$-intercept is $-(\hbar\kappa_n-g)r_n$.
The parameters are $L_x=500a$, $\Delta/t=1$, and $g/t=0.6$.
}\label{fig-twodefect}
\end{figure}
\begin{equation}
\min_n f_n=a\left(\frac{\tau}{\tau^\ast}\right)^{1/(d+1)}
\end{equation}
with
\begin{equation}
\tau^\ast=\frac{\hbar g(\mu)}{(\hbar\kappa(\mu)-g)^d}.
\end{equation}
and $a=d^{1/(d+1)}(1+1/d)$.
This gives the stretched exponential form~(\ref{61010}).
Since the displacement at the surface, or the surface transverse 
magnetization, is given by 
$\xi_\perp\sim M_{\perp s}\sim(\hbar\kappa-g)^{-1}$, 
we have $\tau^\ast\sim\xi_\perp^d$.

We obtain the stretched exponential form only in the limit $\tau\to\infty$.
Deep in the Bose glass phase, 
it may be difficult to observe this form experimentally, because the amplitude
of the surface displacement is small in this limit.
However, the characteristic scale $\tau^\ast$ in the $\tau$ 
direction diverges, $\tau^\ast\sim(\hbar\kappa(\mu)-g)^{-d}$,
where stretched-exponential relaxation should be relatively easy to observe.

\section{Effect of interactions in the delocalized regime}
\label{sec-interaction}

\subsection{$1+1$ dimensions: Continuous phase transition}

We conclude with a discussion of interaction effects.
The delocalization transitions discussed above were treated using an 
independent particle (or \lq\lq independent vortex line'') picture.
As outlined above, interactions\linebreak 

\noindent
\hskip 0.2875in
\epsfxsize=2.8in
\epsfbox{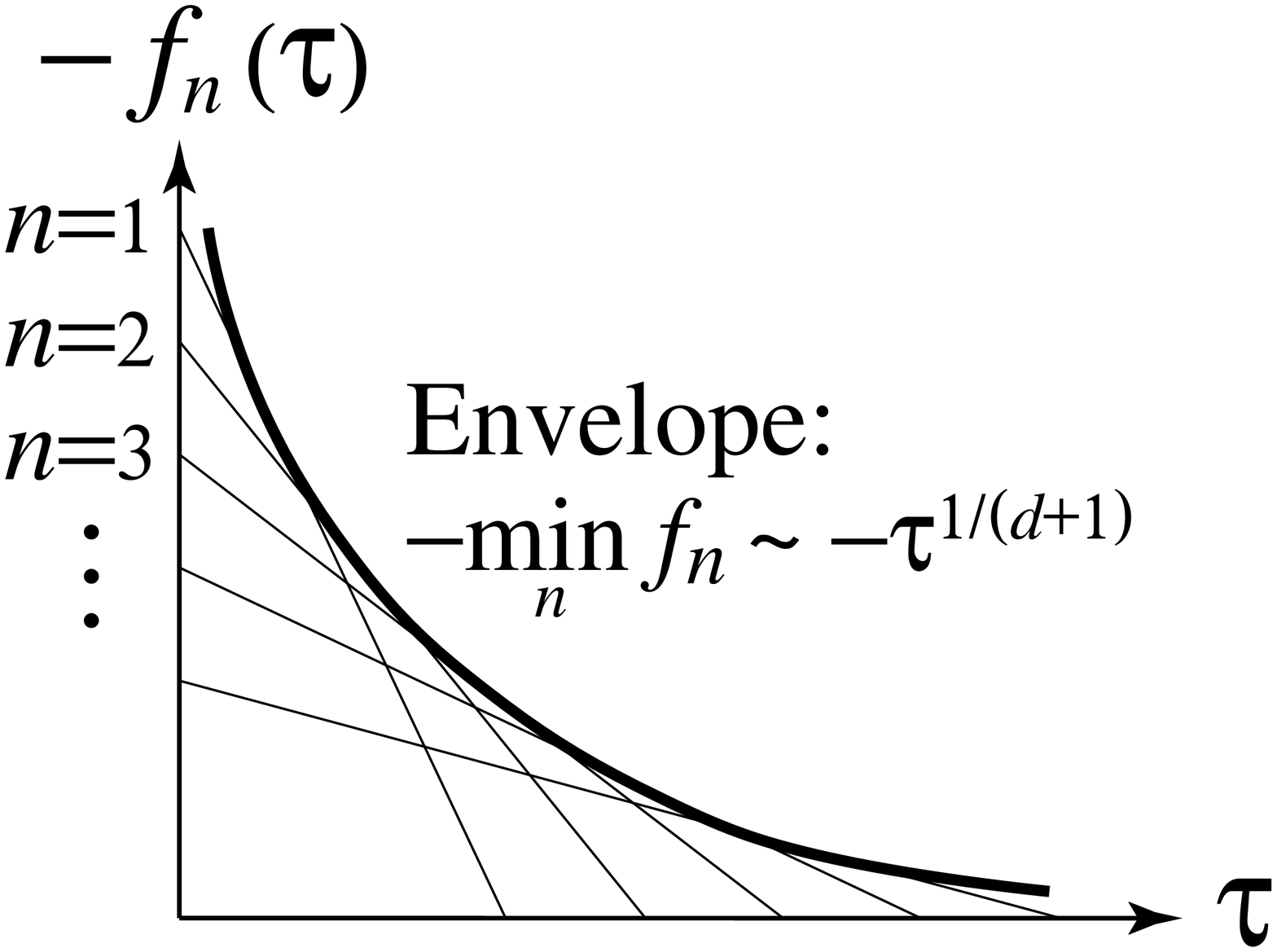}
\begin{figure}
\figcaption{
Schematic of the situation where many different exponential terms
contribute to the summation in~(\protect\ref{61050}).
}
\label{fig-manydefect}
\end{figure}
can be taken into account in the 
Bose-glass phase, provided that we forbid multiple occupancy of the 
localized states.
The physics here is similar to the bands of localized impurity states 
describing electrons in disordered semiconductors.\cite{Shklovskii84}
In both Bose and Fermi glasses, localized states are filled in 
order of increasing energies up to a chemical potential.
In the localized regime, the differences between Bose statistics of 
repelling flux lines interacting with columnar defects and 
Fermi statistics of electrons in disordered semiconductors are 
not expected to be important.\cite{Nelson93a}

Interactions must be handled differently in the delocalized phase.
Consider what happens to the states described by the $(1+1)$-dimensional 
non-Hermitian spectra of Fig.~\ref{fig1Drandom} (a) with increasing 
field $H_{\parallel}$ along the $\tau$ axis.
We assume that the tilt field lies in a range such that there is a 
mobility edge separating low-energy localized states from high-energy 
delocalized ones.
The field $H_{\parallel}$ controls the chemical potential of the equivalent 
disordered boson system.\cite{Nelson93a}
As $H_{\parallel}$ increases, we fill the unoccupied levels in order of 
increasing energy to obtain the ground state.
Eventually all states below the mobility edge are filled,
and additional vortices must then go into extended states above this 
boundary.
These extended states describe macroscopically tilted vortex lines, 
and as we can see from Fig.~\ref{fig1Dcurrent}, the corresponding 
tilt slope is finite at the mobility edge.
Interactions have a weaker effect on delocalized, tilted lines, so 
we expect that {\em many} lines can be accommodated by only a few 
delocalized states with energies just above the mobility edge.
In the presence of thermal fluctuations, there is not a sharp 
distinction between tilted and untilted lines;
any individual line will be both tilted or localized in different 
regions along the $\tau$ axis.
Alternatively, we can imagine that the delocalization transition happens 
at fixed $H_{\parallel}$, with increasing external tilt field $H_{\perp}$, 
as in Fig.~\ref{phase-diagram}.

A physical picture of the delocalized phase in $1+1$ dimensions in the 
presence of interactions has been developed by 
Hwa {\em et al}.\cite{Hwa93}.
Tilted lines are represented by chains of kinks, and the density of 
these chains goes {\em continuously} to zero with decreasing tilt 
field.
Thus we predict that a continuous phase transition is possible, 
in contrast to arguments based on depinning of a single line, which
suggest a first-order transition.\cite{Chen96}
A phenomenon like Bose condensation describes the physics of the tilted 
fraction of the lines, consistent with many lines entering just a 
few extended states above the mobility edge.
However, because phase fluctuations are so strong in $1+1$ dimensions, 
correlations of the boson parameter decay {\em algebraically} to zero 
at large distances instead of approaching a nonzero constant.
There is also algebraic order in a translational order parameter 
associated with the density of tilted lines;
the tilted phase is in fact a \lq\lq supersolid.''\cite{Hwa93}

\subsection{$2+1$ dimensions: Existence of a phase transition}

The physical picture of lines in the delocalized regime is more 
complicated in $2+1$ dimensions, as suggested by the intricate 
spectra for finite $\mbox{\boldmath $g$}$ 
displayed in Fig.~\ref{fig2Drnd}. 
However, here again there is a range of intermediate tilt fields where 
only localized states exist at low energies.
A \lq\lq mobility edge'' now separates real eigenvalues 
in the localized regime 
from a region along the real energy axis where extended and localized 
states {\em coexist}.
Building up a ground state as in $1+1$ dimensions, we would expect 
some vortices to enter extended states describing tilted lines once 
the chemical potential (controlled by $H_{\parallel}$) crosses this 
mobility edge.
A bosonic \lq\lq superfluid'' fraction of tilted lines will coexist 
with a \lq\lq normal fluid'' fraction of lines localized on columns.
As in $1+1$ dimensions, however, any single line will participate in 
both \lq\lq fractions'' as it crosses a sufficiently thick sample.
Although a {\it single} tilted line probably goes into a glassy 
ground state described by Burgers' equation,\cite{Shnerb97}
interactions in a dense liquid of tilted lines can screen out random 
pinning potentials in a weakly perturbed \lq\lq superfluid'' phase of 
entangled lines.\cite{Nelson90}.
The tilted liquid in this dense regime should be pinned only weakly,
and exhibit a linear resistivity.

If the tilted regime is entered by increasing $H_{\perp}$ at fixed 
$H_{\parallel}$, one is shifting the \lq\lq mobility edges'' in the 
$(2+1)$-dimensional spectra at fixed chemical potential.
The sequence of possible phases probably resembles vortex matter 
subjected to point disorder near $H_{\rm c1}$.\cite{Nelson90-1}
For small $H_{\perp}$, the Bose-glass with its transverse Meissner 
effect resembles the usual Meissner phase for $H<H_{\rm c1}$.
Suppose now that $H_{\perp}$ is increased until $H_{\perp}>H_{\perp c}$, 
such that the lower mobility edge drops below the chemical potential.
As illustrated in Fig.~\ref{phase-diagram}, one probably encounters 
a flux-liquid phase (with a finite fraction of tilted, entangled lines),
if $H_{\perp}$ is increased at {\it high} temperatures.
At low temperatures and large tilt fields, however, the stable phase 
is probably a crystalline vortex-smectic phase, where the tilted lines 
are arranged in sheets periodically spaced along the column direction 
to minimize the interaction energies.\cite{Nelson93a-1}
Hwa {\it et al.}\cite{Hwa93} have discussed a theory of how a 
continuous transition directly from this vortex smectic into the Bose 
glass might proceed.
However, it is hard to rule out the possibility of a melted silver of 
tilted flux liquid at {\em all} nonzero temperatures when the density of 
tilted lines becomes very small near the transition curve 
$H_{\perp c}(T)$, similar to what happens in vortices subjected to 
point disorder near $H_{c1}(T)$.
It is not yet known if a point disorder is sufficient to produce a 
distinct \lq\lq vortex glass'' phase 
with a finite vortex density interposed between 
the flux liquid and the Meissner phase.\cite{Nelson90}
There are similar uncertainties about a distinct glassy phase near 
$H_{\perp c}(T)$  in the present problem.

Although we will not resolve these uncertainties here, we can show 
that there must be at least {\it one} phase transition separating a 
Bose glass at small $H_{\perp}$ from a tilted \lq\lq superfluid'' of 
entangled lines when $H_{\perp}$ is large.
In other words, the transition suggested by the existence of a sharp 
\lq\lq mobility edge'' in the $(2+1)$-dimensional single-line spectra 
survives the imposition of repulsive inter-line interactions.
We proceed by first noting that the tilt modulus,
\begin{equation}\label{david-7.1}
c_{44}\equiv\frac{B^2}{4\pi}
\left(\frac{\partial B_{\perp}}{\partial H_{\perp}}\right)^{-1},
\end{equation}
is divergent in the Bose-glass phase which results for small 
$H_{\perp}$.\cite{Nelson93a}
This infinity is a consequence of the transverse Meissner effect:
Although the lines tilt near the sample surfaces in response to a 
perpendicular external field (as in Fig.~\ref{fig-kink}),
there is no macroscopic response in an infinite sample, so
$\partial B_{\perp}/\partial H_{\perp}=0$.
We will show that the tilt modulus $c_{44}$ 
remains {\em finite} in a \lq\lq superfluid'' liquid of entangled, 
tilted lines which results for 
$H_{\perp}>H_{\perp c}$, so that there {\em must} be at least one 
sharp phase transition as one increases $H_{\perp}$.
The tilt modulus is proportional to the inverse superfluid density of the 
equivalent boson system, and the corresponding results for the superfluid
density $\rho_s$ are indicated in Fig.~\ref{phase-diagram}.

We start with the generalization of Eq.~(\ref{11010}) for $N$ lines 
$\{\mbox{\boldmath $x$}_{j}(\tau)\}$, namely,\cite{Nelson93a}
\end{multicols}
\begin{multicols}{2}
\hrule
\end{multicols}
\begin{equation}\label{david-7.2}
E_{\rm flux}
\left[\left\{\mbox{\boldmath $x$}_{j}(\tau)\right\}\right]
\equiv
\frac{\tilde{\varepsilon}_{1}}{2}
\sum_{j=1}^N\int_{0}^{L_{\tau}}d\tau
\left[
\left(\frac{d\mbox{\boldmath $x$}_{j}}{d\tau}\right)^2
-\mbox{\boldmath $g$}\cdot\frac{d\mbox{\boldmath $x$}_{j}}{d\tau}
+V\left[\mbox{\boldmath $x$}_{j}(\tau)\right]
\right]
+\frac{1}{2}
\sum_{i\neq j}\int_{0}^{L_{\tau}}d\tau
V_{\rm int}\left(
\left|\mbox{\boldmath $x$}_{i}(\tau)-\mbox{\boldmath $x$}_{j}(\tau)\right|
\right),
\end{equation}
\begin{multicols}{2}
\noindent
where 
$\mbox{\boldmath $g$}\equiv\phi_{0}\mbox{\boldmath $H$}_{\perp}/(4\pi)$
and we have added a repulsive potential $V_{\rm int}$ between the lines.
We neglect point disorder and
assume for simplicity that the columnar pinning potential is weak 
or the temperature is fairly high, so that the transition out of the 
Bose glass occurs for relatively small values of 
$\mbox{\boldmath $g$}\propto\mbox{\boldmath $H$}_{\perp}$.
The assumptions of small line tilts and local (in $\tau$) interactions 
between flux lines which justify Eq.~(\ref{david-7.2})\cite{Nelson93a} are 
then satisfied even for $H_{\perp}>H_{\perp c}$. 
The many-line partition function associated with Eq.~(\ref{david-7.2})
is
\begin{equation}\label{david-7.3}
{\cal Z}(\mbox{\boldmath $g$})\equiv
\prod_{j=1}^N\int{\cal D}\mbox{\boldmath $x$}_{j}(\tau)
^{-E_{\rm flux}\left[\left\{\mbox{\mibscriptsize x}_{j}(\tau)\right\}\right]
/(k_{B}T)}.
\end{equation}
Upon defining an operator 
$\mbox{\boldmath $X$}_{\rm op}\equiv
\sum_{j=1}^N\mbox{\boldmath $x$}_{j}^{\rm op}$
acting on position eigenstates 
$|\mbox{\boldmath $x$}_{1},\ldots,\mbox{\boldmath $x$}_{N}\rangle$
such that
\begin{equation}\label{david-7.4}
\mbox{\boldmath $X$}_{\rm op}
\left|\mbox{\boldmath $x$}_{1},\ldots,\mbox{\boldmath $x$}_{N}\right\rangle
=\left(\sum_{j=1}^N\mbox{\boldmath $x$}_{j}\right)
\left|\mbox{\boldmath $x$}_{1},\ldots,\mbox{\boldmath $x$}_{N}\right\rangle,
\end{equation}
standard manipulations allow us to write this multidimensional 
path integral in terms of a quantum mechanical matrix element,
\end{multicols}
\begin{multicols}{2}
\hrule
\end{multicols}
\begin{equation}\label{david-7.5}
{\cal Z}(\mbox{\boldmath $g$})=
\int d^2\mbox{\boldmath $r$}'_{1}\cdots d^2\mbox{\boldmath $r$}'_{N}
\int d^2\mbox{\boldmath $r$}_{1}\cdots d^2\mbox{\boldmath $r$}_{N}
\left\langle
\mbox{\boldmath $r$}'_{1}\cdots d\mbox{\boldmath $r$}'_{N}
\Biggm|
e^{\mbox{\mibscriptsize g}\cdot\mbox{\mibscriptsize X}_{\rm op}/(k_{B}T)}
e^{-{\cal H}(\mbox{\bfscriptsize 0})L_{\tau}/(k_{B}T)}
e^{-\mbox{\mibscriptsize g}\cdot\mbox{\mibscriptsize X}_{\rm op}/(k_{B}T)}
\Biggm|
\mbox{\boldmath $r$}_{1}\cdots d\mbox{\boldmath $r$}_{N}
\right\rangle,
\end{equation}
\begin{multicols}{2}
~\\

\hrule
\end{multicols}
\begin{multicols}{2}
\noindent
where the \lq\lq Hamiltonian'' ${\cal H}(\mbox{\bf 0})$ is
\begin{eqnarray}\label{david-7.6}
{\cal H}(\mbox{\bf 0})&=&-\frac{(k_{B}T)^2}{2\tilde{\varepsilon}_{1}}
\sum_{j=1}^N \mbox{\boldmath $\nabla$}_{j}^2
+\sum_{j=1}^N V\left[\mbox{\boldmath $x$}_{j}\right]
\nonumber\\
&&+\frac{1}{2}\sum_{i\neq j}V_{\rm int}
\left(
\left|\mbox{\boldmath $x$}_{i}-\mbox{\boldmath $x$}_{j}\right|
\right).
\end{eqnarray}
Upon noting that
\begin{eqnarray}\label{david-7.7}
\lefteqn{
e^{\mbox{\mibscriptsize g}\cdot\mbox{\mibscriptsize X}_{\rm op}/(k_{B}T)}
e^{-{\cal H}(\mbox{\bfscriptsize 0})L_{\tau}/(k_{B}T)}
e^{-\mbox{\mibscriptsize g}\cdot\mbox{\mibscriptsize X}_{\rm op}/(k_{B}T)}
}\nonumber\\
&&=
\sum_{n=0}^\infty\frac{1}{n!}
e^{\mbox{\mibscriptsize g}\cdot\mbox{\mibscriptsize X}_{\rm op}/(k_{B}T)}
\left(-\frac{{\cal H}(\mbox{\bf 0})L_{\tau}}{k_{B}T}\right)^n
e^{-\mbox{\mibscriptsize g}\cdot\mbox{\mibscriptsize X}_{\rm op}/(k_{B}T)}
\nonumber\\
&&=
e^{-{\cal H}(\mbox{\mibscriptsize g})L_{\tau}/(k_{B}T)},
\end{eqnarray}
where
\begin{eqnarray}\label{david-7.8}
{\cal H}(\mbox{\boldmath $g$})&=&
e^{\mbox{\mibscriptsize g}\cdot\mbox{\mibscriptsize X}_{\rm op}/(k_{B}T)}
{\cal H}(\mbox{\bf 0})
e^{-\mbox{\mibscriptsize g}\cdot\mbox{\mibscriptsize X}_{\rm op}/(k_{B}T)}
\nonumber\\
&=&
\frac{1}{2\tilde{\varepsilon}_{1}}\sum_{j=1}^N
\left(\frac{k_{B}T}{i}\mbox{\boldmath $\nabla$}_{j}
+i\mbox{\boldmath $g$}\right)^2
+\sum_{j=1}^N V\left[\mbox{\boldmath $x$}_{j}\right]
\nonumber\\
&&
+\frac{1}{2}\sum_{i\neq j}V_{\rm int}
\left(
\left|\mbox{\boldmath $x$}_{i}-\mbox{\boldmath $x$}_{j}\right|
\right).
\end{eqnarray}
Upon referring to Table~\ref{corres}, we see that this Hamiltonian is the
generalization of Eq.~(\ref{12020}) for many interacting vortex lines.

Note that if a many-body eigenfunction 
$\Psi_{n}(\mbox{\boldmath $x$}_{1},\ldots,\mbox{\boldmath $x$}_{N};
\mbox{\bf 0})$ of ${\cal H}(\mbox{\boldmath $g$}=\mbox{\bf 0})$ 
for localized lines in the Bose glass phase is 
known, a potentially exact right-eigenfunction with the same energy
for $\mbox{\boldmath $g$}\neq\mbox{\bf 0}$ is then
\begin{equation}\label{david-7.9}
\Psi^R_{n}\left(\mbox{\boldmath $x$}_{1},\ldots,\mbox{\boldmath $x$}_{N};
\mbox{\boldmath $g$}\right)
=e^{\mbox{\mibscriptsize g}\cdot\mbox{\mibscriptsize X}_{\rm op}/(k_{B}T)}
\Psi_{n}\left(\mbox{\boldmath $x$}_{1},\ldots,\mbox{\boldmath $x$}_{N};
\mbox{\bf 0}\right)
\end{equation}
as follows immediately from the first line of Eq.~(\ref{david-7.8}).
As in the single-particle case, this gauge transformation connecting 
the $\mbox{\boldmath $g$}=\mbox{\bf 0}$ and
$\mbox{\boldmath $g$}\neq\mbox{\bf 0}$ problems only works provided 
that the new eigenfunction is normalizable.\cite{Hwa93b}

To treat the tilted phase of entangled lines, it is convenient to 
modify the original path-integral partition function via the change of 
variables,
\begin{equation}\label{david-7.10}
\mbox{\boldmath $x$}_{j}(\tau)=\mbox{\boldmath $x$}'_{j}(\tau)
+\frac{\mbox{\boldmath $g$}}{\tilde{\varepsilon}_{1}}\tau.
\end{equation}
Upon transforming to a quantum-mechanical matrix element as before,
we arrive at
\end{multicols}
\begin{multicols}{2}
\hrule
\end{multicols}
\begin{equation}\label{david-7.11}
{\cal Z}(\mbox{\boldmath $g$})=
\int d^2\mbox{\boldmath $r$}'_{1}\cdots d^2\mbox{\boldmath $r$}'_{N}
\int d^2\mbox{\boldmath $r$}_{1}\cdots d^2\mbox{\boldmath $r$}_{N}
\left\langle
\mbox{\boldmath $r$}'_{1}\cdots \mbox{\boldmath $r$}'_{N}
\Biggm|
{\rm T}
\exp\left(-\frac{1}{k_{B}T}\int_{0}^{L_{\tau}}{\cal H}'
(\mbox{\boldmath $g$};\tau)d\tau\right)
\Biggm|
\mbox{\boldmath $r$}_{1}\cdots \mbox{\boldmath $r$}_{N}
\right\rangle
\end{equation}
\begin{multicols}{2}
\noindent
with
\begin{eqnarray}\label{david-7.12}
{\cal H}'(\mbox{\boldmath $g$};\tau)
&=&-\frac{(k_{B}T)^2}{2\tilde{\varepsilon}_{1}}
\sum_{j=1}^N \mbox{\boldmath $\nabla$}_{j}^2
+\sum_{j=1}^N V'\left[\mbox{\boldmath $x$}_{j},\tau\right]
\nonumber\\
&&
+\frac{1}{2}\sum_{i\neq j}V_{\rm int}
\left(
\left|\mbox{\boldmath $x$}_{i}-\mbox{\boldmath $x$}_{j}\right|
\right),
\end{eqnarray}
where the constant imaginary vector potential is missing, but 
$V[\mbox{\boldmath $x$}_{j}]$ has been replaced by a $\tau$-dependent 
disorder potential
\begin{equation}\label{david-7.13}
V'\left[\mbox{\boldmath $x$}_{j},\tau\right]\equiv
V\left[\mbox{\boldmath $x$}_{j}
+\frac{\mbox{\boldmath $g$}}{\tilde{\varepsilon}_{1}}
\tau\right].
\end{equation}
The symbol ${\rm T}$ in front of the exponential in 
Eq.~(\ref{david-7.11}) stands for time ordering.
The above new Hamiltonian (related to Eq.~(\ref{david-7.8}) via an 
imaginary Galilean transformation\cite{Nelson90}) describes a set of 
vortex lines moving along the $\tau$ axis in the presence of a set of 
parallel, tilted columnar defects.
The response functions for a liquid of interacting entangled lines in 
the presence of tilted disorder have been discussed and reviewed 
by T\"{a}uber and 
Nelson.\cite{Uwe97}
The vortex tilt moduli both parallel and perpendicular to the plane 
of tilt in this case are different.
However, both are explicitly found to be {\em finite}, in contrast to 
the infinite tilt moduli in the Bose glass.
Thus there must indeed be at least one genuine phase transition for the 
$(2+1)$-dimensional system with increasing 
$\mbox{\boldmath $H$}_{\perp}\propto\mbox{\boldmath $g$}$ in the presence 
of interactions and disorder.

\section*{Acknowledgments}
The authors are grateful to Nadav Shnerb for many helpful comments 
and discussions, as well as for conversations with B. I. Halperin and 
J. Wang.
This research was supported by the National Science Foundation through
Grand No.~DMR94-17047 and by the Harvard Materials Research Science
and Engineering Laboratory through Grant No.~DMR94-00396.
One of the authors (N.H.) wishes to thank the Nishina 
Foundation for financial support during his stay at Harvard University.

\appendix
\renewcommand{\theequation}{\Alph{section}.\arabic{equation}}
\setcounter{section}{0}

\section{Exact solution of the one-dimensional Schr\"odinger equation 
with a point impurity}
\label{app-1D}

In this Appendix, we describe the derivation of the single-impurity
results given in Subsec.~\ref{sec-1D1imp-exact}.
The derivation is subtle even in the Hermitian case, since 
we are interested in finite-size effects as well as the thermodynamic limit.
Throughout this Appendix, we assume $g>0$ without loss of generality.

We have found it useful to solve the Schr\"odinger equation 
${\cal H}\psi^R(x)=\varepsilon\psi^R(x)$ 
in a way different from that sketched in Ref.~\cite{Hatano96}.
Because of the periodic boundary conditions~(\ref{31015}), 
we define the Fourier transformation in the form
\begin{equation}\label{A31021}
\psi^R(x)=\frac{2\pi}{L_{x}}\sum_{k}\tilde{\psi}^R(k)e^{ikx},
\end{equation}
where the summation runs over $k=2\pi n/L_{x}$ for integer $n$.
The inverse transformation is given by
\begin{equation}\label{A31031}
\tilde{\psi}^R(k)=\frac{1}{2\pi}\int_{0}^{L_{x}}\psi^R(x)e^{-ikx}dx.
\end{equation}
The Schr\"{o}dinger equation becomes
\begin{equation}\label{A31041}
(\hbar k+ig)^2\tilde{\psi}^R(k)-\frac{mV_{0}}{\pi}\psi^R(0)
=2m\varepsilon\tilde{\psi}^R(k).
\end{equation}
Upon assuming that $C\equiv\psi^R(0)$ is finite, we have
\begin{equation}\label{A31051}
\tilde{\psi}^R(k)=\frac{mV_{0}C}{\pi\hbar^2}
\left[\left(k+i\frac{g}{\hbar}\right)^2-K^2\right]^{-1},
\end{equation}
where $K(\varepsilon)\equiv\sqrt{2m\varepsilon}/\hbar$.
The real-space wave function is given by substituting 
$\tilde{\psi}^R(k)$ in Eq.~(\ref{A31021}) with Eq.~(\ref{A31051}).

The condition $\psi^R(0)=C$ results in the equation for the
energy spectrum $\varepsilon$; namely
\begin{equation}\label{A31061}
\frac{2\pi}{L_{x}}
\sum_{k}\left[\left(k+i\frac{g}{\hbar}\right)^2-K^2(\varepsilon)\right]^{-1}
=\frac{\pi\hbar^2}{mV_{0}}.
\end{equation}
Let us first analyze this equation in the limit $L_{x}\to\infty$:
\begin{equation}\label{A31071}
\int_{-\infty}^{\infty}dk
\left[\left(k+i\frac{g}{\hbar}\right)^2-K^2(\varepsilon)\right]^{-1}
=\frac{\pi\hbar^2}{mV_{0}}.
\end{equation}
The integrand has two poles at $k=\pm K-ig/\hbar$ in the complex plane
of $k$.
For $|{\rm Im}\;K|<g/\hbar$, the integral~(\ref{A31071}) vanishes
when we close the integration contour in the upper half plane.
Thus, nonzero solutions arise only for $|{\rm Im}\;K|\geq g/\hbar$.
In the following, we derive solutions for $|{\rm Im}\;K|>g/\hbar$ and
for $|{\rm Im}\;K|=g/\hbar$ separately.

For $|{\rm Im}\;K|>g/\hbar$, the poles straddle the real axis, and 
we have
\begin{equation}\label{A31081}
\frac{\pi i}{K}=\frac{\pi\hbar^2}{mV_{0}},
\end{equation}
or
\begin{equation}\label{A31091}
K=i\frac{mV_{0}}{\hbar^2}\equiv i\kappa_{\rm gs}.
\end{equation}
Thus we arrive at the unique localized ground-state solution for 
$|{\rm Im}\;K|>g/\hbar$ given as Eq.~(\ref{31101}).
The boundary for the existence of this localized state is the critical 
field, which leads to Eq.~(\ref{31111}).

For ${\rm Im}\;K=g/\hbar$, 
one of the poles of the integrand of Eq.~(\ref{A31071}) approaches 
the real $k$ axis, which makes the\linebreak 

\noindent
\hskip 0.2875in
\epsfxsize=2.8in
\epsfbox{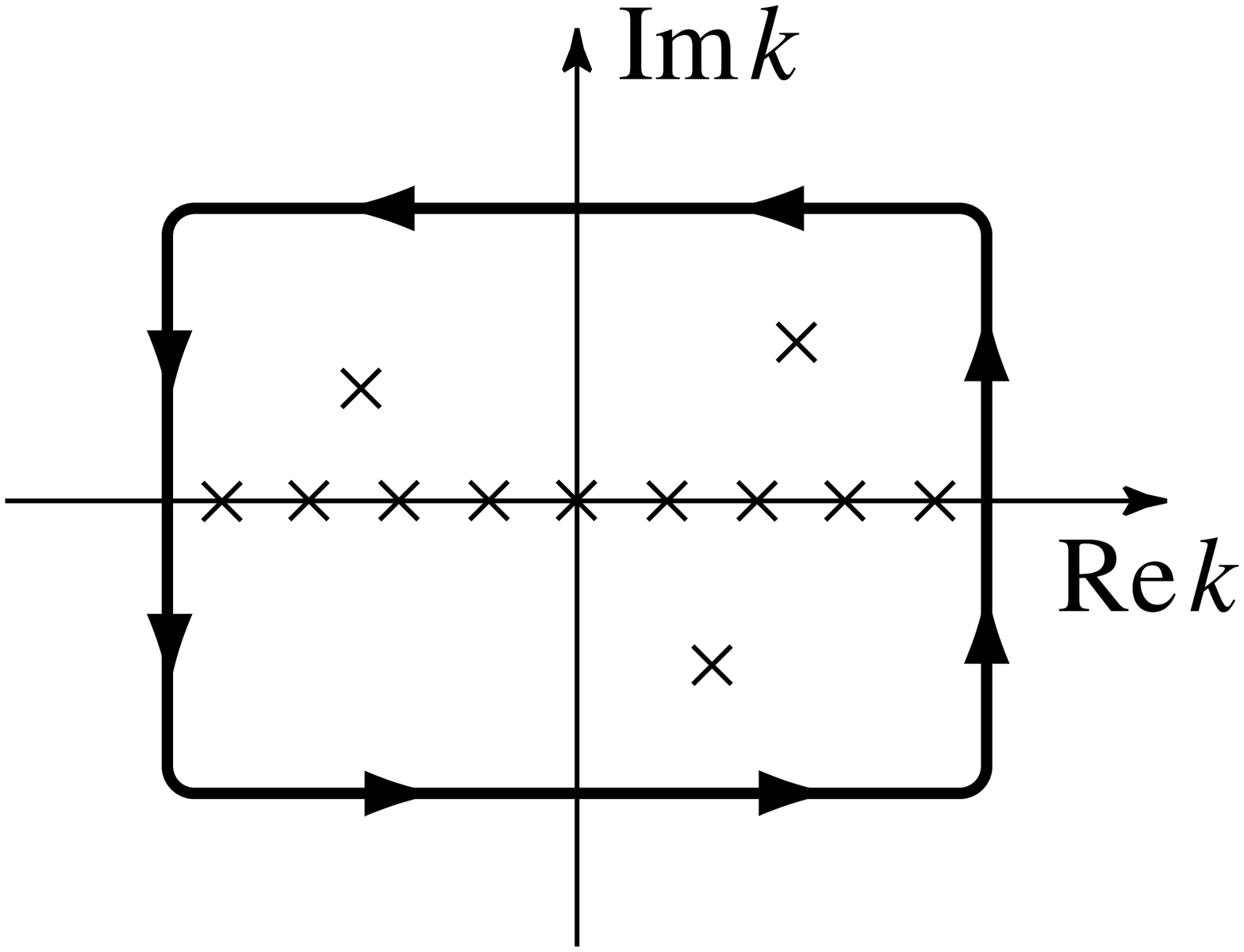}
\begin{figure}
\figcaption{
The integration contour used in the evaluation 
of Eq.~(\protect\ref{A31132}).
The crosses on the horizontal axis indicate the poles of the factor
$(e^{iLz}-1)^{-1}$, while the other two crosses indicate the poles of 
the function $R(z)$.
}
\label{figsum}
\end{figure}
evaluation of the integral 
in the limit $L_{x}\to\infty$ difficult.
Hence we return to Eq.~(\ref{A31061}), and explore the solutions 
including finite-size corrections, using
the formula
\begin{equation}\label{A31131}
\frac{2\pi}{L}\sum_{n=-\infty}^\infty
R\left(\frac{2\pi}{L}n\right)
=-\sum_{\zeta} {\rm Res}\;\left(R(z)\frac{2\pi i}{e^{iLz}-1};\zeta\right),
\end{equation}
where $R(z)$ is a rational function with the conditions
that (i) the order in $z$ of the denominator is greater than the
order of the numerator at least by two, and (ii) the function does not have 
any poles at integral points.
The summation with respect to $\zeta$ runs over the residues arising 
from all poles of $R(z)$.
The formula follows from evaluating
\begin{equation}\label{A31132}
\oint dz R(z)\frac{2\pi i}{e^{iLz}-1},
\end{equation}
with the integration contour as in Fig.~\ref{figsum}.
Thus Eq.~(\ref{A31061}) becomes
\begin{equation}\label{A31141}
\left(e^{iKL_{x}+L_{x}g/\hbar}-1\right)^{-1}
-\left(e^{-iKL_{x}+L_{x}g/\hbar}-1\right)^{-1}
=\frac{i\hbar^2K}{mV_{0}},
\end{equation}
or
\begin{equation}\label{A31151}
K\left[
\cosh \left(L_{x}\frac{g}{\hbar}\right) -\cos \left(L_{x}K\right)
\right]
+\frac{mV_{0}}{\hbar^2}\sin \left(L_{x}K\right)=0.
\end{equation}
The wave function corresponding to each solution of 
Eq.~(\ref{A31141}) is
\begin{equation}\label{A31161}
\psi^R(x)\propto
\frac{e^{iKx+xg/\hbar}}{e^{iKL_{x}+L_{x}g/\hbar}-1}
-\frac{e^{-iKx+xg/\hbar}}{e^{-iKL_{x}+L_{x}g/\hbar}-1}
\end{equation}
for $0\leq x <L_{x}$.
In the case of 
$\lim_{L_{x}\to\infty}{\rm Im}\;K(L_{x})>\frac{g}{\hbar}$,
putting $K$ to be pure imaginary results in the ground-state
solution~(\ref{A31091}) and~(\ref{31121}) 
for $g<g_{c}$ in the limit $L_{x}\to\infty$.

We are now in a position to discuss the case
\begin{equation}\label{A31163}
\lim_{L_{x}\to\infty}{\rm Im}\;K(L_{x})=\frac{g}{\hbar},
\end{equation}
where $K(L_{x})$ denotes a solution of Eq.~(\ref{A31151}).
This case includes all delocalized wave functions.
By setting 
\begin{equation}\label{A31171}
K(L_{x})=k(L_{x})+i\frac{g}{\hbar}+i\delta\kappa(L_{x})
\end{equation}
with
\begin{equation}\label{A31181}
\lim_{L_{x}\to\infty}\delta\kappa(L_{x})=0,
\end{equation}
Eq.~(\ref{A31141}) becomes
\begin{equation}\label{A31191}
L_{x}\frac{g}{\hbar}e^{-w}
\simeq L_{x}\left(\frac{g}{\hbar}-\frac{g_{c}}{\hbar}\right)+w
\end{equation}
in the limit $L_{x}\to\infty$, where
\begin{equation}\label{A31201}
w\equiv L_{x}\delta\kappa(L_{x})-iL_{x}k(L_{x}).
\end{equation}
We notice that $\kappa(L_{x})$ and $k(L_{x})$ have the same
order of magnitude with respect to $L_{x}$.

For $g\neq g_{c}$, the second term of the right-hand side of 
Eq.~(\ref{A31191}) is negligible compared to its first term, and
hence
\begin{equation}\label{A31211}
w\simeq\ln\frac{g}{g-g_{c}}.
\end{equation}
This gives a series of solutions in addition to
the ground state~(\ref{A31091}).
For $g<g_{c}$, we have
\begin{eqnarray}
L_{x}\delta\kappa&=&\ln\frac{g}{g_{c}-g},
\\
L_{x}k&=&\ln(-1)=i\pi(2n+1)
\end{eqnarray}
with $n$ an integer, {\em i.e.}\ 
the excited states given in Eqs.~(\ref{31231}) and~(\ref{31241}).
For $g>g_{c}$, we have 
\begin{eqnarray}
L_{x}\delta\kappa&=&\ln\frac{g}{g-g_{c}},
\\
L_{x}k&=&\ln 1=2i\pi n,
\end{eqnarray}
or the excited states given in Eqs.~(\ref{31251}) and~(\ref{31261}).

Finally, for $g=g_{c}$, Eq.~(\ref{A31191}) yields a solution
$w=O(\ln L_{x})$, or Eqs.~(\ref{31281}) and~(\ref{31282}).

\end{multicols}

\end{document}